\documentclass[preprint,10pt,5p]{elsarticle}
\journal{Astroparticle Physics}
\usepackage[utf8x]{inputenc}
\usepackage[toc,page]{appendix}
\usepackage{url}
\usepackage{natbib}
\usepackage{ulem}
\usepackage{graphicx}
\usepackage{mathtools}
\usepackage{epstopdf}
\usepackage{amsmath}
\usepackage{gensymb} 
\usepackage{indentfirst}
\usepackage{hyperref}
\usepackage{mathrsfs}
\usepackage{subcaption}
\usepackage{color}
\usepackage{titlesec}

\begin{document}
\begin{frontmatter}
\date{\today}

\title{Observation of Reconstructable Radio Emission Coincident with an X-Class Solar Flare in the Askaryan Radio Array Prototype Station\\}

\author[OSUPhysics] {P. Allison} 
\author[CHIBA]{S. Archambault} 
\author[UW]{J. Auffenberg} 
\author[UMD]{R. Bard} 
\author[OSUPhysics,OSUAstronomy]{J. J. Beatty} 
\author[UW]{M. Beheler-Amass} 
\author[KU,NRNU]{D. Z. Besson} 
\author[UW]{\\M. Beydler} 
\author[UNL]{C. Bora} 
\author[NTU]{C.-C. Chen} 
\author[NTU]{C.-H. Chen} 
\author[NTU]{P. Chen} 
\author[OSUPhysics]{B. A. Clark\corref{cor1}}
\ead{clark.2668@osu.edu}
\cortext[cor1]{Corresponding Authors}
\author[UNL]{A. Clough} 
\author[OSUPhysics]{A. Connolly\corref{cor1}}
\ead{connolly@physics.osu.edu}
\author[UCL]{J. Davies} 
\author[UC]{\\C. Deaconu} 
\author[UW]{M. A. DuVernois} 
\author[UMD]{E. Friedman} 
\author[HI]{B. Fox} 
\author[HI]{P. W. Gorham}
\author[Whittier]{J. Hanson} 
\author[UW]{K. Hanson} 
\author[UW]{J. Haugen} 
\author[HI]{B. Hill}
\author[UMD]{K. D. Hoffman} 
\author[OSUPhysics]{E. Hong\corref{cor1}} 
\ead{ripple80@gmail.com}
\author[NTU]{S.-Y. Hsu} 
\author[NTU]{L. Hu} 
\author[NTU]{J.-J. Huang} 
\author[NTU]{M.-H. Huang} 
\author[CHIBA]{A. Ishihara} 
\author[UW]{A. Karle} 
\author[UW]{J. L. Kelley} 
\author[UW]{ \\R. Khandelwal} 
\author[CHIBA]{M. Kim} 
\author[UNL]{I. Kravchenko} 
\author[UNL]{J. Kruse} 
\author[CHIBA]{K. Kurusu} 
\author[Weizmann]{H. Landsman} 
\author[KU] {U. A. Latif} 
\author[UW]{A. Laundrie} 
\author[NTU]{C.-J. Li} 
\author[NTU]{\\T. C. Liu} 
\author[UW]{M.-Y. Lu} 
\author[CHIBA]{K. Mase} 
\author[UMD]{R. Maunu} 
\author[UW]{T. Meures} 
\author[HI]{C. Miki} 
\author[NTU]{J. Nam} 
\author[UCL]{R. J. Nichols} 
\author[Weizmann]{G. Nir} 
\author[UC]{E. Oberla} 
\author[UW]{\\A. \'{O}Murchadha} 
\author[UD]{Y. Pan} 
\author[OSUPhysics]{C. Pfendner\corref{cor1}}
\ead{pfendner.1@osu.edu}
\author[KU]{K. Ratzlaff} 
\author[UMD]{M. Richman} 
\author[UD]{J. Roth} 
\author[HI]{B. Rotter} 
\author[UW]{P. Sandstrom} 
\author[UD]{D. Seckel} 
\author[NTU]{\\Y.-S. Shiao} 
\author[UNL]{A. Shultz} 
\author[UMD]{M. Song} 
\author[KU]{J. Stockham} 
\author[KU]{M. Stockham} 
\author[Moscow]{M. Sullivan}
\author[UMD]{J. Touart} 
\author[NTU]{H.-Y. Tu} 
\author[HI]{G. S. Varner} 
\author[UC]{\\A. G. Vieregg} 
\author[NTU]{M.-Z. Wang} 
\author[NTU]{S.-H. Wang} 
\author[CalPoly]{S. A. Wissel} 
\author[CHIBA]{S. Yoshida} 
\author[KU]{R. Young} 

\address[OSUPhysics] {Dept. of Physics and Center for Cosmology and AstroParticle Physics, The Ohio State University, Columbus, OH, USA}
\address[CHIBA]{Dept. of Physics, Chiba University, Tokyo, Japan}
\address[UW]{Dept. of Physics and Wisconsin IceCube Particle Astrophysics Center, University of Wisconsin, Madison, WI, USA}
\address[UMD]{Dept. of Physics, University of Maryland, College Park, MD, USA}
\address[OSUAstronomy] {Dept. of Astronomy, The Ohio State University, Columbus, OH USA}
\address[KU]{Dept. of Physics and Astronomy and Instrumentation Design Laboratory, University of Kansas, Lawrence, KS, USA}
\address[NRNU]{National Research Nuclear University, Moscow Engineering Physics Institute, Moscow, Russia}
\address[UNL]{Dept. of Physics and Astronomy, University of Nebraska-Lincoln, Lincoln, NE, USA}
\address[NTU]{Dept.~of~Physics,~Grad.~Inst.~of Astrophys.,~Leung Center for Cosmology and Particle Astrophys., National Taiwan Univ., Taipei, Tawian}
\address[UCL]{Dept. of Physics and Astronomy, University College London, London, UK}
\address[UC]{Dept. of Physics and Kavli Institute for Cosmological Physics, The University of Chicago, Chicago, IL, USA}
\address[HI]{Dept. of Physics and Astronomy, University of Hawaii-Manoa, Honolulu, HI, USA}
\address[Whittier]{Dept. of Physics and Astronomy, Whittier College, Whittier, CA, USA}
\address[Weizmann]{Weizmann Institute of Science, Rehovot, Israel}
\address[UD]{Dept. of  Physics  and  Astronomy,  University of Delaware, Newark, DE, USA}
\address[Moscow]{Moscow Engineering and Physics Institute, Moscow, Russia}
\address[CalPoly]{Dept. of Physics, California Polytechnic State University, San Luis Obispo, CA, USA}

\begin{abstract}
The Askaryan Radio Array (ARA) reports an observation of radio
emission coincident with the ``Valentine's Day'' solar flare
on Feb. 15$^{\rm{th}}$, 2011 in the prototype ``Testbed'' station.
We find $\sim2000$ events that passed our neutrino search criteria
during the 70 minute period of the flare,
all of which reconstruct to the location of the sun.
A signal analysis of the events reveals them to be consistent 
with that of bright thermal noise correlated across antennas.
This is the first natural source of radio emission reported by ARA that is tightly
reconstructable on an event-by-event basis.
The observation is also the first for ARA to point radio from individual events to an extraterrestrial source on the sky.
We comment on how the solar flares, coupled with improved systematic uncertainties in reconstruction algorithms, could aid in a mapping
of any above-ice radio emission, such as that from cosmic-ray air showers, to 
astronomical locations on the sky.
\end{abstract}


\end{frontmatter}


\section{Introduction}
The Askaryan Radio Array (ARA) is a radio array designed to detect
ultra-high energy (UHE) neutrinos via their radio-Cherenkov emission in ice \cite{Askaryan:1962,Askaryan:1965}.
The array consists of 150-850~MHz antennas buried in the ice at 200~m depth.
The signature of a UHE neutrino would be
an impulsive signal that originates from a neutrino-induced
cascade of charged particles in the otherwise quiet ice.

The ARA Collaboration plans to deploy a 100-km$^2$ array at $200$~m depth
in the ice near the South Pole.
So far, five deep stations have been deployed--one in 2012, two in 2013, and another two in 2018.
In 2011, prior to the deep deployments, an initial prototype ``Testbed''
station was deployed at $\sim30$~m depth to assess the level of anthropogenic 
and natural backgrounds at radio frequencies, clarity of ice near the South Pole in our frequency band,
and to test potential designs.

We report on an observation of radio emission from the sun that is coincident in time with the Feb. 15$^{\rm{th}}$, 2011 ``Valentine's Day'' solar flare. The radio emission arrives with relative time delays among our antennas that indicate a direction to its origin that tracks the sun's movement--with a 2$\degree$ degree systematic offset in azimuth, and 10$\degree$ systematic offset in elevation--over the 70 minute duration of the flare. The radiation was generated at the sun and propagated to earth, and is not the neutrino-associated Askaryan radiation the array was designed to detect.

The events are the first reported by ARA
to reconstruct radio on an event-by-event basis to an extraterrestrial source on the sky. 
As systematic uncertainties in reconstruction algorithms -- such as errors in antenna positions and modeling of the depth-dependent index of refraction -- are reduced, such events from a moving high-statistics, above ice emitter can be used to improve ARA's mapping of above-ice RF sources to locations on the sky. This will be especially applicable to studies of the direction of cosmic rays which produce geomagnetic radio emission through extensive air showers.

The solar flare events reported here are different from solar-related emission previously reported by ARA from two days earlier on Feb.~13$^{\rm{th}}$~\cite{Allison2011wk} in that they have tight reconstructions and track the sun for an extended period of time; a more extensive discussion of the 
Feb.~13$^{\rm{th}}$ flare 
can be found in Appendix~\ref{app:otherflares}.

These events also aid ARA's understanding of the radio sky. Sources of radio emission, both natural and human-made, can interfere with both triggering and data analysis.  
We would also like to be open to observations of astrophysical radio sources given ARA's unusually large field of view with a broad bandwidth.
To this end, we are able to build spectrograms that demonstrate the observation of type-II and type-IV solar radio emission that agrees with other radio observatories.  

The paper is organized as follows. In Sec. 2, we describe the ARA Testbed instrument. In Sec. 3, we outline the main features of the radio emission observed from the flare, in terms of spectra and reconstructions. In Sec. 4, we present a signal analysis of the events demonstrating evidence for the hypothesis that they are correlated thermal noise from a point source. In Sec. 5, we discuss the potential of this and future solar flare detections in ARA, namely their potential role in understanding ARA's response to above-ice RF sources like cosmic-ray air showers. In Sec. 6, we discuss a search for other solar flares with the Testbed and with deep ARA stations. In Sec. 7, we summarize our results.

\section{The ARA Testbed Instrument}
The Askaryan Radio Array (ARA) ``Testbed" instrument is an array of 14 ``high"-frequency (150-850 MHz) 
broadband antennas buried up to 30 m deep, and two ``low"-frequency (30-300 MHz)
antennas placed at the surface.
All are located at the South Pole with the goal of detecting ultra-high energy 
neutrinos via their radio-Cherenkov emission in ice. 
The Testbed was a prototype instrument deployed in January 2011 
and operated until January 2013. 
The station's main purpose was to assess the radio noise environment,
measure ice properites, and explore design choices later utilized in a mature station;
a full instrument description of the Testbed is available in \cite{Allison2011wk}. 
Of the 14 higher-band antennas,
the analysis presented in this paper primarily uses data from eight which were buried at a depth of $\sim$30~m in ``boreholes":
four vertically polarized (VPol) bicone antennas and four horizontally polarized (HPol) bowtie-slotted cylinder antennas.
These antennas were selected for the analysis because, 
(a) for each polarization, the antennas have the same basic dipole-like design 
(bicone for VPol antennas, bowtie-slotted cylinder for HPol antennas), 
and (b) they have the same sampling rate of 2 GHz.

The station consists of two primary elements;
``downhole" components which receive and transmit the signal,
and ``surface" components which provide data acquisition and system control. 
Downhole components include the antennas, signal conditioning, and transmission to the surface. 
Surface components include additional signal conditioning,
and triggering and digitization electronics. 

The RF signal chain provides amplification and filtering 
in preparation for digitization and triggering. 
In the ice, the signal undergoes $\sim$ 40~dB of low-noise amplification,
a 450 MHz notch filter to remove South Pole communications signals,
and is transmitted to surface electronics via radio-over-fiber (RFoF). 
At the surface, a transceiver returns it to electronic RF,
and the signal undergoes an additional 40 dB of second stage amplification,
a 150-850 MHz band-pass filter to remove out-of-band amplifier noise,
and a splitting to trigger and digitizing paths. 

In ARA, the trigger is designed to identify transient impulsive events 
and the digitizer records these signals with high fidelity. 
The trigger path routes to a tunnel diode, 
which functions as a $\sim$ 5 ns power integrator. 
Any signal exceeding 5-6 times the mean thermal noise power 
in 3 out of 8 borehole antennas within a 110 ns time window 
triggers the digitization of all 16 antennas in the array.
Signal to be digitized is sampled at 2 (1) GHz for 
borehole (shallow) antennas with a 12-bit custom digitizer.
The digitization window is 250 ns wide,
centered to within 10 ns on the trigger time. 
Impulsive events, which produce large integrated power over short time windows,
naturally satisfy this trigger. 
The trigger will also fire if the blackbody thermal noise background of the ice 
randomly fluctuates high in a sufficient number of channels,
or if some radio loud source illuminates the array.

\section{Characteristics of Observed Radio Emission}
\label{characteristics}

The ARA team first noticed the events of interest in March~2014 during 
the final stages of a data analysis aimed at searching for UHE neutrinos 
in the ARA prototype Testbed station~\cite{Allison201562}. 
In that analysis, we searched for events that did not have strong 
continuous-wave (CW) contamination, and that originated in the ice below the Testbed station.  
Their location of origin was determined through a peak in the cross-correlation of 
waveforms from different antennas across the station, 
accounting for delays in any hypothesized direction of origin.

The events of interest were discovered during final tests performed in the neutrino search,
and were determined to originate from the Feb.~15$^{\rm{th}}$ flare due 
to time coincidence and directional reconstruction that tracked the sun's location.  
By removing the original search requirement that the event's origin be in the ice below the station,
we found 244 events were recorded in the time period from 02:04-02:58  Feb. 15$^{\rm{th}}$ UTC 2011 in only 10\% 
of the dataset used until ``unblinding'' the final  stage of analysis.
When the full 100\% ublinded data set is examined, 2323 events between 02:01-02:58 AM UTC pass cuts.
This time period was found to be coincident with the February 15$^{\rm{th}}$, 2011 X-2.2 solar flare,
which the RHESSI Flare Survey webpage~\cite{RHESSI} reports to have begun at 01:43:44 UTC the same day in X-ray measurements. These events on Feb. 15$^{\rm{th}}$ follow closely events from an M-class flare on Feb. 13$^{\rm{th}}$, which were reported on in the prototype instrument paper and are discussed further in Appendix~\ref{app:otherflares}.

\subsection{Trigger Rates and Spectra}
\begin{figure}[ht]
\centering
\includegraphics[width=\linewidth]{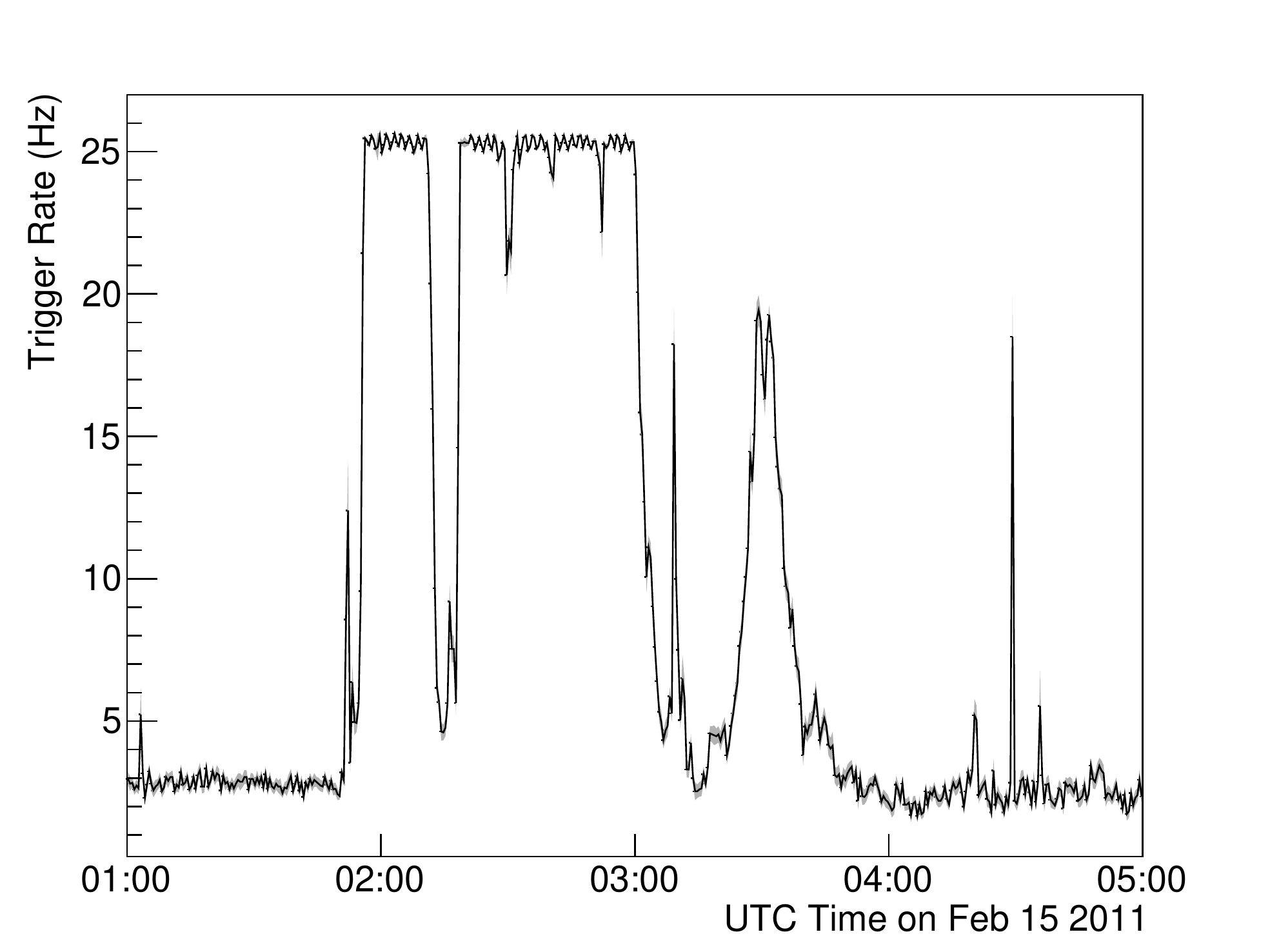}
\caption{The Testbed global trigger rate during the solar flare. Before and after the flare, the instrument triggers near the pre-determined ``thermal-noise riding" rate  \citep{Allison2011wk}. The black line is the mean trigger rate per thirty second time bin, and the grey band is the standard error on that mean (that is, the square root of the number of entries per bin).}
\label{fig:trigrates}
\end{figure}

Figure~\ref{fig:trigrates} shows that during the hour over which these 244 events were observed in
10\% of the data, our trigger rates rose from a few~Hz during a normal period to as much as 25~Hz.  
This maximum trigger rate is set by the deadtime during readout whenever a trigger is fired.

\begin{figure}[ht]
\centering
\includegraphics[width=\linewidth]
{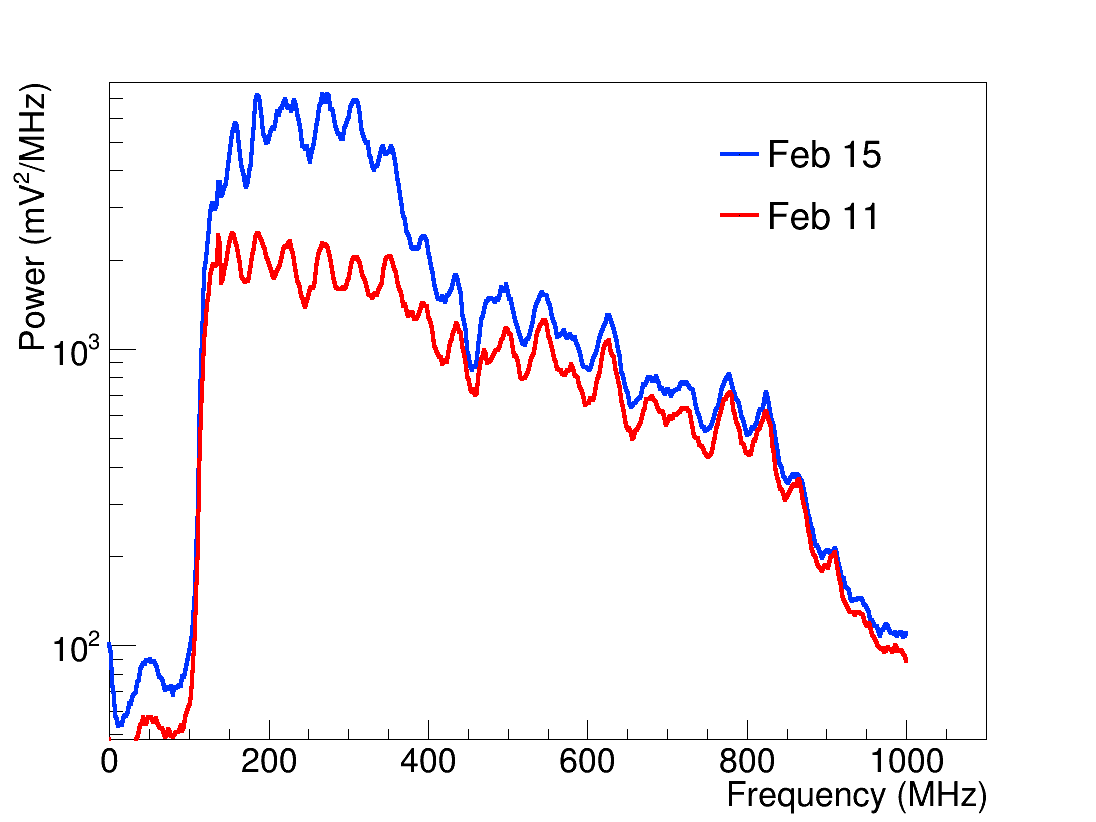}
\caption{Average spectra of Channel 2 (VPol) for the 2323 events that pass analysis cuts designed for the neutrino search on Feb 15$^{\rm{th}}$ (blue) and 2300 events between 2 AM and 3:00 AM that are thermal noise forced triggers (red) from Feb 11$^{\rm{th}}$. This shows the power excess observed between 200-400 MHz during this particular time of the flare. The small scale oscillations, present in both the flare and thermal noise sample, are attributed to instrumental characteristics. The lines are the mean of the distribution of spectral amplitudes in a given frequency bin; the distribution is an exponential, as is expected for the power spectrum of thermal noise plotted in the units of mV${^2}$/MHz.}
\label{fig:average_spectra-flarevsquiet}
\end{figure}

In Fig.~\ref{fig:average_spectra-flarevsquiet}, we show average spectra from one antenna (VPol) for
the 2323~events observed in 100\% of the data,
compared with an average spectra over 2300 events from the forced trigger sample on Feb 11,
dominated by thermal noise.
One can see that over the period of the flare, the excess emission is predominantly
in the 200-400~MHz portion of the band.

\begin{figure}
\centering
\includegraphics[width=\linewidth]{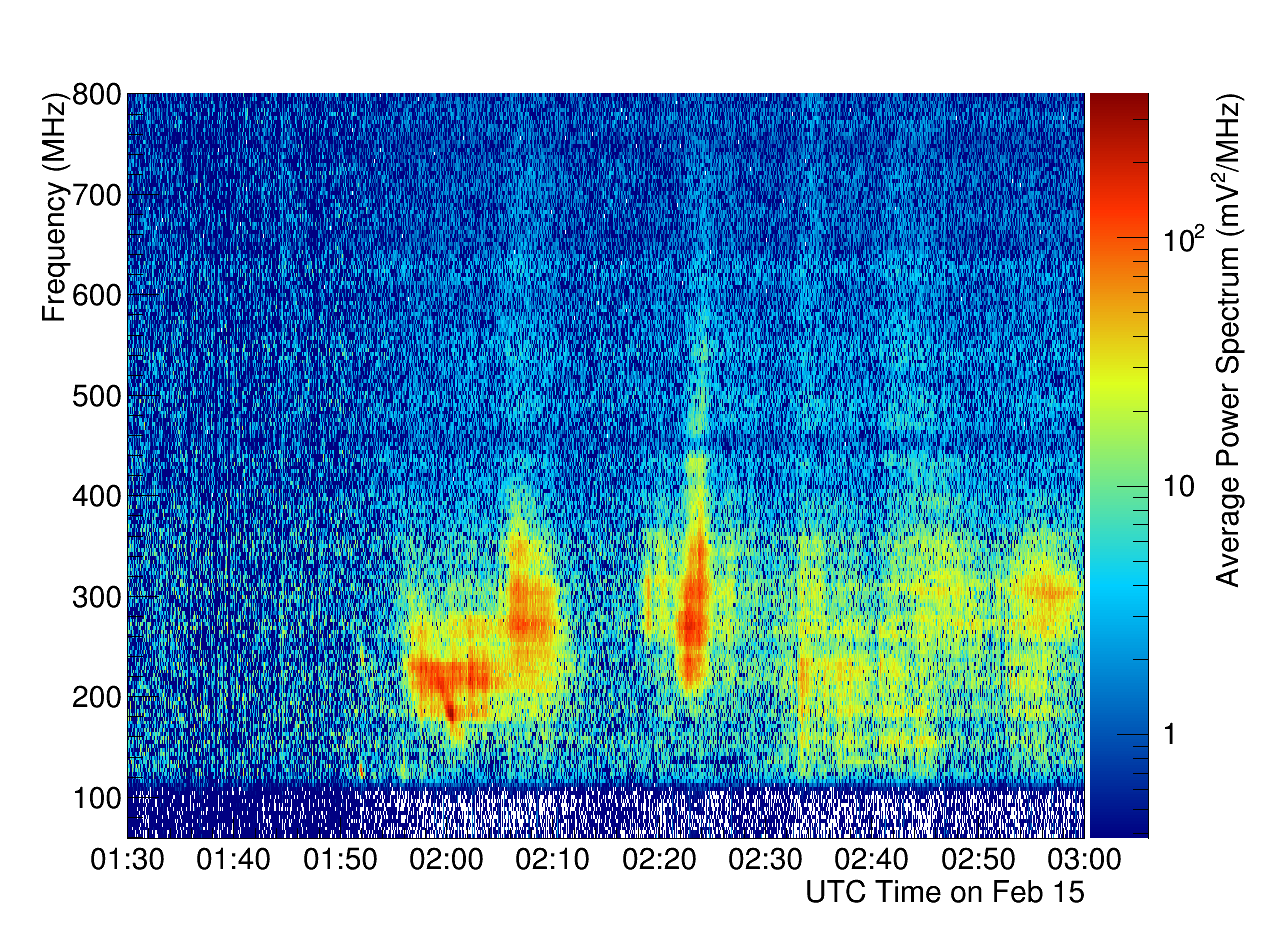}\hfill
\includegraphics[width=\linewidth]{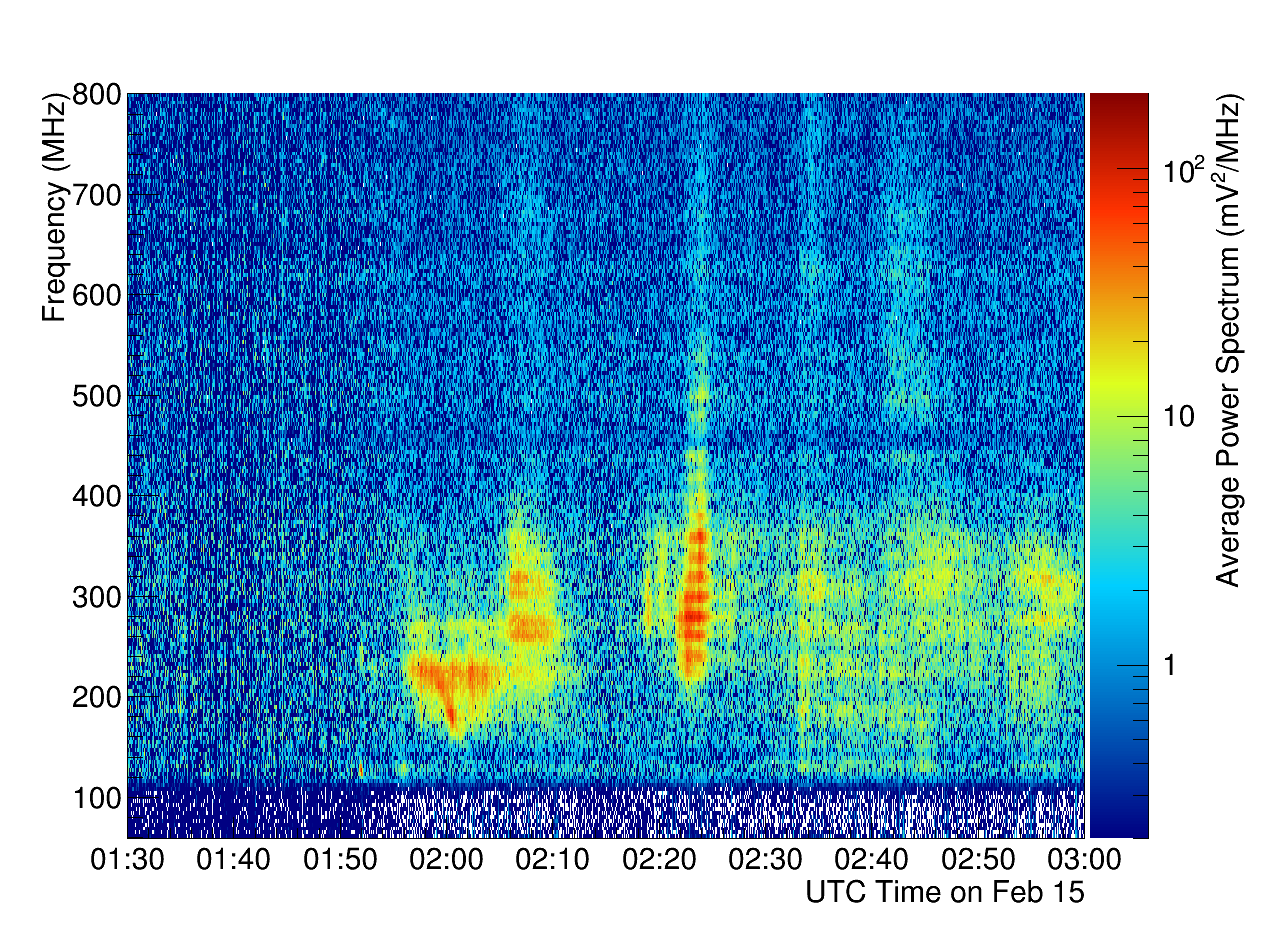} 
\caption{\label{fig:spectrograms_flareperiod} (Top) A background-subtracted spectrogram of Channel 2 of the Testbed, for the hour from the beginning of the flare. The background sample is taken from an identical time period on Feb. 11$^{\rm{th}}$ (four days prior) when the sun was not flaring. (Bottom) A background-subtracted spectrogram produced from the coherently-summed waveform from two channels (Channels 3 and 4) given the delay relative to one another that gives the strongest cross-correlation value. Note that the coherently-summed spectrogram demonstrates that many features are shared between antennas. We only use two channels, as opposed to four channels combined with a directional hypothesis, to avoid a reduction in the signal strength due to any timing misaligments for a given directional hypothesis.
}
\end{figure}

We create a spectrogram for the 1:30-3:00~AM UTC period and 
compare with those made by solar radio astronomers e.g. \cite{White, Liu2007}. 
In Fig.~\ref{fig:spectrograms_flareperiod},
we plot a background-subtracted spectrogram of a borehole 
VPol channel over the entire hour of the flare.
The background spectrogram is constructed from an identical time window on 
February 11$^{\rm{th}}$, 2011, 
when the sun would have been in approximately the same position,
but was not active.

The radio emission begins to saturate our trigger at approximately 1:57~AM UTC,
roughly coinciding with the peak of the soft x-ray flux.
Such coincidence is a feature typical of type-II flare radiation \cite{White}.
Two reference measurements of the peak soft x-ray flux are given by the RHESSI instrument (3-10~keV),
which reports peaking at 1:55~AM UTC \cite{RHESSI} and by the Fermi 
GBM ($>$10~keV coinciding with GOES observations) \cite{FermiExplanation},
which reports peaking at 1:59~UTC \cite {FermiGBM}.

The spectrograms, and their associated features between 150 and 500 MHz, agree with
other radio observatories that classify this flare emission as type-II followed by type-IV.
The Culgoora telescope, with sensitivity from 18-1800 MHz,
records spectrograms comparable to our own~\cite{CulgooraSpectrograph}
and lists this flare in their type-II catalog \cite{CulgooraType2}.

\subsection{Directional Reconstruction}

\label{sec:directional_reconstruction}

The most striking feature of these events is how well they ``reconstruct'' uniquely to the sun,
tracking the motion of the sun in azimuth during that hour.  
In other words, considering all hypothesized directions across the sky,
we found the highest cross-correlation values in a direction within 2$^{\circ}$ in azimuth 
of the sun, without distinctly different directions also giving competitive cross-correlations for the same event. The azimuthal angle of the reconstruction peak has a 2$^{\circ}$ systematic offset from the true value of the sun's azimuth.
The events also track the solar position in elevation, but with a significant systematic offset, appearing $\sim 10 \degree$ higher in the sky than the true solar elevation.
In this section, first we will review the method that we use to calculate cross-correlation values associated with
positions on the sky before showing cross-correlation maps for a typical flare event.
We also describe our calculations of coherently summed waveforms that we use to investigate the nature of the correlated
noise component of these events.

\subsubsection{Reconstruction in ARA Analyses}
In order to determine the direction of the source of radio signals,
we use an interferometric technique similar to the one used in a search for a flux of diffuse neutrinos using data from the ARA Testbed
station that takes into account the index of refraction of the ice surrounding the antennas~\cite{Allison201562}.
We first map the cross-correlation function from pairs of waveforms 
from two different antennas to expected arrival delays from different putative source directions.
For each direction on the sky, the mapped cross-correlations from many pairs of antennas 
are added together and where the delays between different pairs of antennas signals have the strongest agreement,
there is a peak in the correlation map.

\begin{figure*}[ht]
\centering
 \includegraphics[width=.48\linewidth]{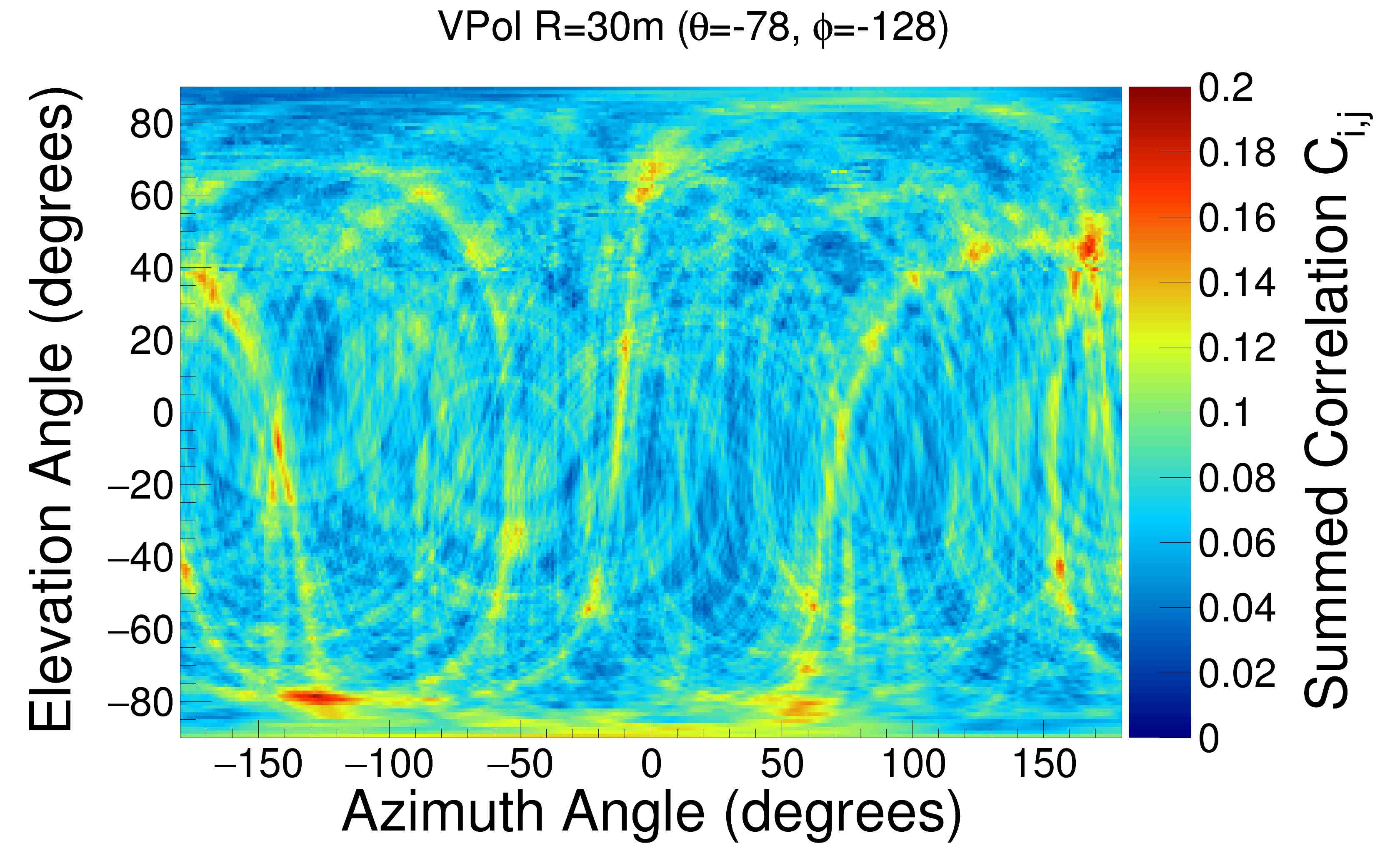}\hfill
 \includegraphics[width=.48\linewidth]{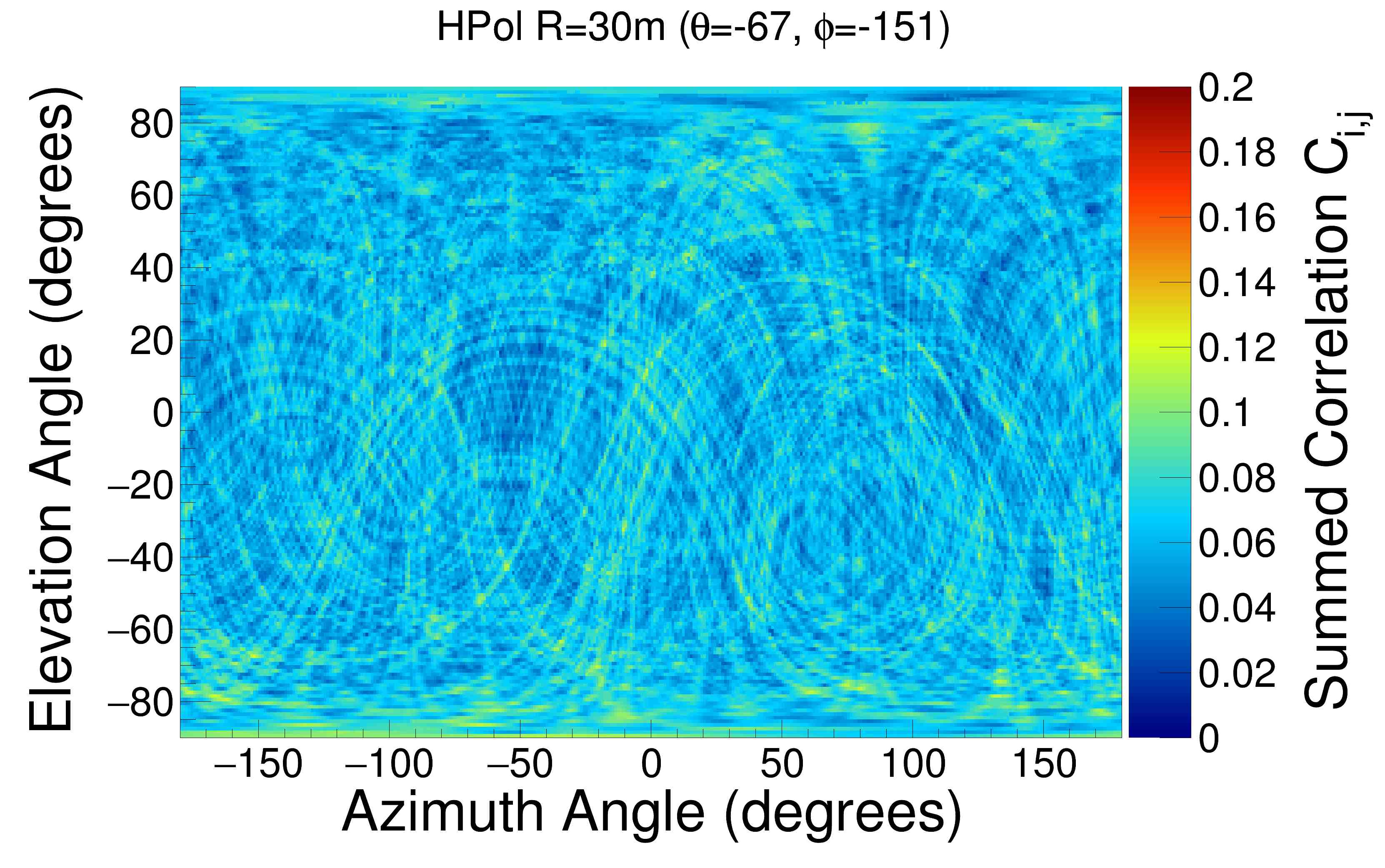}\hfill
 \includegraphics[width=.48\linewidth]{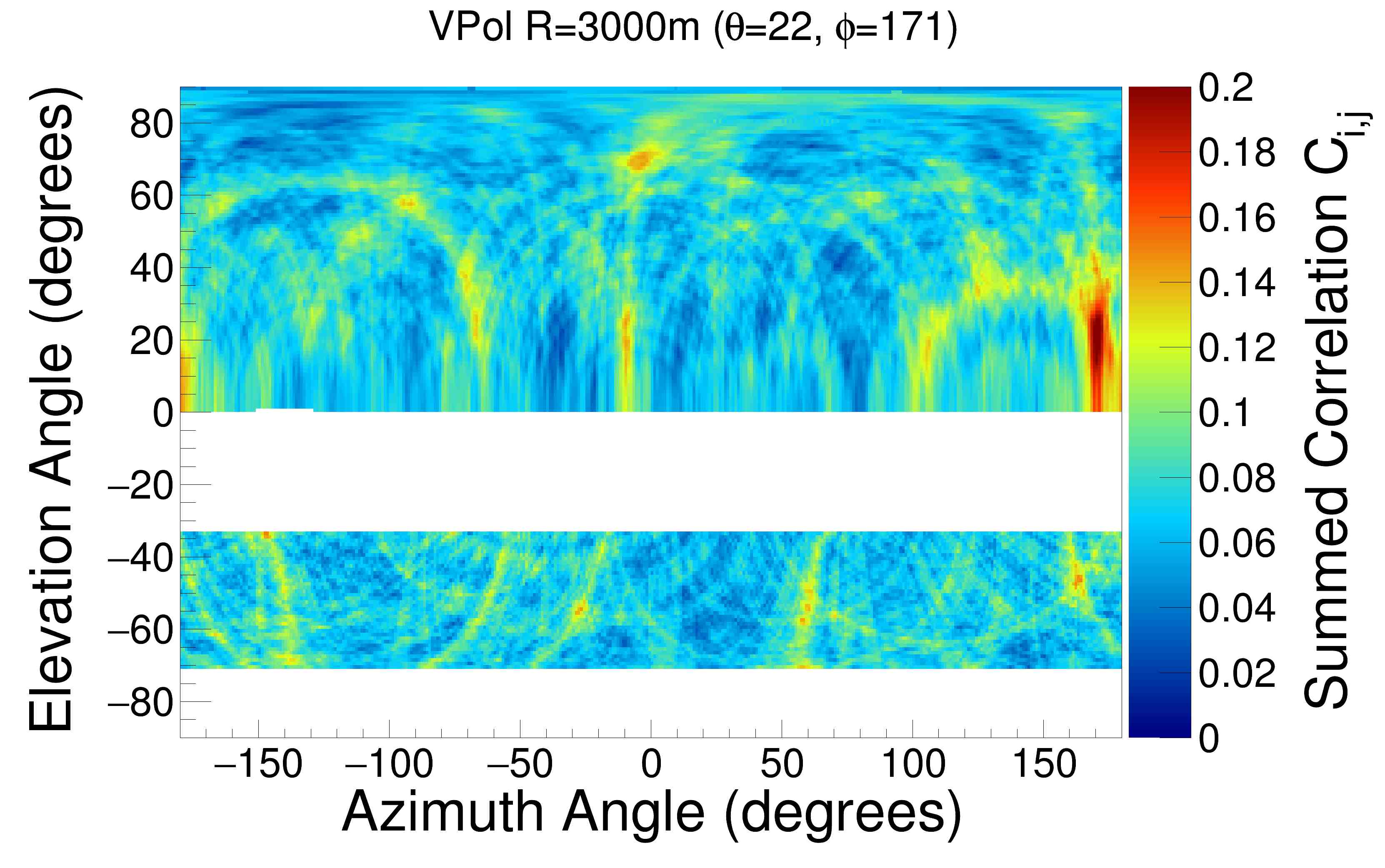}\hfill
 \includegraphics[width=.48\linewidth]{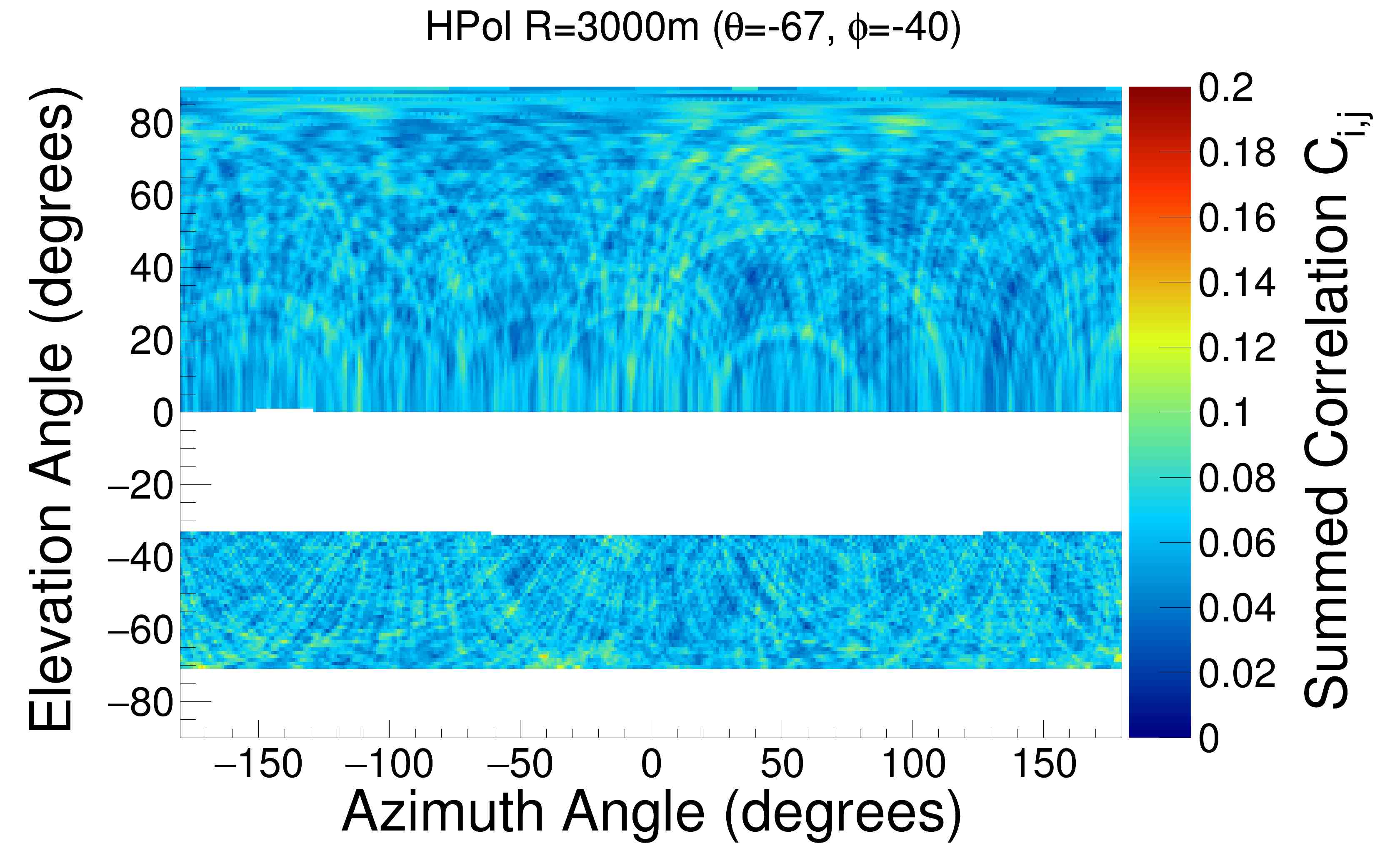}
 \caption{Map of cross-correlation values associated with directional hypotheses across the sky for the same event 
from the flare period
as the one shown above in Figs.~\ref{fig:wf_flareevent_1} and~\ref{fig:sp_flareevent_1}.  The upper two figures show reconstruction maps for
the hypothesis of 30~m distance, while the lower two figures show reconstruction maps for the 3000~m distance hypothesis. The maps are created in the Testbed local coordinate system. The white gaps in the 3000~m maps correspond to areas of ice where there is either no ray tracing solution (upper white band) or where there is bedrock (lower white band). }
\label{fig:reconstructions_flareevent_1}
\end{figure*}

For a given pair of antenna waveforms, the cross-correlation between the voltage waveform on the $i$-th antenna ($f_i(t)$) and the voltage waveform on the $j$-th antenna (($g_j(t)$) as a function of time lag $\tau$ can be found from Eq. \ref{equ:singlecorr}:
\begin{equation}
C_{i,j}(\tau)=\sum_{t=-\infty}^{\infty}f_i(t)g_j(t+\tau)
\label{equ:singlecorr}
\end{equation}
The time lag $\tau$ depends on the position of the source relative to the array, characterized as
elevation angle $\theta$, azimuthal angle 
$\phi$, and the distance $R$ to the source;
the origin of this coordinate system is defined as the average of the positions of the antennas contributing to the map.
As in~\cite{Allison201562}, we only consider two hypothesized distances, 
30~m and 3000~m. In the original diffuse analysis, these were chosen because 30~m is roughly the distance of the local calibration pulser, and 3000~m is a estimate for a far-field emitter like a neutrino interaction.
The total cross-correlation strength for a given point on the
 sky ($\theta$,$\phi$) is given by summing over all like polarization pairs of antennas as in Eq. \ref{eq:skycorr}
\begin{equation}
C(\theta,\phi; R)=\sum_{i=1}^{n_{ant}-1}\sum_{j=i+1}^{n_{ant}}C_{i,j}[\tau(\theta,\phi; R)]
\label{eq:skycorr}
\end{equation}
Plotting this total cross-correlation strength for all points on the sky 
yields a correlation map, as is shown in Fig.~\ref{fig:reconstructions_flareevent_1}. The azimuth and elevation correspond to the position on the sky of a putative emitter relative to the station center.

The arrival delays $\tau$ are found by calculating a path from a hypothesized source location 
to an antenna through an ice model that accounts for the changing index of the Antarctic firn.
We consider a constant $n=1$ index of refraction in the air.
The ray-tracing method models the changing index of refraction as follows:
\begin{equation}
n(z)=1.78 - 1.35 e^{0.0132z}
\end{equation}
This index of refraction model is taken from a fit to data taken by the RICE experiment \cite{2004JGlac..50..522K},
and is the same as the one used in \cite{Allison201562}.
To smooth uncertainties in this ice model and other systematics, 
we calculate the Hilbert envelope of the cross-correlation function before summing over pairs,
as is done in previous analyses \cite{Allison201562, Lu2017}.

\subsubsection{Reconstructed Directions Track the Sun}

Figs.~\ref{fig:reconstructions_flareevent_1}
shows maps of cross-correlation values obtained by 
considering directions across the sky for the same event
as shown in Figs.~\ref{fig:wf_flareevent_1} in the Appendix.
Comparison plots from a quiescent period can be found in 
Appendix~\ref{sec:supporting}.
Note that we find the strongest cross-correlation values 
when we consider the $R=30$~m hypothesis rather than the $R=3000$~m hypothesis. Although, the peak in the cross-correlation map has an elevation that is closer to the true elevation of the sun in the $R=3000$~m case.

We attribute the offsets in the azimuth and elevation of the reconstruction peak, as well as the higher correlation peak value at the near radius, to coupled systematics uncertainties associated with our reconstruction algorithms. This includes the orientation of the local surface normal $\hat{n}$, the parameterization of the depth-dependent index of refraction $n(z)$, variations of the index-of-refraction at the surface due to blowing snow and human activity, unknown cable delays at the level of 1-2 ns, and surveying uncertainties at the level of 10-20 cm. The relative importance of these various systematics is under ongoing study.

\begin{figure*}
\centering
\hspace{0.2in} 
\includegraphics[width=\linewidth]{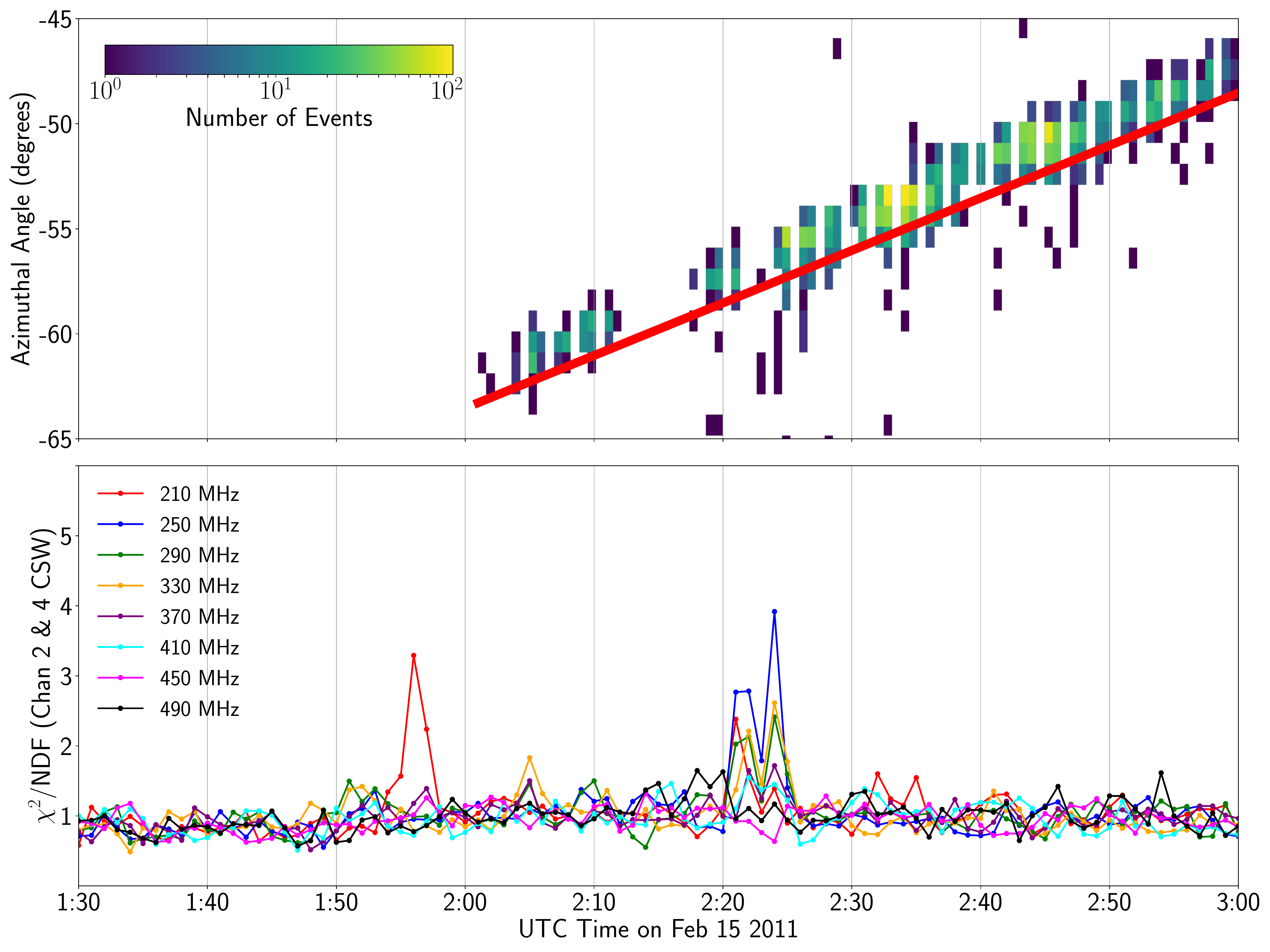}
\caption{(Top panel) Reconstructed direction in azimuth of events occurring during the flare, and the expected location of the sun drawn as a red line. The azimuthal angle here follows the continent global coordinate system described in the Appendix. (Bottom panel) The $\chi^2$/NDF values for fits of the spectral amplitudes to Rayleigh distributions of the CSW formed from channel 2 and 4 (see section  \ref{sec:thermal_noise_correlations})  given the delay relative to one another that gives the strongest cross-correlation value. We only use two channels to avoid a reduction in the signal strength due to any timing misalignments for a given directional hypothesis. Note the increased deviation from the Rayleigh fit in the few minutes after 2:20 UTC which matches a period when the events no longer pass the analysis cuts. }
\label{fig:solar_tracking}
\end{figure*}

The top panel of Fig.~\ref{fig:solar_tracking} show the reconstructed directions 
for all events passing cuts as a function of time for the duration of the flare,
taken from the peak of the maps for the $R=3000$~m hypothesis. 
The events during this time period reconstructed 
to a direction within a few degrees in azimuth
and approximately 10$^{\circ}$ in elevation angle of the sun's position in the sky,
appearing higher than the known location of the sun (13.3$^{\circ}$).
We note that the second increase in trigger rates in Fig. \ref{fig:trigrates} around 3:30 AM UTC does include a set of $\sim200$ events that also reconstruct to the sun, but do not pass any analysis cuts. We do not study them in depth, and instead focus on the emission from 1:50-3:00 AM UTC where many events pass analysis cuts and we have direct overlap with results from other solar observatories.

\subsection{Coherently Summed Waveforms}
In addition to building a correlation map, we can construct coherently summed waveforms (CSWs) 
from antennas of like polarization by first delaying the waveforms with respect to each other by a $\tau$
that is specific to an antenna pair, and then summing.
For a set of $\tau$'s corresponding to the direction of the source, 
the CSW should have uncorrelated noise suppressed relative to any correlated component.

If we wish to observe this CSW under a source direction hypothesis, 
we derive the lags $\tau$ from the ray-tracing method described above.
We can also find the CSW in a way that does not depend on the index of refraction model.  
To do this for a pair of antennas $i$ and $j$, 
we consider all potential lags $\tau$ and choose the one that provides the highest $C_{i,j}$.

\section{Correlated Thermal Noise from the Sun}
\label{sec:thermal_noise_correlations}
We conclude that the source of the reconstruction of the events
is the presence of correlated thermal noise between antennas.
This is consistent with the sun acting as a bright thermal point source
that illuminated all antennas.
The observation that a thermal source generates fields that are correlated between spatially separated detectors is expected and understood from classical theories of statistical optics; for example, see \textit{Goodman} \cite{goodman} Chapters 5 and 6.

Thermal noise would be characterized by spectral amplitudes, 
in a given frequency bin, whose distribution over many events would follow Rayleigh distributions~\cite{goodman}.
To verify this thermal nature, we looked at the spectral amplitudes for different 
frequency bins and attempted to fit each against a Rayleigh distribution.
For a sliding three-minute time bin across the duration of the flare,
we fit the distribution of spectral amplitudes $x$ in a given frequency bin to a Rayleigh:
\begin{equation}
H(x)=\dfrac{Ax}{\sigma^2} \exp{-\dfrac{x^2}{2\sigma^2}}.
\label{eq:rayleighpt1}
\end{equation} 
as was done to characterize the noise environment in other analyses \cite{Allison201562, Allison:2015eky}.
The normalization of the fit $A$ and the characteristic width parameter $\sigma$ are allowed to float.

\begin{figure}[ht]
\centering
  \includegraphics[width=0.5\linewidth]{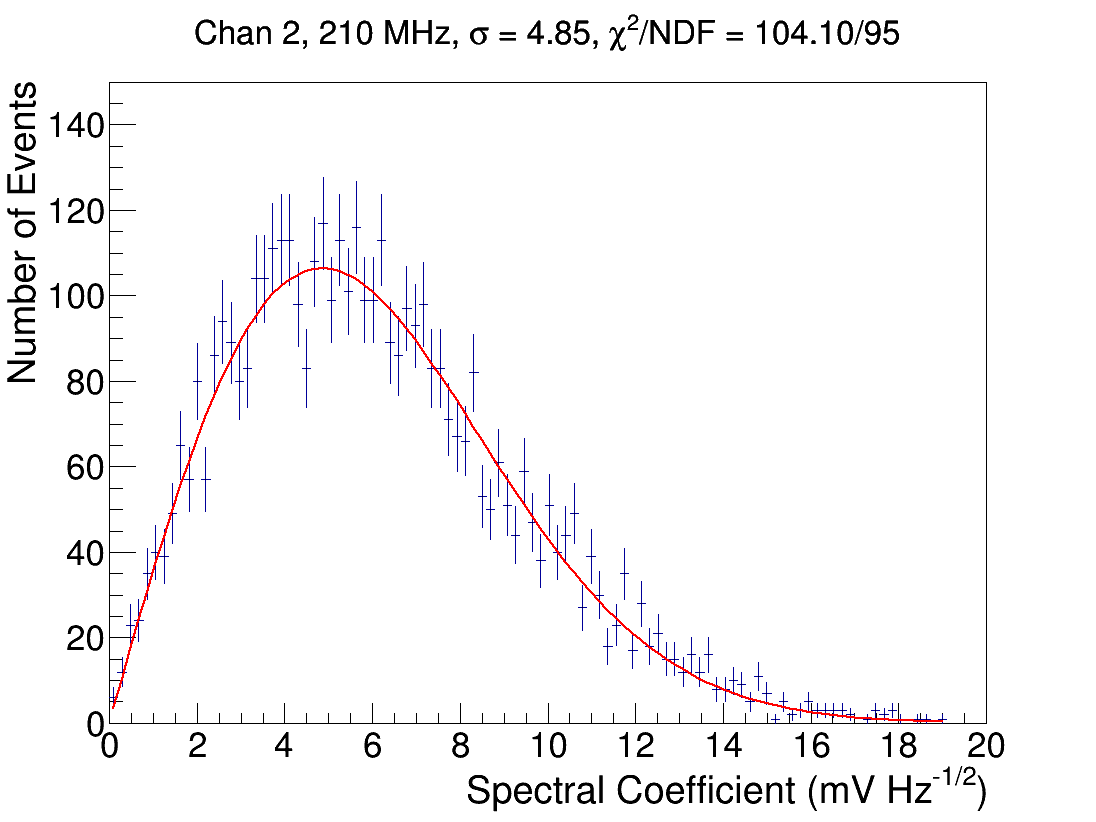}\hfill
 \includegraphics[width=.5\linewidth]{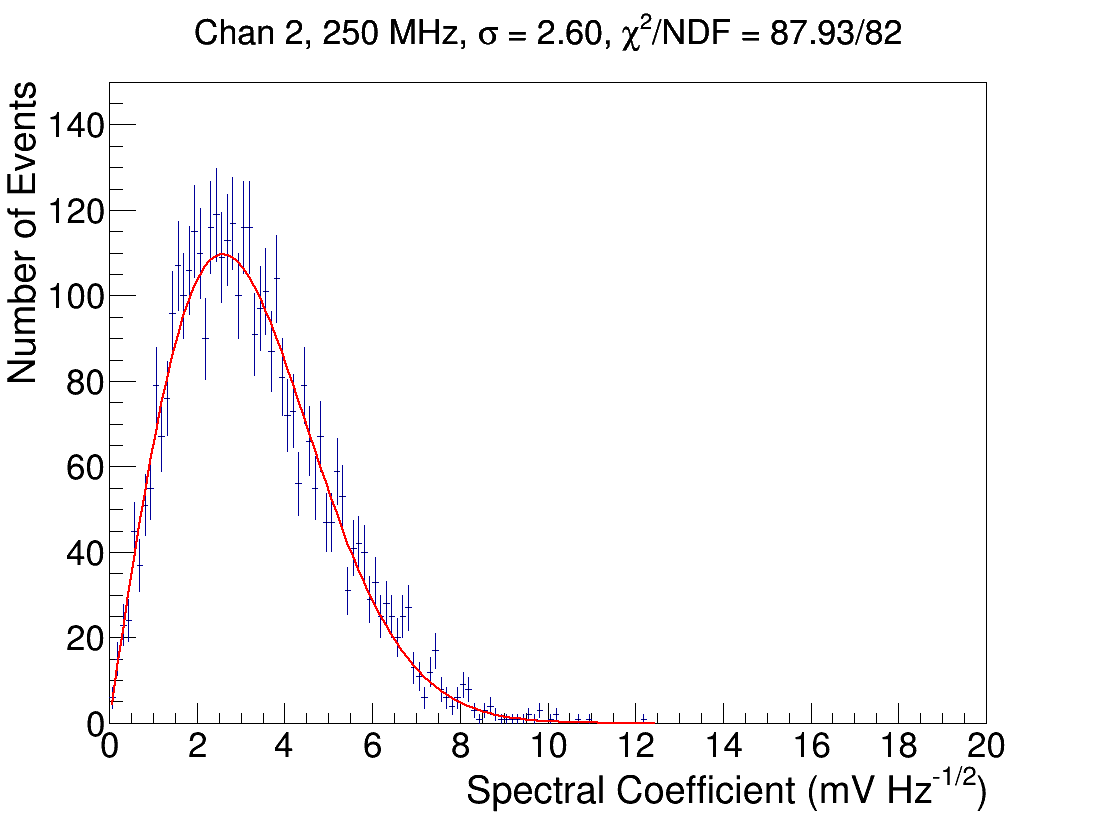}\hfill \\
  \includegraphics[width=.5\linewidth]{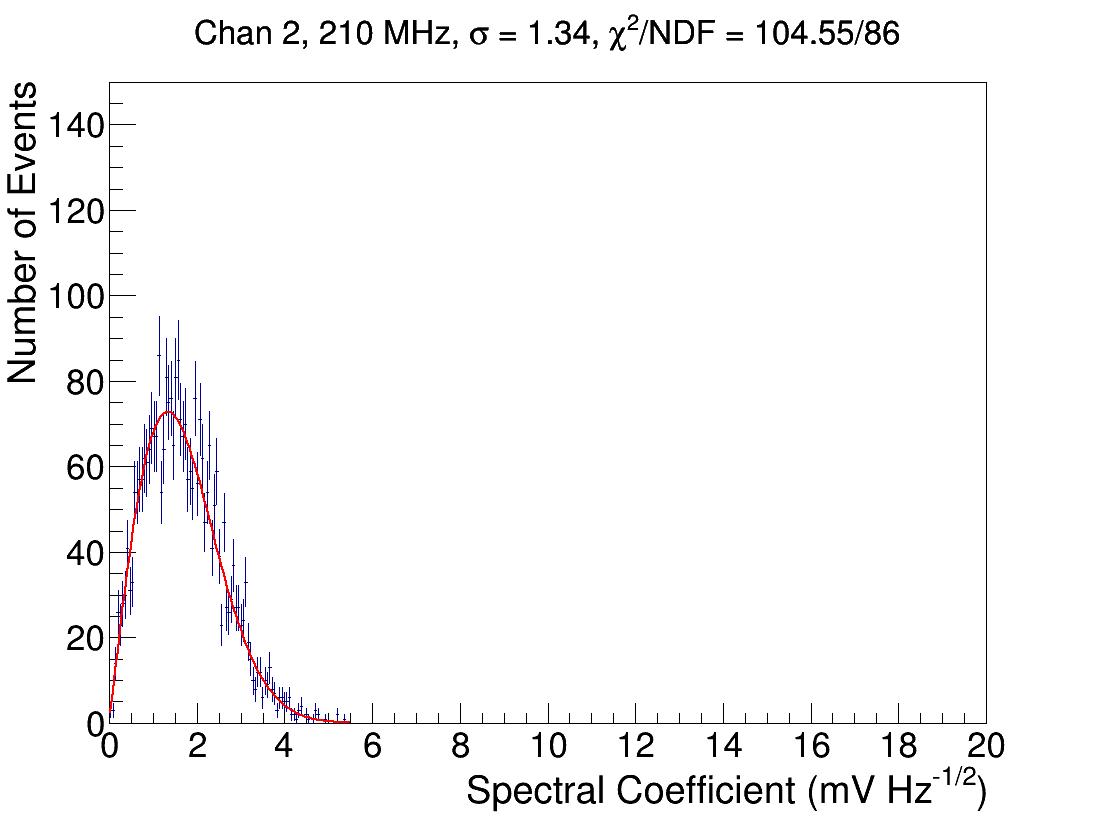}\hfill
 \includegraphics[width=.5\linewidth]{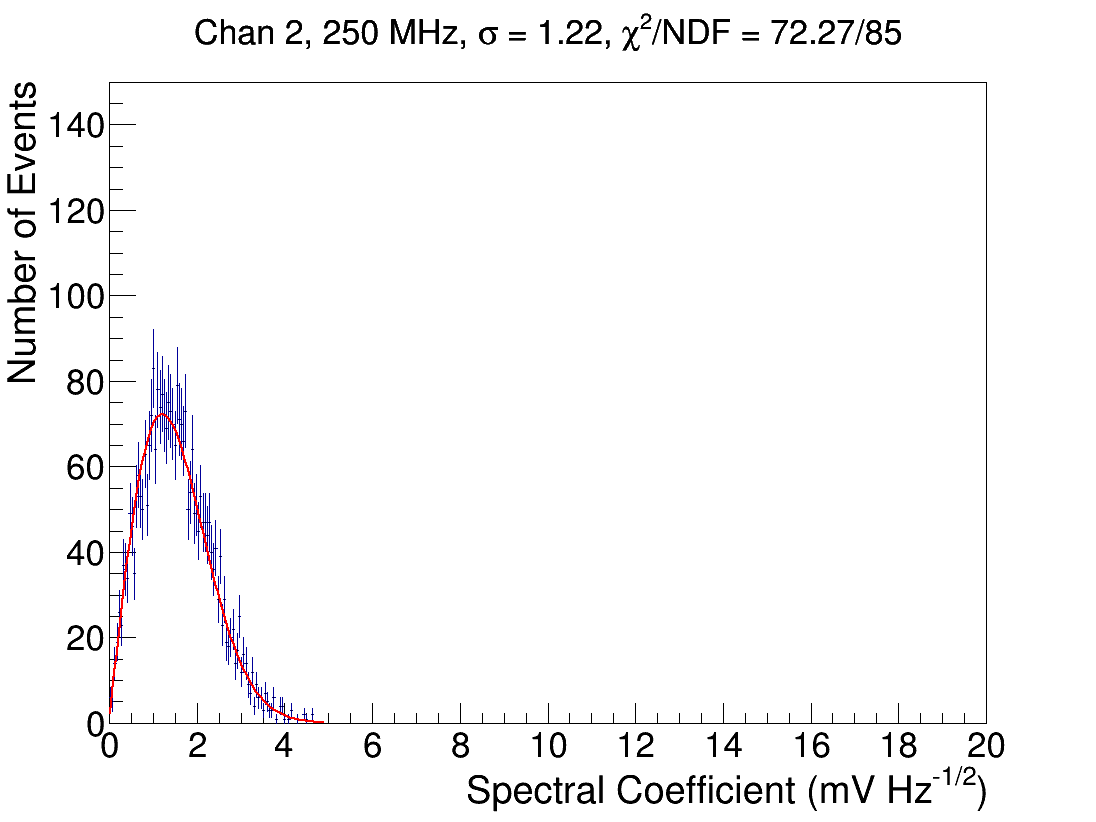}\hfill
 \caption{(Top): Distributions of amplitude spectra at two frequencies where the emission was strong, 210~MHz and 250~MHz, during the period of flare. These events  satisfied the RF trigger between 1:59 and 2:01 AM UTC. (Bottom): The same as the top figure, but during a time when the sun was not active.   These events are from the forced triggered data sample recorded on February 11$^{\rm{th}}$, 2011 between 1:40~AM and 3~AM.}
\label{fig:rayleighs_flare_210_250}
\end{figure}

The top panel of Fig.~\ref{fig:rayleighs_flare_210_250} shows two spectral amplitude distributions 
for 210 and 250 MHz with their best fit Rayleigh superimposed. 
For comparison, the bottom panel of Fig.~\ref{fig:rayleighs_flare_210_250} shows a set of Rayleigh distributions from February 11$^{\rm{th}}$, 2011 when the sun was not active.
The quality of the fit is evaluated by computing the reduced $\chi^2$ test statistic for each spectral and time bin. 
The distribution of the reduced $\chi^2$ test statistic is Gaussian and centered about 1 for all time bins except for the five minutes after 2:20~AM.
This is shown in the bottom panel of Fig.~\ref{fig:solar_tracking} for a two-channel CSW (described below).
Note that this short period coincides with a gap in the passing events that track with the Sun's position.
Note also that the $\sigma$ parameter, which represents the power in a spectral bin,
is higher on Feb 15 than on Feb 11, consistent with the brightened thermal emission over background
observed in Sec. \ref{characteristics}. Fits for a larger variety of frequencies on February 15$^{\rm{th}}$ are available in Appendix \ref{fig:many_rayleigh}.

\begin{figure}[ht]
\centering
\includegraphics[width=\linewidth]{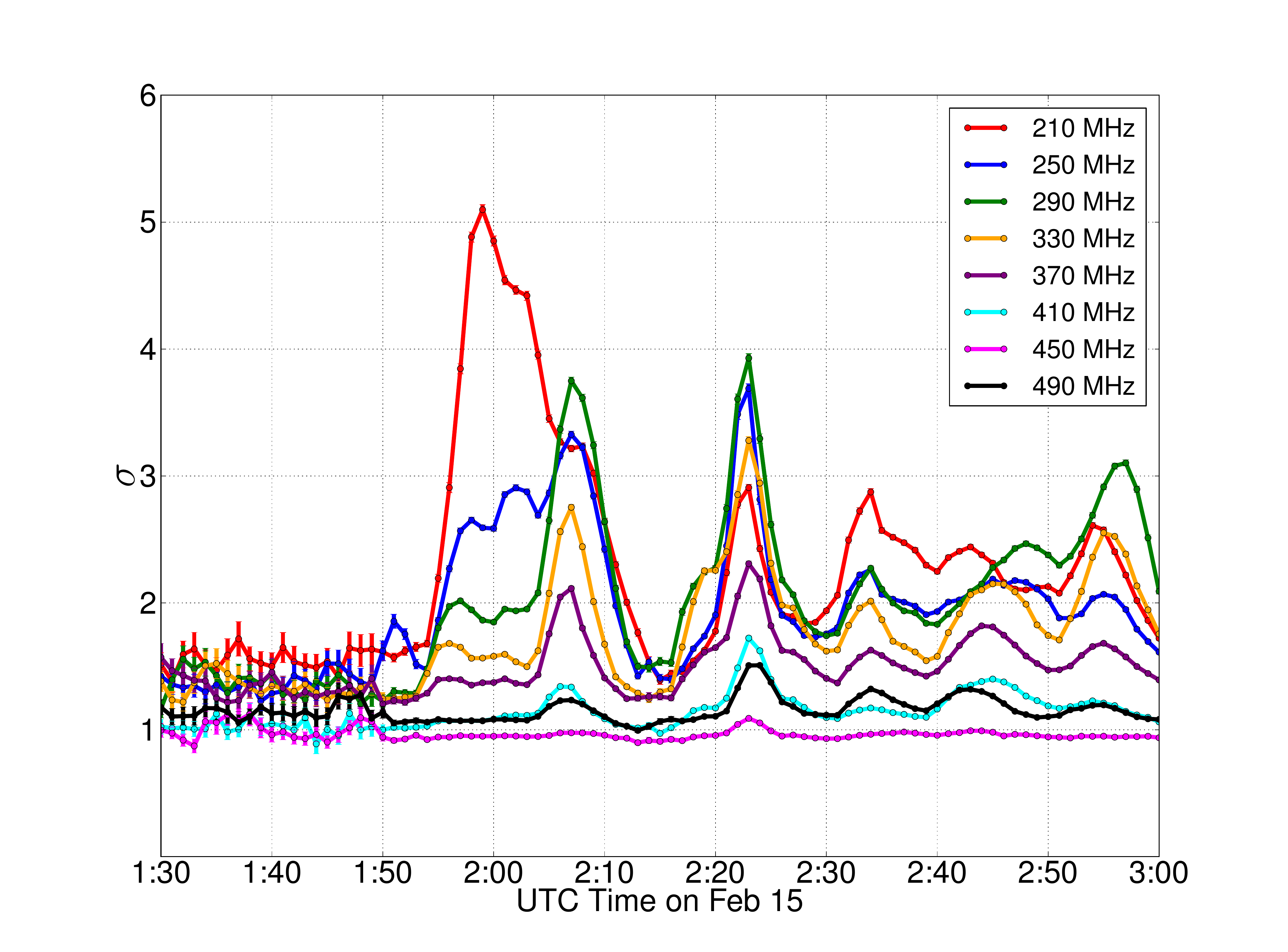}\hfill
\includegraphics[width=\linewidth]{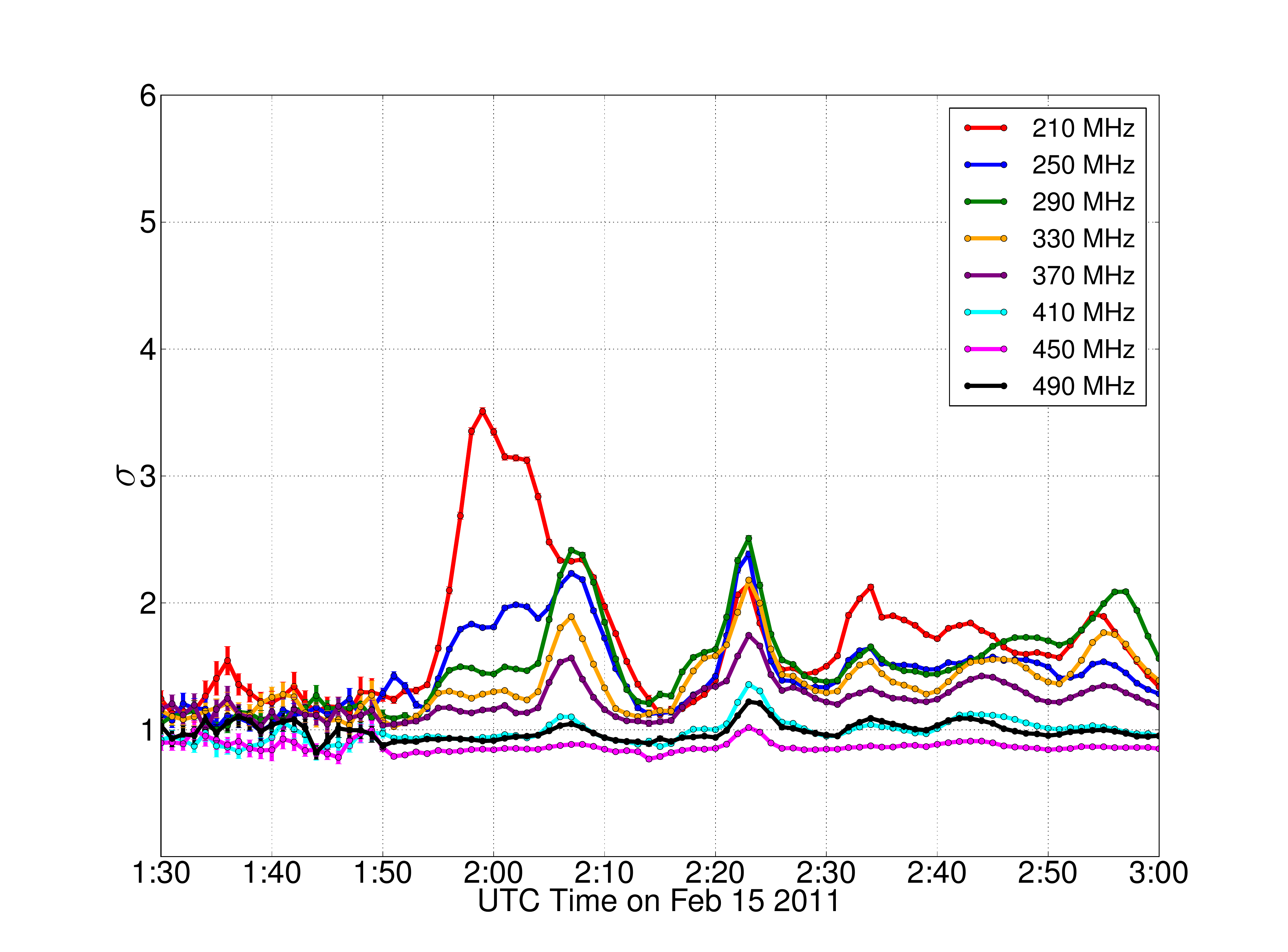}\hfill
\caption{\label{fig:rayleigh_evolution} (Top) The time evolution of the Rayleigh fit parameter for channel two as a function of time. (Bottom) The same for the  CSW from channels 2 and 4. In both, the error bars are the one-standard deviation errors on the fit parameter. The larger error bars before 1:50 are due to lower statistics before the trigger saturates after $\sim$1:50AM.}
\label{fig:rayleighs_evolution}
\end{figure}

In the top panel of Fig.~\ref{fig:rayleigh_evolution}, we plot the evolution of the fit parameter $\sigma$ as a function of time. 
We find that for frequencies between 200-400~MHz, $\sigma$ increases at times coincident with the brightening of the flare observed in spectrograms. 
Based on the quality of the single-channel fits, we find the excess power during the flare in the 200-400 MHz band to be consistent with that of thermal emission. We do not mean that the observed emission has a relationship between the intensity of spectral modes that follows a blackbody spectrum.

We also perform the Rayleigh fits for the CSWs described in Sec. \ref{sec:directional_reconstruction},
and best fit values of the $\sigma$ parameter are shown as a function of 
time on the bottom of Fig.~\ref{fig:rayleigh_evolution} for comparison to those derived 
from the single-channel fits.
These particular CSWs are formed out of only two antennas of like polarization and using the time lag that gives 
the highest cross-correlation value. 
This is likely to give us a time lag that is more
appropriate for picking out the correlated component 
than a time lag that is derived from a directional hypothesis, given that we observe timing uncertainties associated with our directional hypotheses.
By fitting Rayleigh distributions to the spectral amplitudes derived from the CSWs,
we are investigating the nature of the emission that is correlated between the antennas,
as the CSW suppresses the uncorrelated component relative to the correlated component.
The radiation that is correlated between the antennas is also consistent with thermal, and looks 
similar to the distribution of total excess emission seen in a single channel in the left panel. 

In the progress of determining the thermal nature of the events, 
we also investigated the possibility that the reconstructability of the events
was due to the presence of (1) low signal-to-noise ratio transients, (2) continuous-wave (CW) contamination,
(3) time-dependent features like chirps that might be expected in solar radio
emission. 
We found evidence for only the thermal emission described in this section,
though we review the findings of these other investigations in Appendix \ref{app:alt_hyp}.

\section{Implications}
\label{sec:implications}

Although the detection of solar flares by ARA is not the experiment's purpose, such events provide the opportunity to enhance the information extracted from above-ice RF sources, especially the astronomical coordinates of cosmic-ray air showers. Being sourced from the sun, the solar flare events enable validation of software used to project an event's RF reconstruction direction onto celestial coordinates.

The power of these flare events is, at present, limited by systematic uncertainties associated with ARA reconstruction algorithms. Each antenna has both an associated surveying uncertainty and also an associated cable delay uncertainty. Additionally, there is an overall uncertainty in the details of the ice medium transporting the radio signal - specifically, the refractive index profile as well as the surface slope at the optical entry point into the ice. Once those uncertainties can be narrowed to the level of the statistical error on the ARA channel-to-channel timing cross-correlation ($\sim$100 ps), solar flare events can then be used to help point reconstructed ultra-high energy cosmic rays. Specifically, the solar flare data is useful because it contains many events ($>1500$ events/min) which have clear reconstructions on an event-by-event basis. The sun is a true far-field plane wave emitter at a known location, and moves slowly enough to be treated as stationary while statistics accumulate, and still sweeps out a series of azimuthal angles. 

Finally, if events like these are observed again, the flare can serve as a calibration source for the entire array. As future ARA stations are deployed, 
they will be positioned further from pulsers deployed near South Pole infrastructure (like the pulser atop the IceCube Counting Laboratory, or ``ICL rooftop pulser"). As this distance increases, 
the effectiveness of the ICL Pulser will diminish as 
the signal will be considerably weaker at the far stations.
With a solar flare as a calibration source, 
the strength of the signal remains consistent across all stations.

\section{Other Solar Flares in ARA Data}
\subsection{Search for Other Solar Flares with the Testbed}

In addition to the Feb.~15$^{\rm{th}}$ events found through the interferometric analysis described here, and the Feb.~13$^{\rm{th}}$ events discussed in Appendix \ref{app:otherflares}, we also located bursts of radio emission coincident with three other solar flares in 2012. Events were found to correlate with solar flares from March 5 2012, March 7 2012, and Nov 21 2012.

We checked the reason these other three flare do not pass the interferometric analysis. For this purpose, we again removed the cut which required the events come from within the ice. For the March 5 flare, no single cut is responsible for the rejection of all events. For March 7 and November 21, all events are rejected by a cut designed to remove events contaminated by continuous-wave (CW) emission; more than 86\% of recorded events are also simultaneously rejected by the ``Reconstruction Quality Cut'' which imposes requirements on the height and size of the peak in the interferometric maps. We do not explore these events further here.

\subsection{Search for Solar Flares with Deep ARA Stations}
A search was performed to attempt to detect similar solar flare emission in the 
deep stations around the times of the X-class solar flares listed in table \ref{tab:flares_summary} in Appendix \ref{app:otherflares}.
In examining data from one of the deep stations (A2) taken during the 2013 season 
(data already analyzed for purposed of a diffuse search \cite{Allison:2015eky}), 
the cut criteria were altered to attempt to identify clusters of events that reconstruct to the Sun's position.
Because of the differences between the Testbed station and the deep station, 
the same set of cuts were not used but rather only one cut was implemented: 
the event must have a reconstructed correlation value greater than 0.18 with a assumed reconstruction radius of 3000 m. 
To test the validity of this method's ability to reconstruct events above the ice, 
it was applied to ICL rooftop pulser events (described in Sec.~\ref{sec:polarization}), which reconstructed regularly to the correct position above the ice.
After applying this cut on the data taken during the 2013 solar flare periods, we find no events passing these cuts. However, this search was not explicitly designed to detect solar events, and it is also the case that ARA has yet to analyze data from 2014 forward, which was a more active part of the solar cycle than 2013 \cite{SolarCycle}, and so some deep station cases may still be in archival data.

\section{Conclusions and Discussion}

The ARA Testbed has recorded events during a solar flare on February 15$^{\rm{th}}$, 
2011 that contain a broadband signal and uniquely and tightly reconstruct within a few degrees of the Sun.
The excess radio emission is found to contain a component
that is coherent across the antennas, consistent with correlated thermal emission expected from
the sun being a bright thermal point source.  

This is the first observation by ARA, and by an ultra-high energy Antarctic radio neutrino telescope, of emission that can be identified as being produced at an extraterrestrial source on an event-by-event basis.
The events offer a high statistics data set that, coupled with improved systematic uncertainties, can validate the software ARA uses to map above-ice RF sources, such as cosmic-ray air showers, to celestial coordinates.

From the perspective of solar physics, we offer a comparison of spectrograms 
derived from the total observed emission and compare to the spectrograms obtained
with the correlated component amplified.
This may provide a view of the mechanisms through which the correlated
and uncorrelated components of the emission are produced.  
On a basic level, due to our 2~GHz sampling of 
250~ns waveforms, ARA may be able to contribute to the knowledge of solar
flare phenomena on smaller time scales than can be probed elsewhere, and thus
smaller length scales.

ARA's location in the southern hemisphere and radio quiet environment may add data to the worldwide network of solar burst spectrometers.
At present, most spectrometers are located in the Northern Hemisphere (e.g. Greenbanks in Virginia, Palehua in Hawaii, Ondrejov in the Czech Republic). Some reside in the south--for example, Culgoora and Learmouth--but are contaminated by steady-state man-made spectral features, for example the noise lines in Culgoora spectrographs near 180 and 570 MHz. The radio community has already expressed a need for a so-called ``Frequency Agile Solar Radiotelescope'' \cite{1538-3873-119-853-303}, to which ARA bears remarkable resemblance in bandwidth and digitization electronics. This could also be true for other radio neutrino-telescopes deployed to Antarctica, such as ARIANNA \cite{Barwick:2014rca}.

\section{Acknowledgements}

We thank the National Science Foundation for their generous support through Grant NSF OPP-902483 and Grant NSF OPP-1359535.
We further thank the Taiwan National Science Councils Vanguard Program: NSC 92-2628-M-002-09 and the Belgian F.R.S.-FNRS Grant 4.4508.01.
We are grateful to the U.S. National Science Foundation-Office of Polar Programs and the U.S. National Science Foundation-Physics Division.
We also thank the University of Wisconsin Alumni Research Foundation, the University of Maryland and the Ohio State University for their support.
Furthermore, we are grateful to the Raytheon Polar Services Corporation and the Antarctic Support Contractor, for field support.
B. A. Clark thanks the National Science Foundation for support through the Graduate Research Fellowship Program Award DGE-1343012.
A. Connolly thanks the National Science Foundation for their support through CAREER award 1255557, and also the Ohio Supercomputer Center. 
A. Connolly, H. Landsman, and D. Besson thank the United States-Israel Binational Science Foundation for their support through Grant 2012077.
A. Connolly, A. Karle, and J. Kelley thank the National Science Foundation for the support through BIGDATA Grant 1250720.
K. Hoffman likewise thanks the National Science Foundation for their support through CAREER award 0847658.
D. Z. Besson and A. Novikov acknowledge support from National Research Nuclear University MEPhi (Moscow Engineering Physics Institute).
R. J. Nichol thanks the Leverhulme Trust for their support.

We would finally like to thank  Dan Gauthier, John Beacom, Kenny Ng, Dale Gary, and Nita Gelu for helpful discussions.

\section{Bibliography}
\bibliographystyle{unsrt}
\refstepcounter{section}
\bibliography{references}

\begin{appendices}
\section{Supporting Figures}
\label{sec:supporting}

\subsection{Waveforms and Spectra}
\begin{figure}[ht]
\centering
\includegraphics[width=\linewidth]{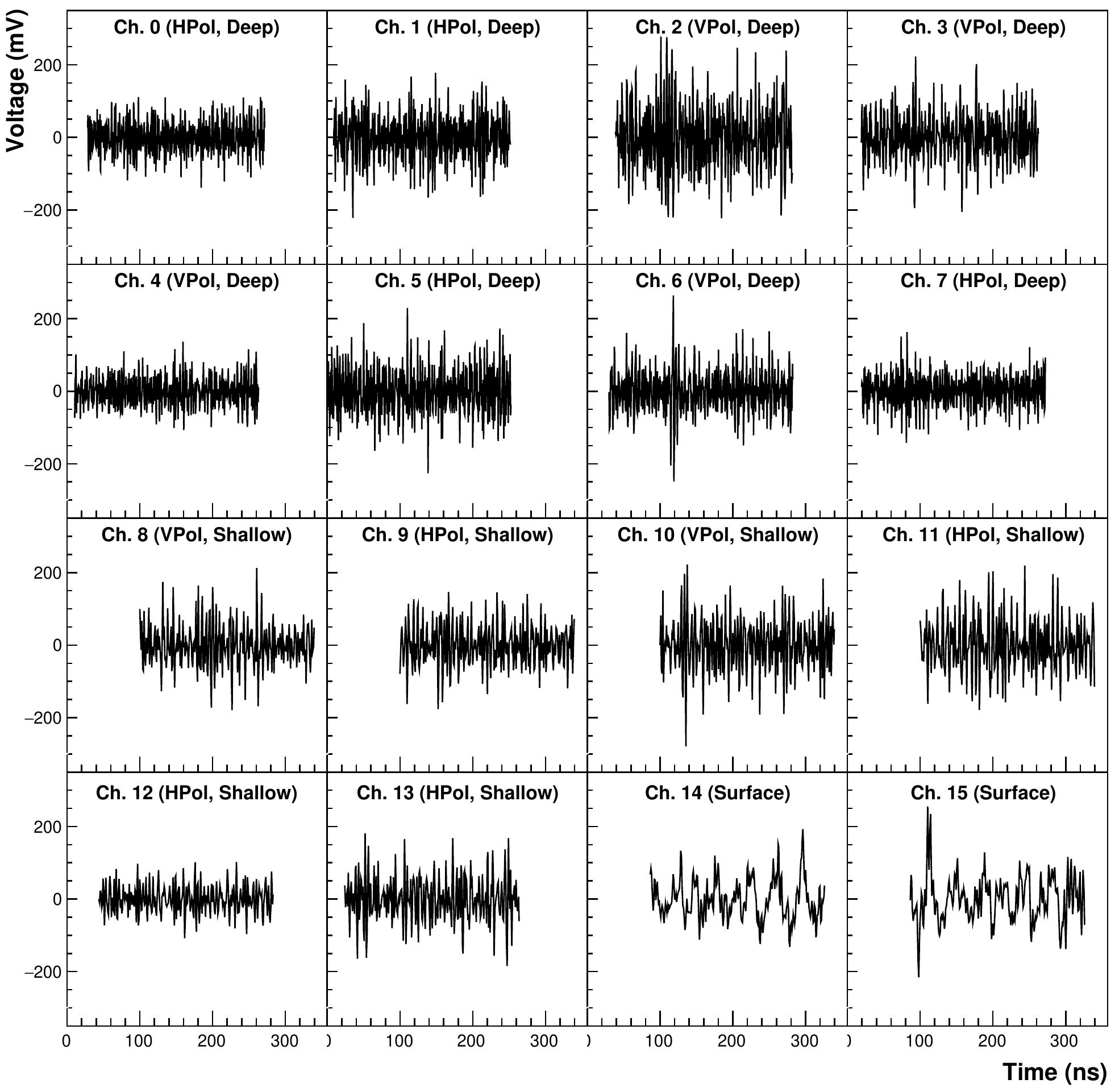}
\caption{Example waveform for a typical event during the flare period, recorded at 2:04 AM, Feb 15$^{\rm{th}}$ 2011 UTC.}
\label{fig:wf_flareevent_1}
\end{figure}

\begin{figure}[ht]
\centering
\includegraphics[width=\linewidth]{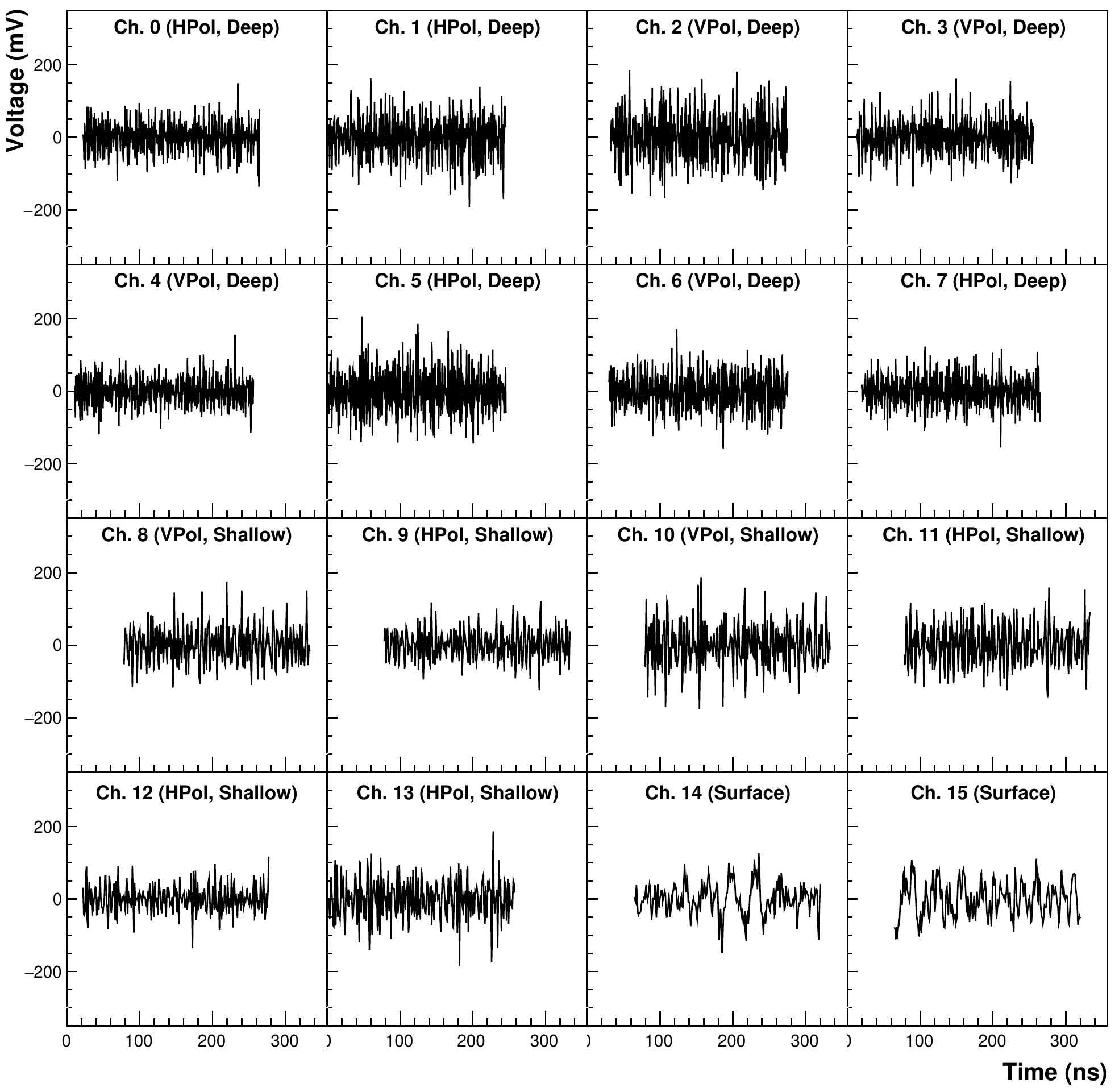}
\caption{Example waveforms for a typical non-flare event, recorded at 2:04 AM, Feb 11$^{\rm{th}}$ 2011 UTC.}
\label{fig:wf_nonflareevent}
\end{figure}

\begin{figure}[ht]
\centering
\includegraphics[width=\linewidth]{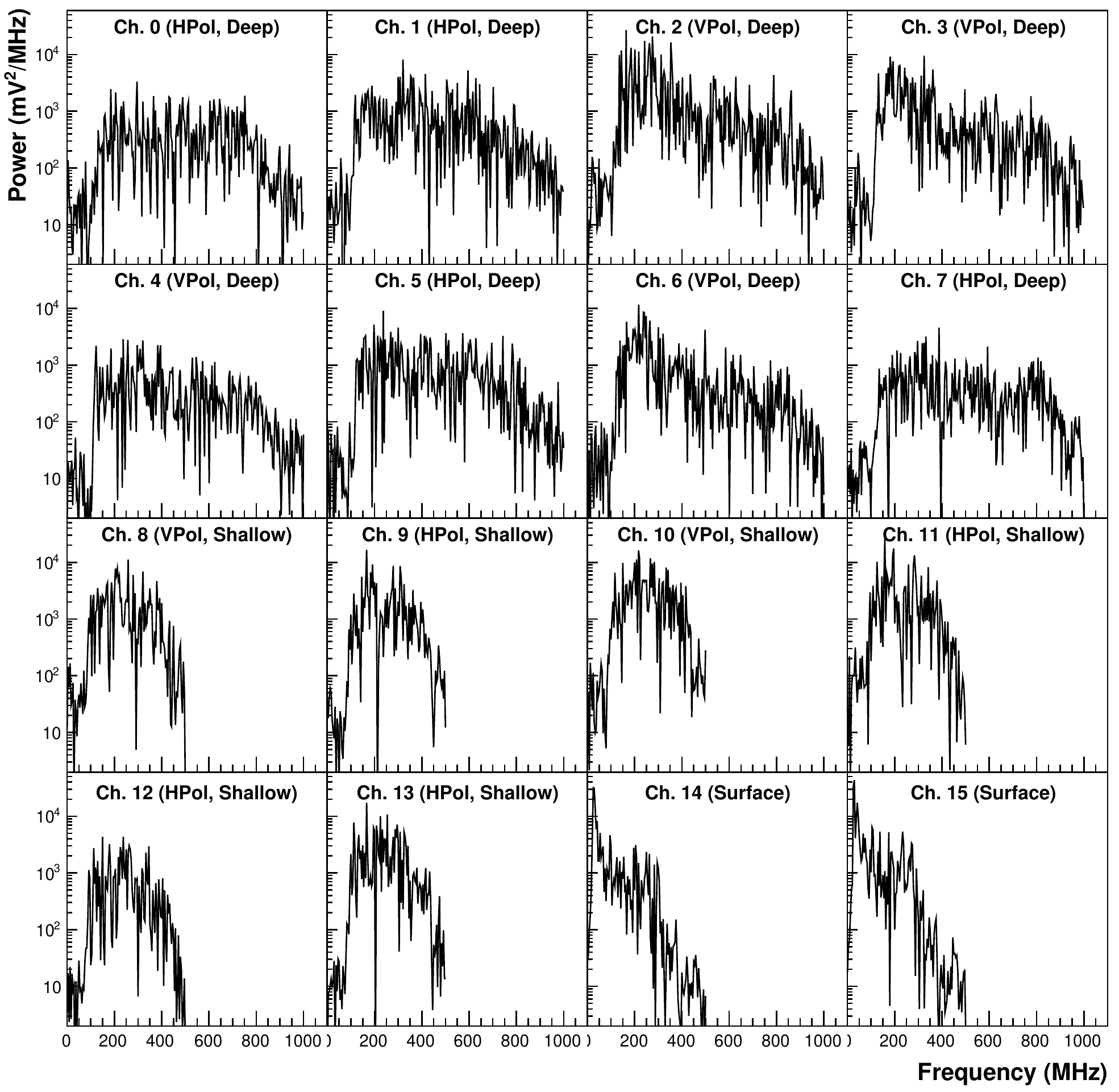}
\caption{Example amplitude spectra for the same event during the flare as in Fig.~\ref{fig:wf_flareevent_1}, recorded at 2:04 AM, Feb 15$^{\rm{th}}$ UTC.}
\label{fig:sp_flareevent_1}
\end{figure}

\begin{figure}[ht]
\centering
\includegraphics[width=\linewidth]{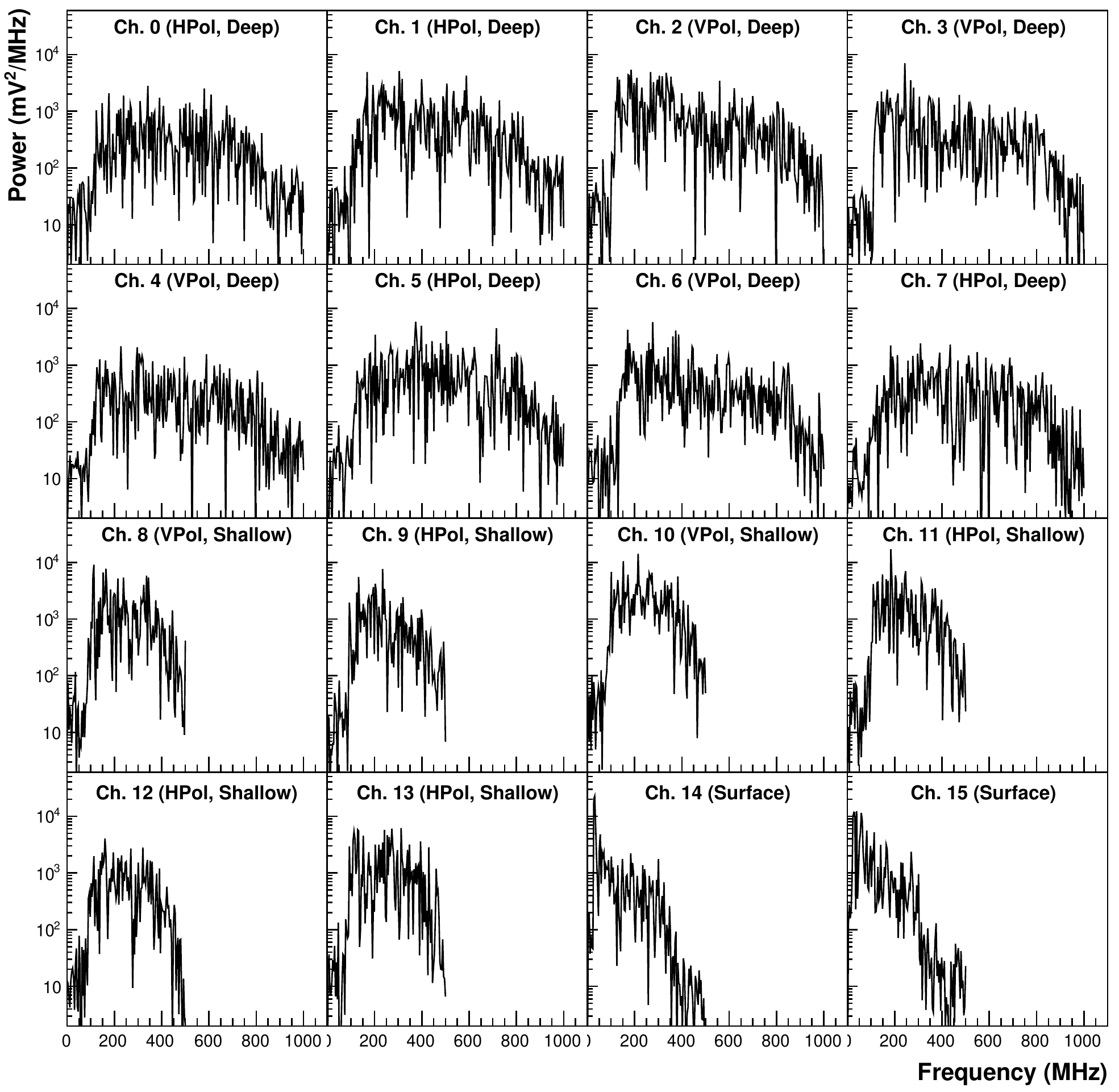}
\caption{Amplitude spectra for the same typical event in Fig.~\ref{fig:wf_nonflareevent} from a time where the sun was quiescent, at 2:04 AM, Feb 11$^{\rm{th}}$ UTC.}
\label{fig:sp_nonflareevent}
\end{figure}

Fig.~\ref{fig:wf_flareevent_1}  shows waveforms from an event at 02:04~GMT, 
near the beginning of the flare period.
For comparison, in Fig.~\ref{fig:wf_nonflareevent}  we show a typical event from February 11, 2011, 
four days before the flare of interest when the sun was not active. 
Figs.~\ref{fig:sp_flareevent_1} and~\ref{fig:sp_nonflareevent} show corresponding spectral amplitudes,
from the Fourier transforms of the waveforms in the previous two figures.

\subsection{Rayleigh Fits}
\begin{figure}
\centering
\includegraphics[width=.48\linewidth]{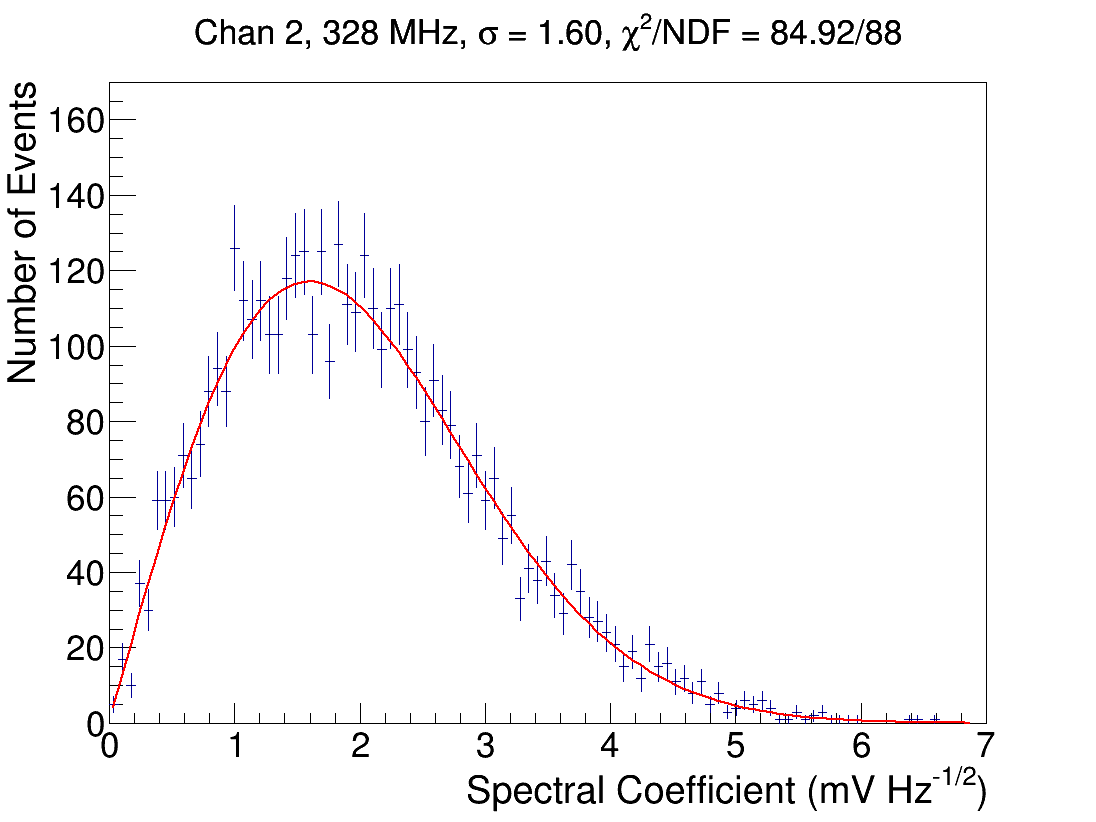}
\includegraphics[width=.48\linewidth]{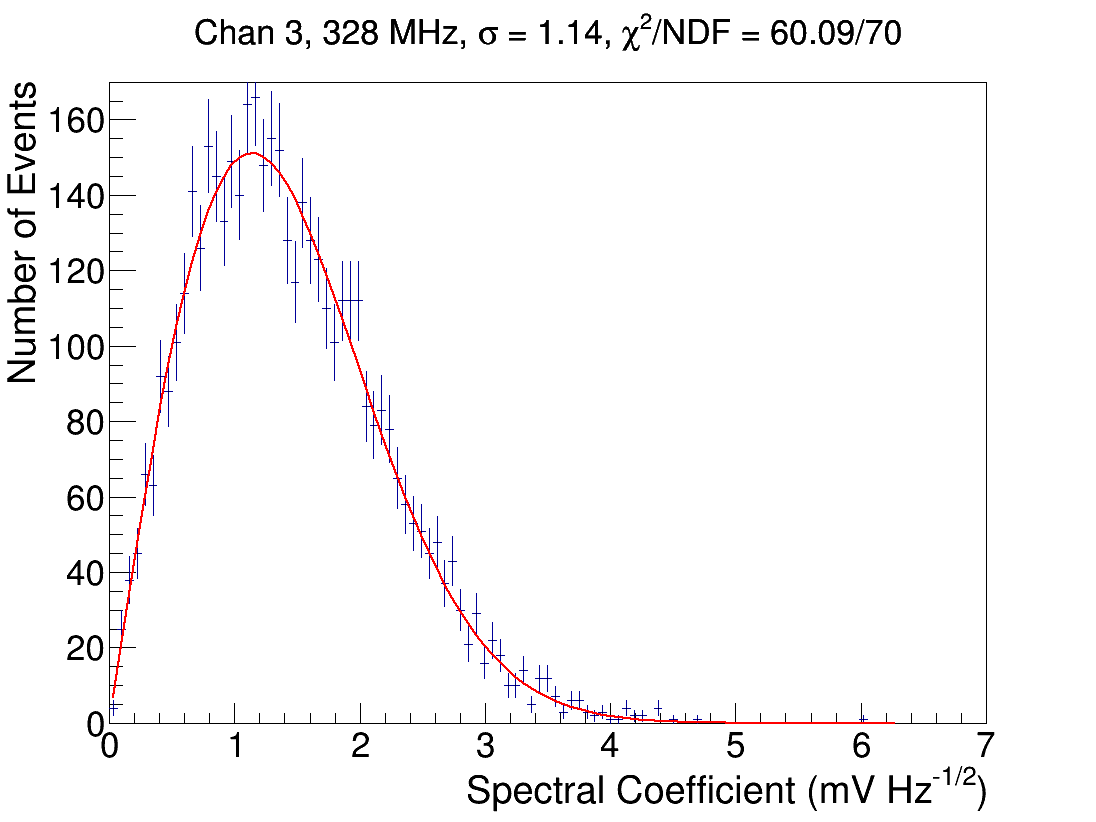}
\includegraphics[width=.48\linewidth]{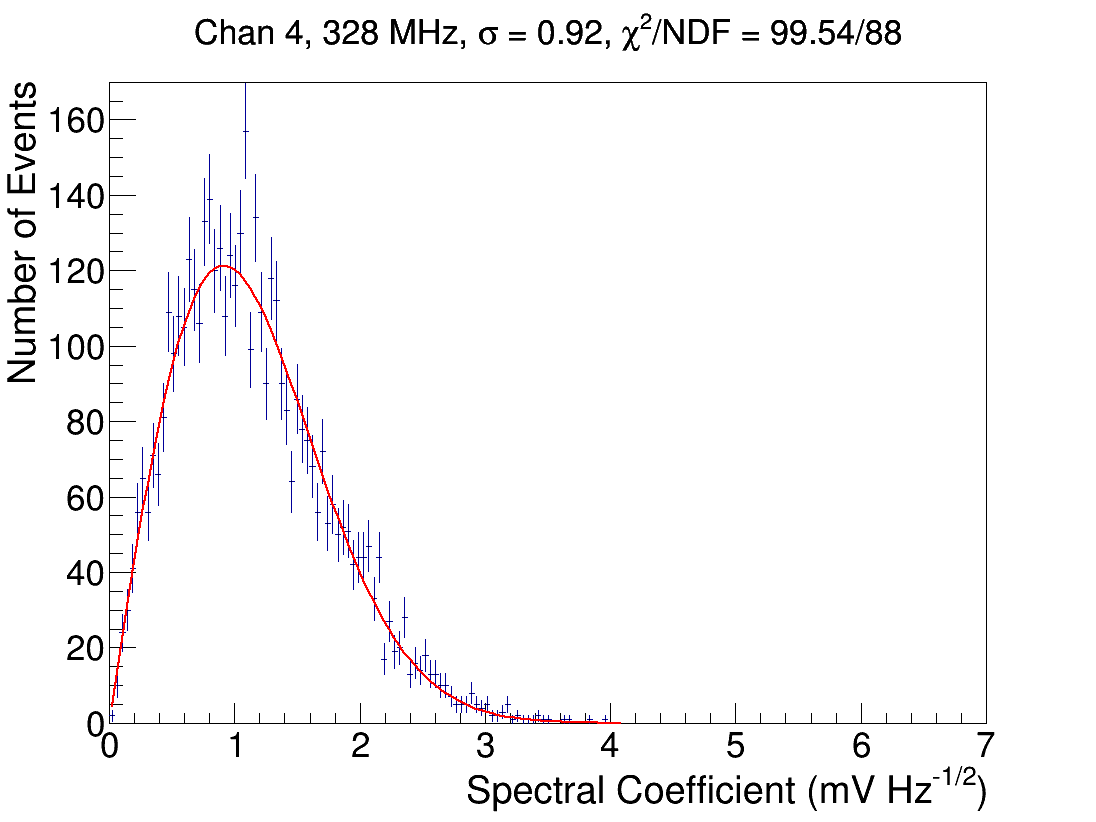}
\includegraphics[width=.48\linewidth]{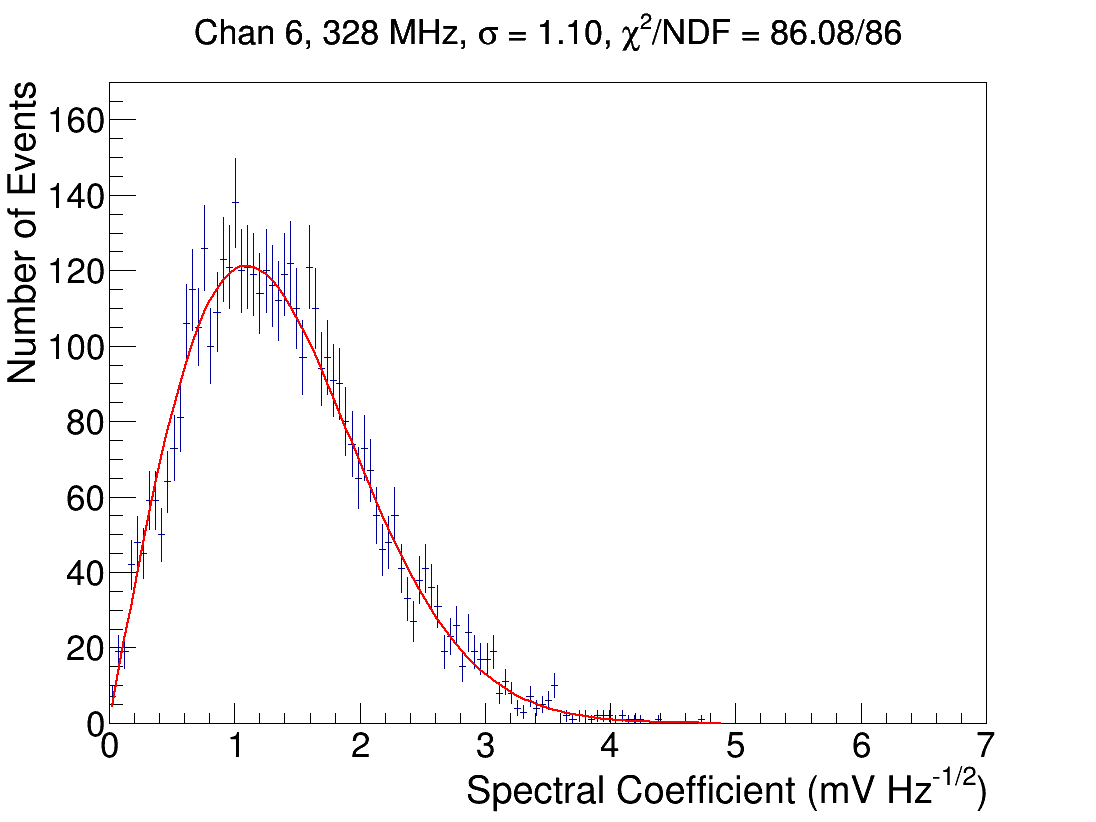}
\caption{Distribution of spectral amplitudes at 328 MHz for the four borehole VPol antennas in the Testbed. }
\label{fig:rayleighs_all}
\end{figure}

\begin{figure}
\centering
\includegraphics[width=.9\linewidth]{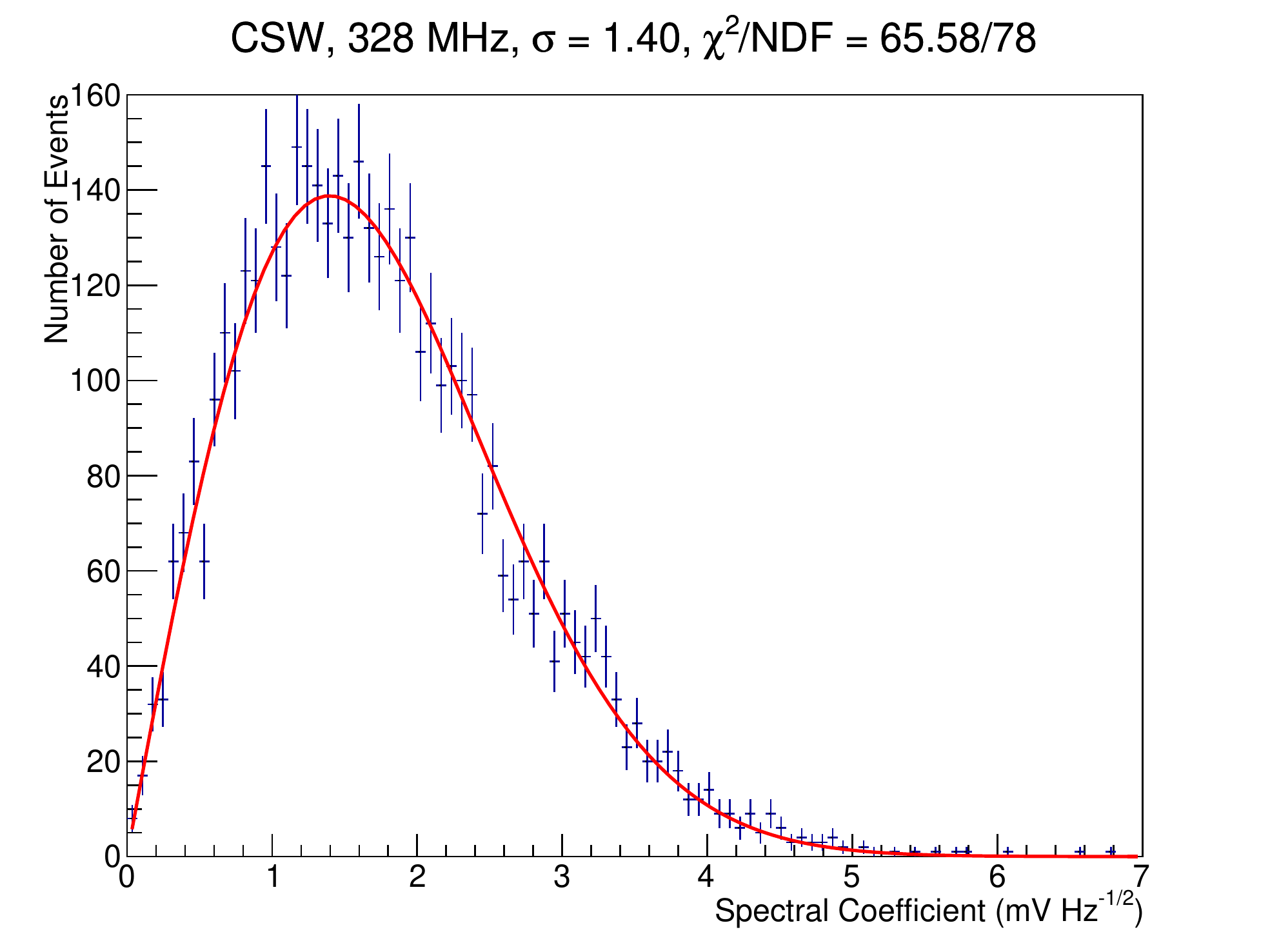}
\caption{Distribution of spectral amplitudes at 328 MHz for the coherently summed waveform derived from the four borehole VPol antennas in that are shown in Fig.~\ref{fig:rayleighs_all}. The CSW is made with directional hypothesis time lags corresponding to the correlation peak on the map for a given event.}
\label{fig:rayleighs_csw} 
\end{figure}

An example of fits for four borehole antennas is given in Fig.~\ref{fig:rayleighs_all}
and the fit to the CSW of those four boreholes is given in Fig.~\ref{fig:rayleighs_csw}.
This demonstrates again that the single channel spectra, and the CSW, are both thermal in nature.

\begin{figure}[ht]
\centering
  \includegraphics[width=.48\linewidth]{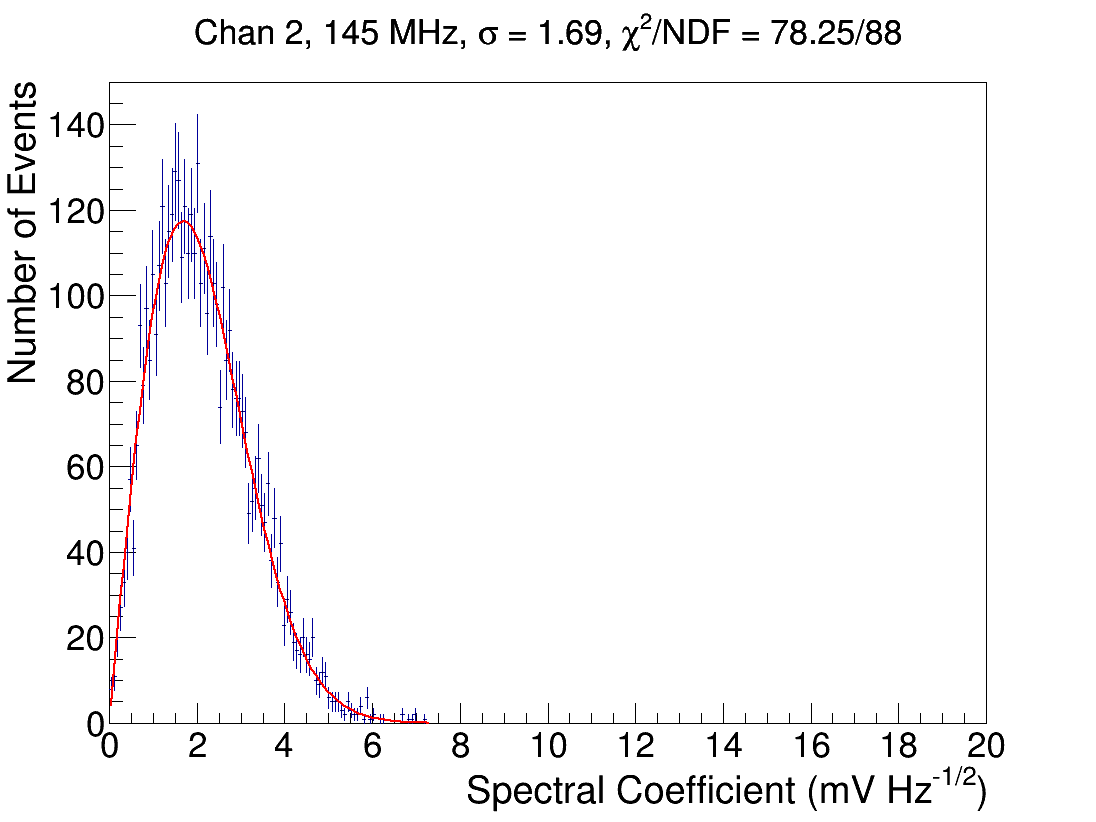}\hfill
 \includegraphics[width=.48\linewidth]{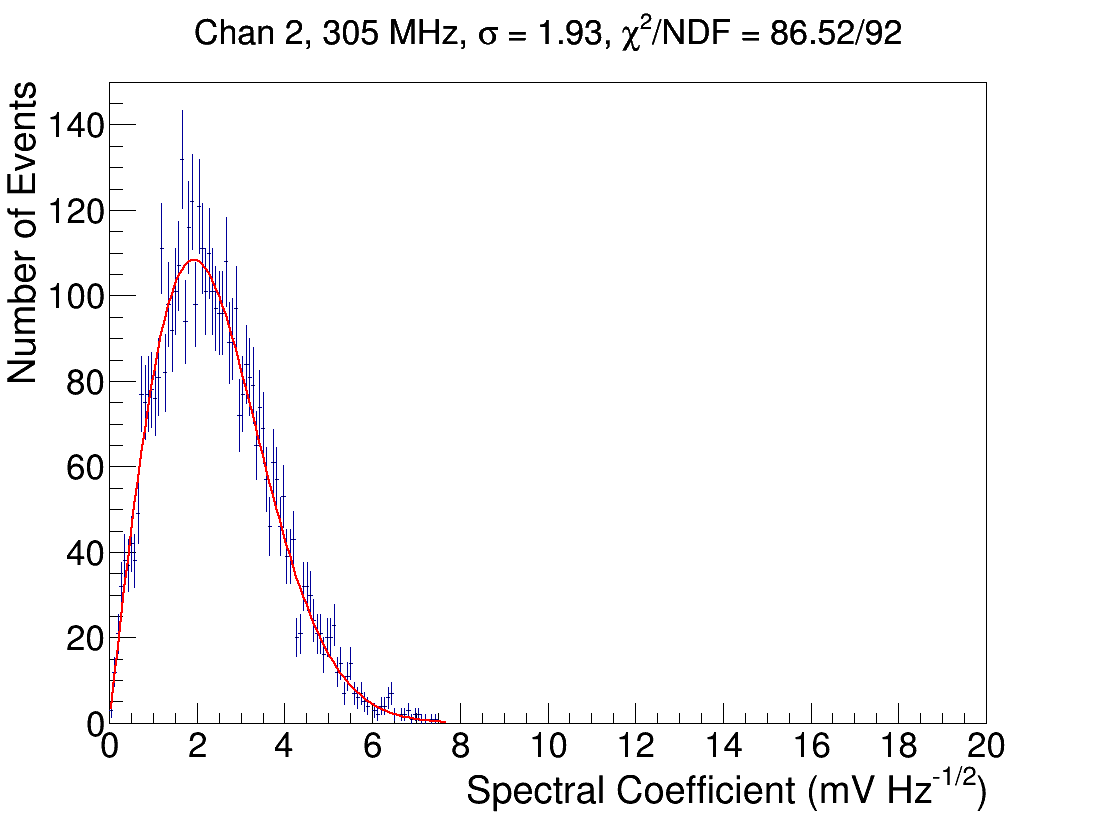}\hfill
 \includegraphics[width=.48\linewidth]{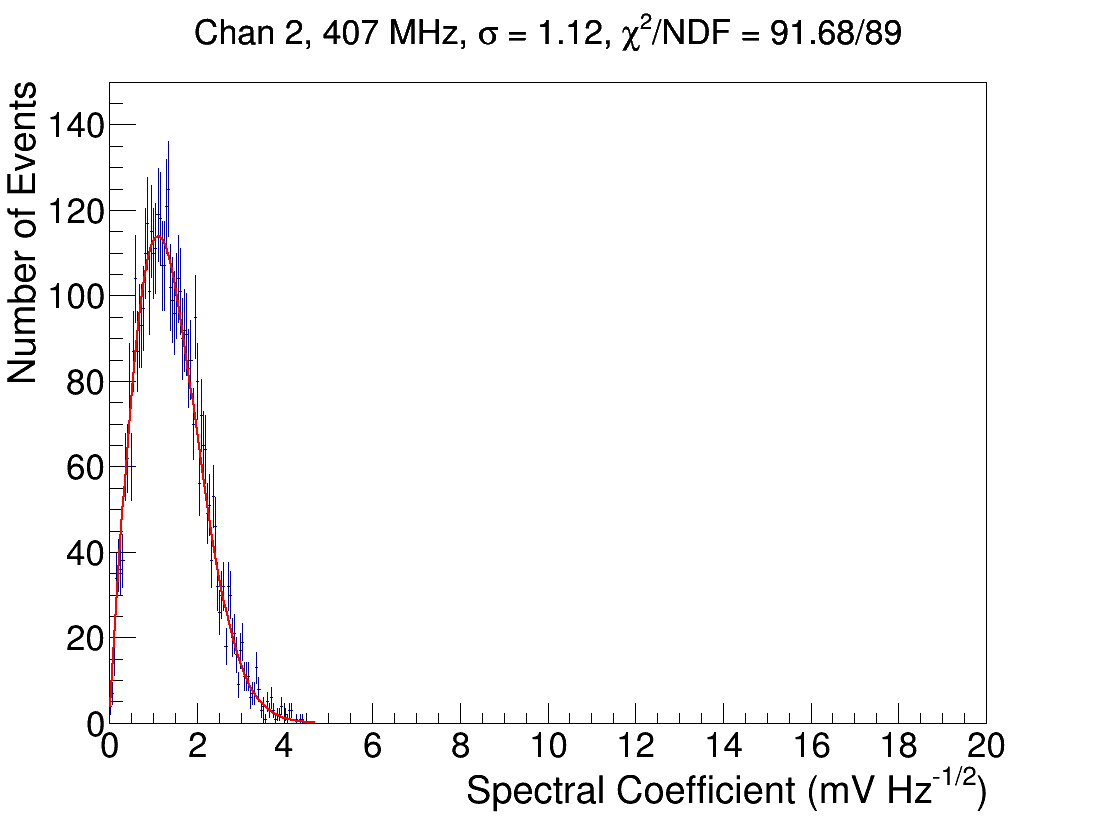}\hfill
 \includegraphics[width=.48\linewidth]{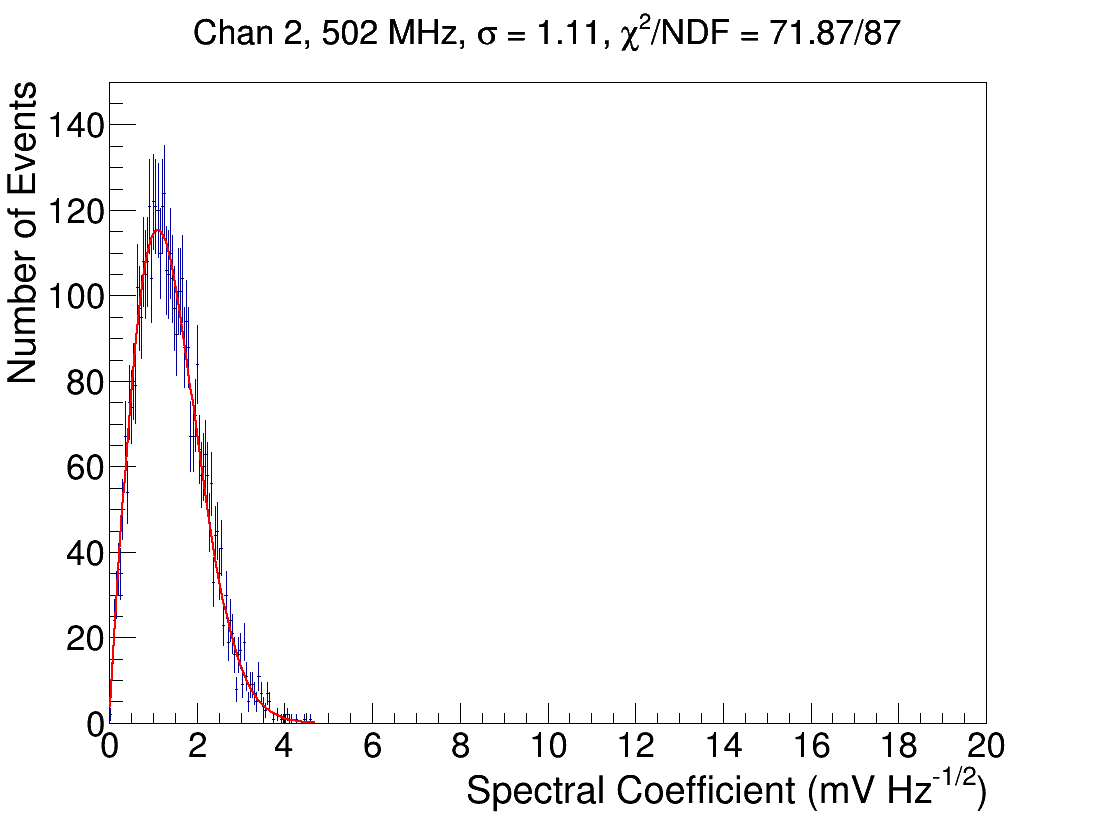}
 \hfill
  \includegraphics[width=.48\linewidth]{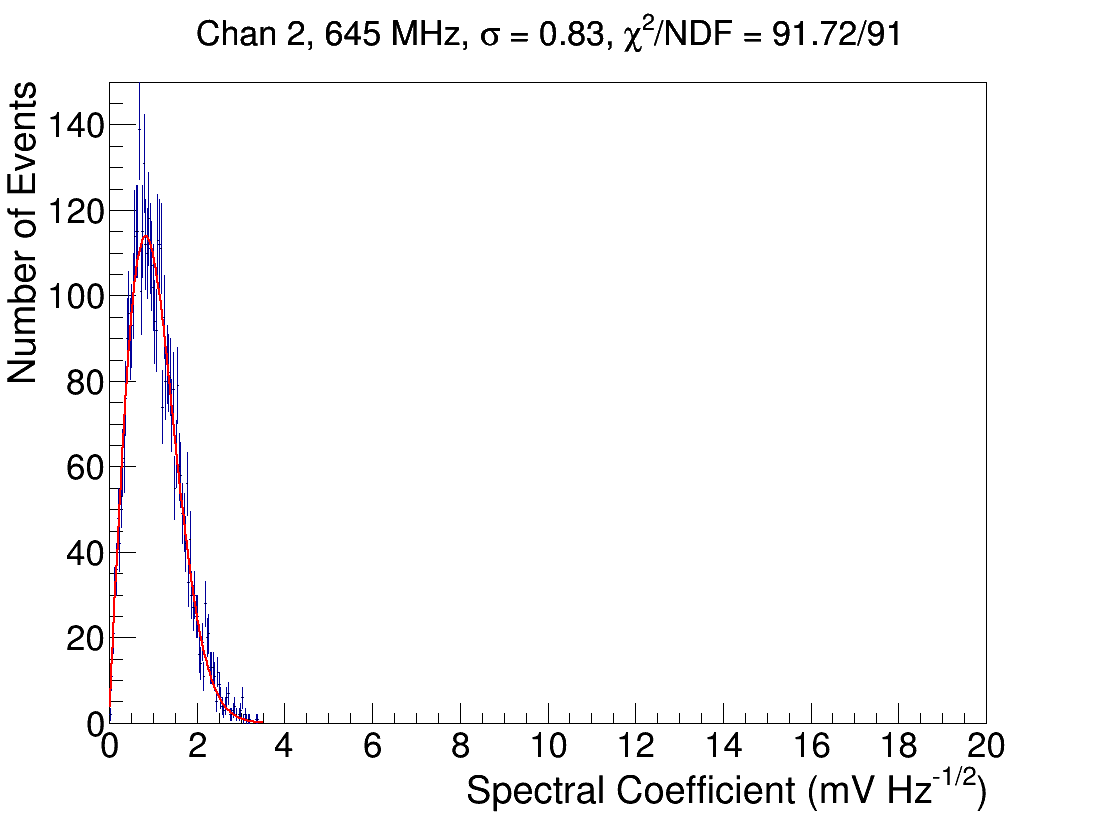}
  \hfill
  \includegraphics[width=.48\linewidth]{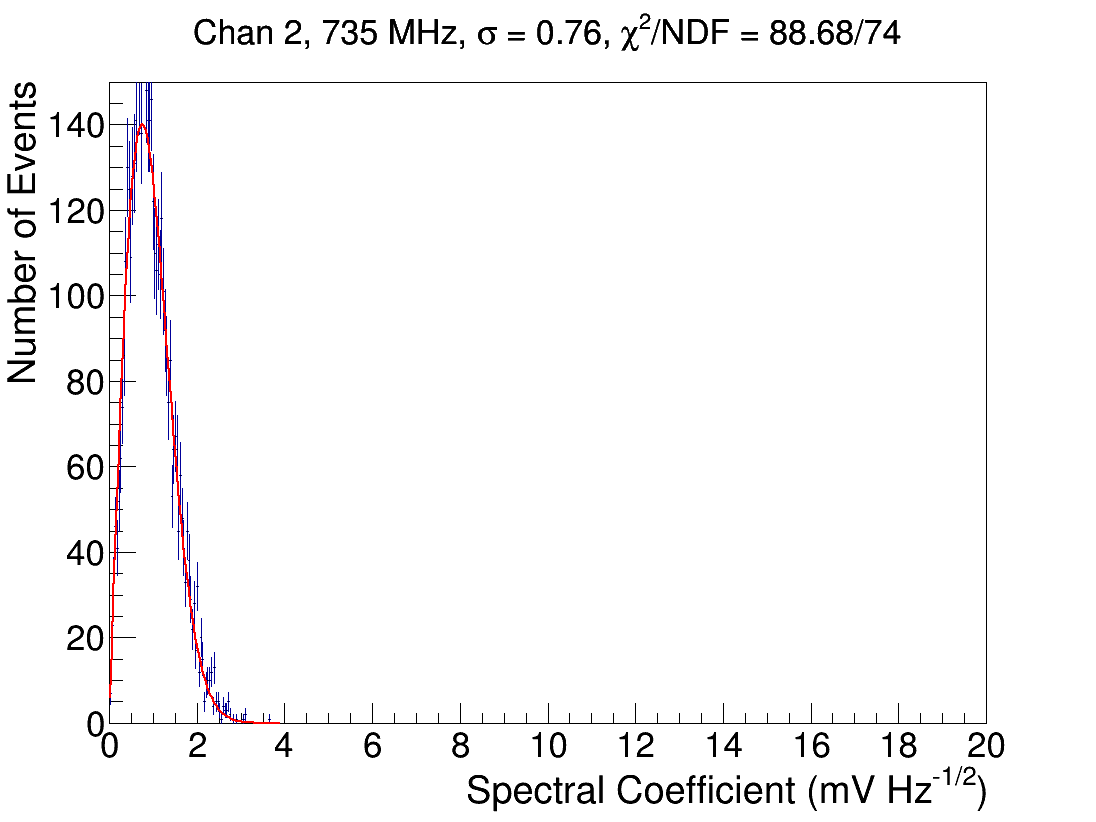}
\caption{A selection of spectral amplitudes for several frequencies across our band with their best fit Rayleigh's superimposed.}
\label{fig:many_rayleigh}
\end{figure}

We also show in figure \ref{fig:many_rayleigh} spectral amplitude distribution and their associated Rayleigh fits for a broader selection of frequencies.

\begin{figure*}[ht]
\centering
  \includegraphics[width=.48\linewidth]{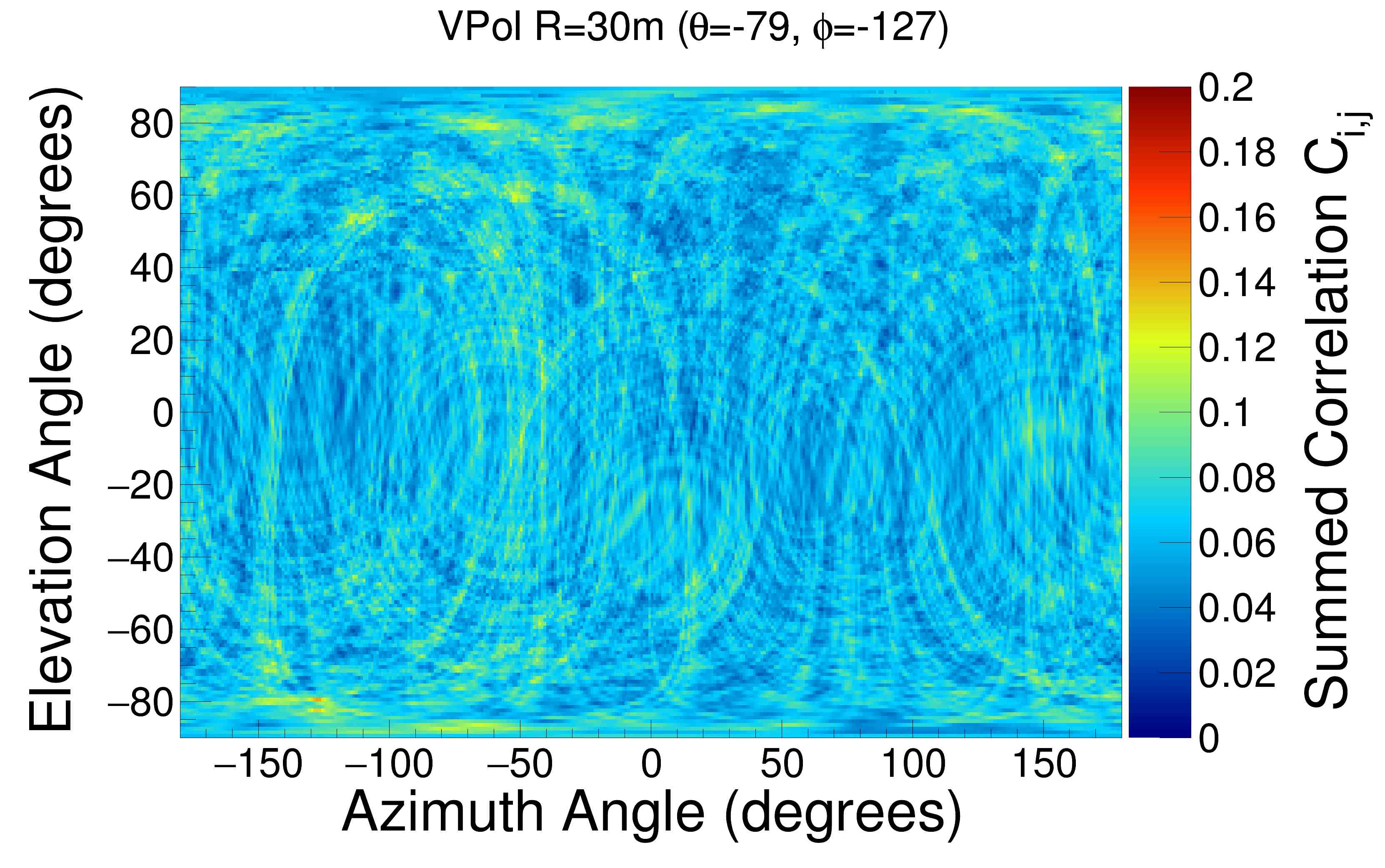}\hfill
 \includegraphics[width=.48\linewidth]{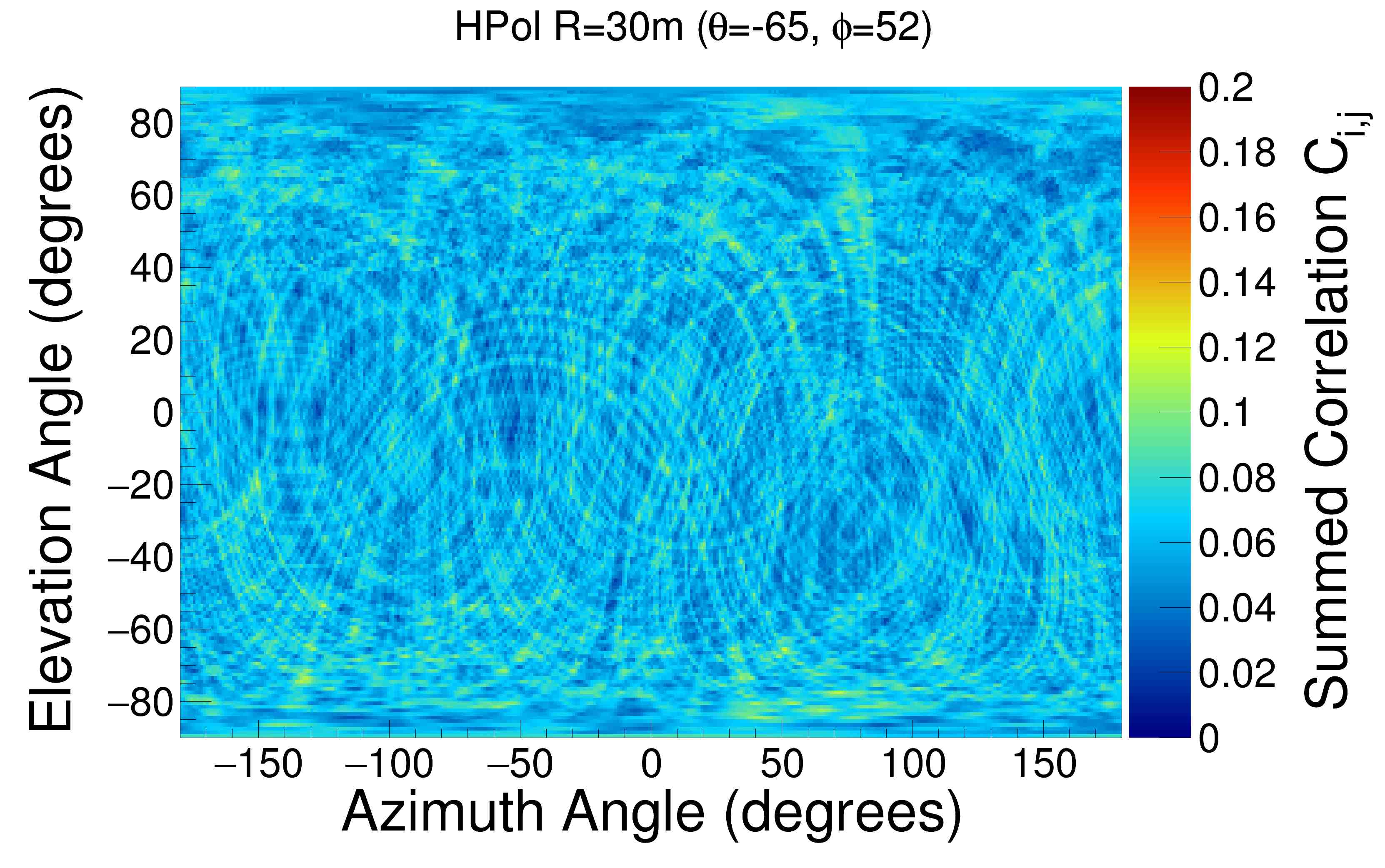}\hfill
 \includegraphics[width=.48\linewidth]{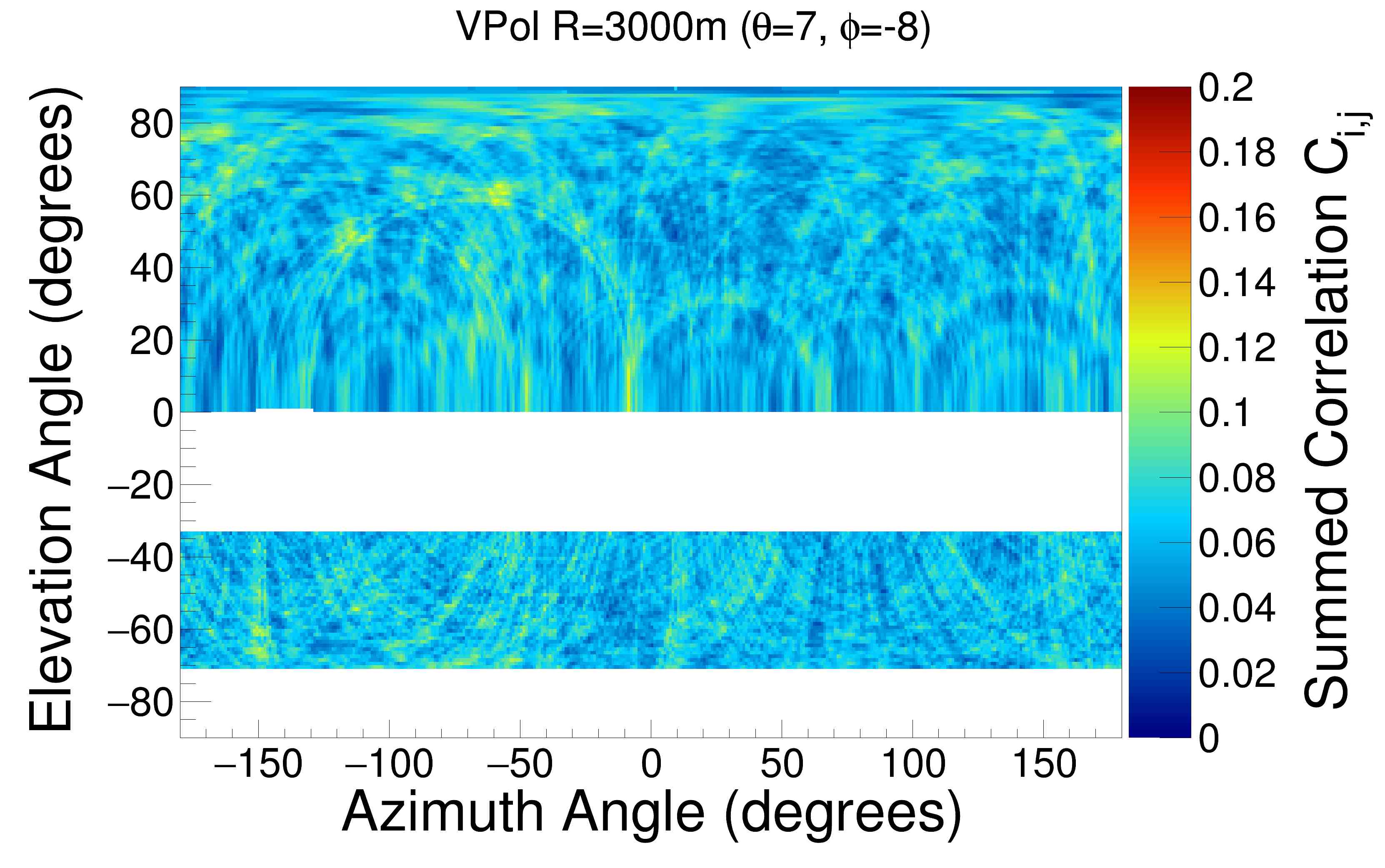}\hfill
 \includegraphics[width=.48\linewidth]{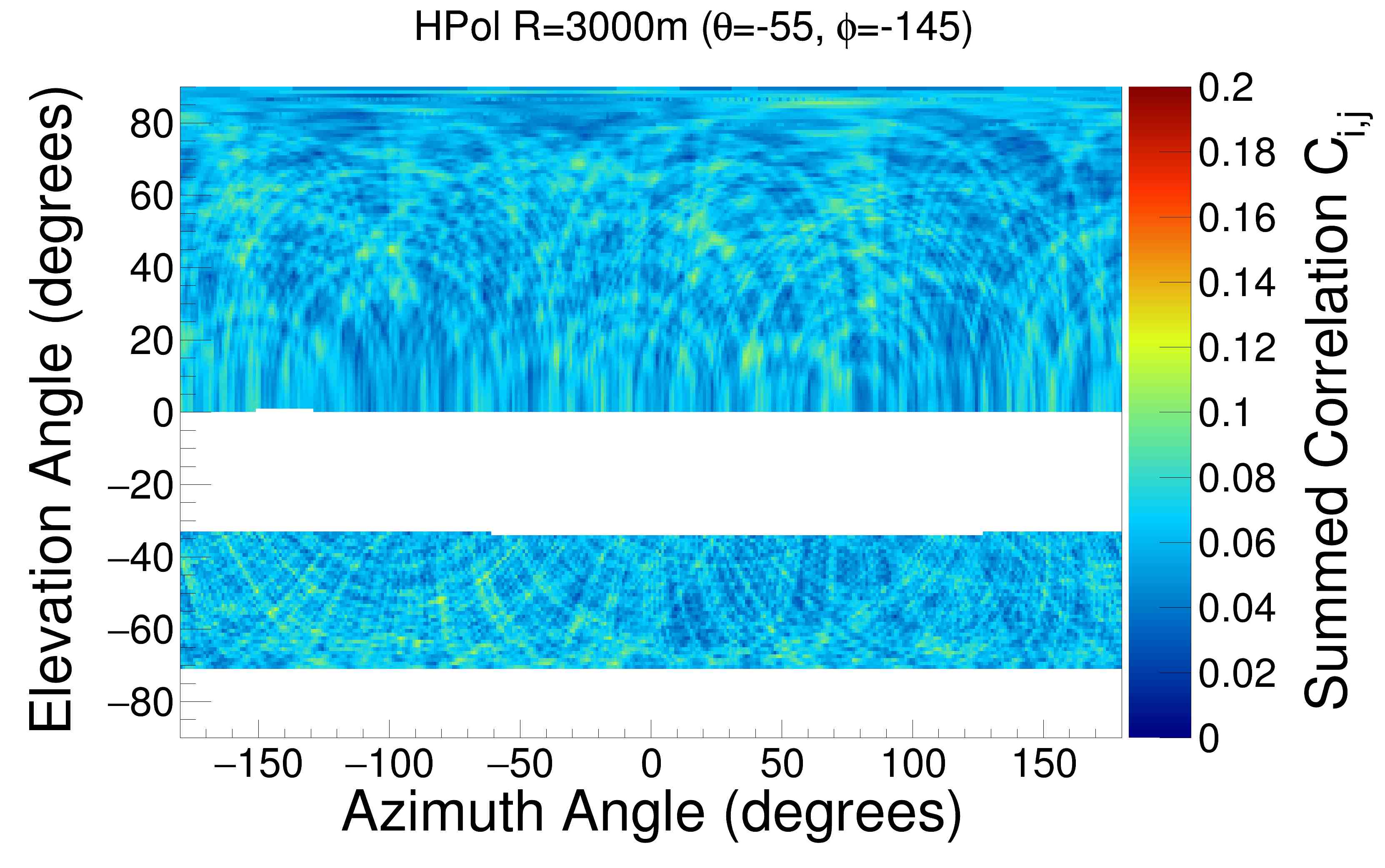}
\caption{Map of cross-correlation values associated with directional hypotheses across the sky for the same event
from a time when the sun was quiescent
as the one shown above in Figs.~\ref{fig:wf_nonflareevent} and~\ref{fig:sp_nonflareevent}.  The upper two figures show reconstruction maps for
the hypothesis of 30~m distance, while the lower two figures show reconstruction maps for the 3000~m distance hypothesis. The maps are created in the Testbed local coordinate system.}
\label{fig:reconstructions_nonflareevent}
\end{figure*}

\subsection{Polarization}
\label{sec:polarization}

As seen in Fig.~\ref{fig:reconstructions_flareevent_1},  
we find higher peak cross-correlation values from waveforms from VPol antennas than from the HPol antennas.  
We decided to investigate whether this VPol dominance had originated in the sun, 
or rather was due to higher gains in VPol antennas than HPol antennas.  
Although some solar flares do emit radio that is circularly polarized--thought to be associated with how plasmas generating the radio emission orient themselves with respect to dynamic magnetic field lines--we would expect it to be a coincidence if the emission would happen to prefer the vertical direction at the site of the ARA Testbed.

In order to measure the relative response of the VPol and HPol antennas in the Testbed {\it in situ}, 
we used recorded waveforms from a pulser mounted to the top of the IceCube Counting Laboratory (ICL) at the South Pole, nicknamed the ``ICL Pulser.''  
This is a Seavey broadband ($\sim200-1200$~MHz), dual-polarized quad-ridged horn antenna, 
the same as is used for the ANITA payload~\cite{Gorham:2008dv}, that transmits pulses in both polarizations.  
The ICL Pulser sits 2.1~km from the Testbed at a 13~m height, 
so that the pulse destined for a Testbed antenna would approach the ice at approximately $\sim 0.36^{\circ}$ above the surface.  
This Seavey antenna is ideally suited for this 
measurement because its response across the band is very similar in both polarizations~\cite{Gorham:2008dv}.

For this study, we used ICL Pulser data taken on January 26$^{\rm{th}}$, 2013,
during two different forty-minute long runs where the pulser was either transmitting in VPol or HPol mode.
We use forty events in each run where we found the highest cross-correlation value for a 
direction within 10$^{\circ}$ in azimuth of the expected ICL Pulser position
and a zenith angle above the surface.  
We take the Fourier Transform of each waveform and, for each antenna, 
average the spectral amplitudes for the forty events in a run.

We use the following procedure to measure the frequency-dependent ratio of HPol to VPol spectral amplitudes.
We considered pairs of antennas that were deployed in the same 
borehole and within 3-6~m of each other in depth, and for each pair, 
took the ratio of the average spectral amplitudes measured in the HPol during HPol 
transmission mode to the average spectral amplitudes 
measured in the VPol during VPol transmission mode, 
after accounting for the different Fresnel coefficients governing the fraction of field-amplitude transmitted at the 
air-ice boundary for each polarization
, $t_H$ and $t_V$.
For the 89.64$^{\circ}$ angle with respect to normal at air-to-ice interface, 
the transmission coefficients 
are $t_{V}=0.0185$ for VPol signals
and $t_{H}=0.0138$ for HPol signals using an index of 
refraction at the surface of 1.35. The four ratios are then averaged together.

To compute the Fresnel coefficients for the antenna calibration, we assume that the ICL roof-top pulser is 13.16m above the ice at a distance of 2099.06m away (accounting for differences in ice elevation at the location of Testbed and the ICL). This leads to a angle from the surface of $\theta_{pulser} = \tan ^{-1} (\frac{13.16\rm{m}}{2099.06\rm{m}}) = 0.36 \degree$ or $89.64 \degree$ from surface normal.

The transmission coefficient in \textit{voltage} for the polarization parallel to the normal (perpendicular or the surface of the ice, our VPol) is given by $t_V$:
\begin{equation}
t_V = \frac{2 n_{\rm{air}} \cos \theta_i}{n_{\rm{air}} \cos \theta_t + n_{\rm{firn}} \cos \theta_i} = \frac{2 \sin \theta_t \cos \theta_i}{\sin (\theta_i + \theta_t) \cos (\theta_i - \theta_t )}
\end{equation}
The transmission coefficient in \textit{voltage} for the polarization perpendicular to the normal (parallel to the surface of the ice, our HPol) is given by $t_H$:
\begin{equation}
t_H = \frac{2 n_{\rm{air}} \cos \theta_i}{n_{\rm{air}} \cos \theta_i + n_{\rm{firn}} \cos \theta_t} = \frac{2 \sin \theta_t \cos \theta_i}{\sin (\theta_i + \theta_t)}
\end{equation}
where $n_{\rm{air}} = 1$ is the index of refraction in air, $n_{\rm{firm}} = 1.35$ is the assumed index of refraction in the shallow ice, $\theta_t$ is the transmission angle with respect to normal, and $\theta_i$ is the incidence angle with respect to normal, and the relationship between $\theta_t$ and $\theta_i$ is found by Snell's law.

\begin{figure*}
\centering
\begin{subfigure}{0.47\textwidth}
\includegraphics[width=\linewidth]{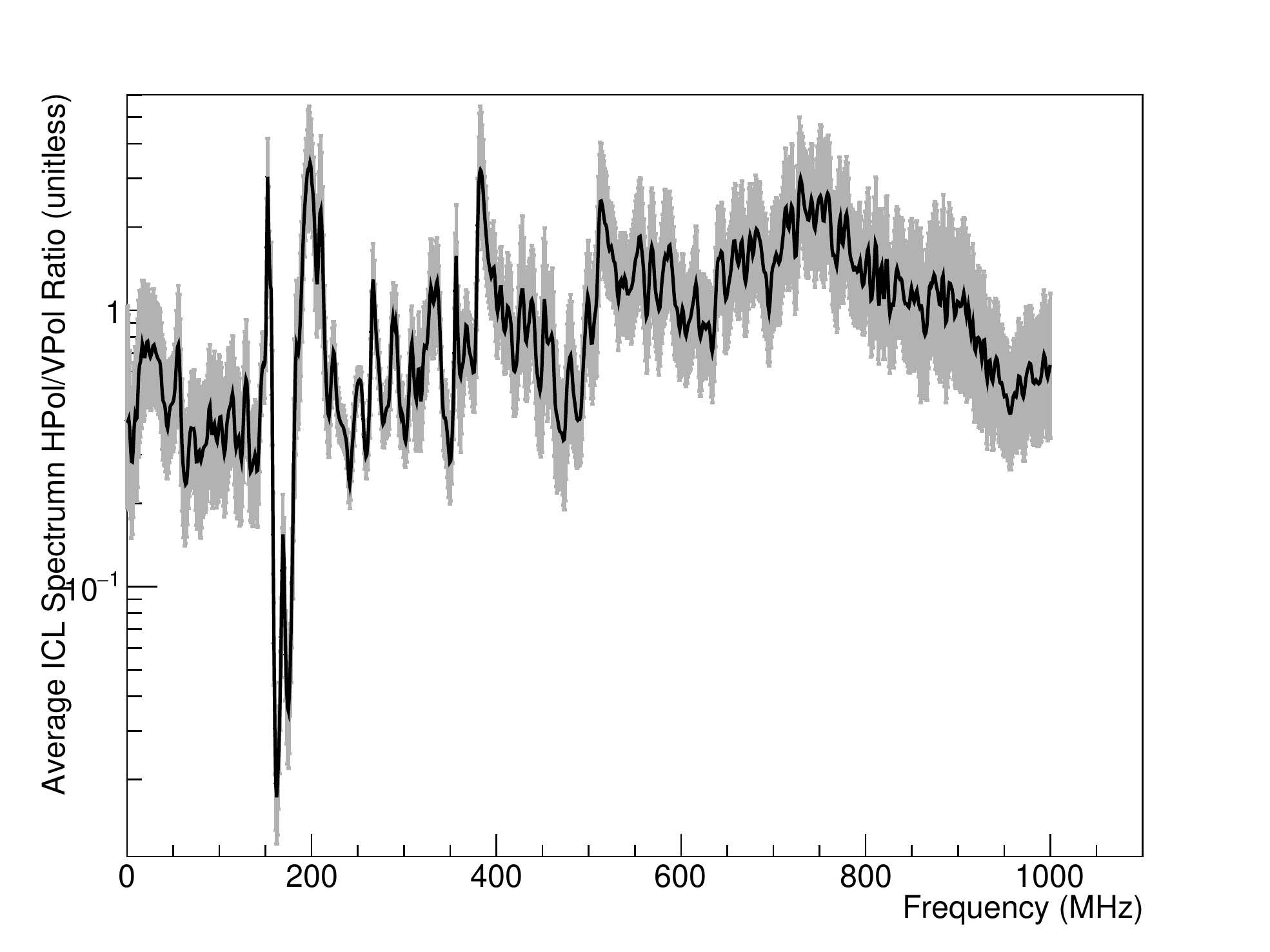}
\caption{H/V ratio of the Antenna Response}
\end{subfigure}
\hfill
\begin{subfigure}{0.47\textwidth}
\includegraphics[width=\linewidth]{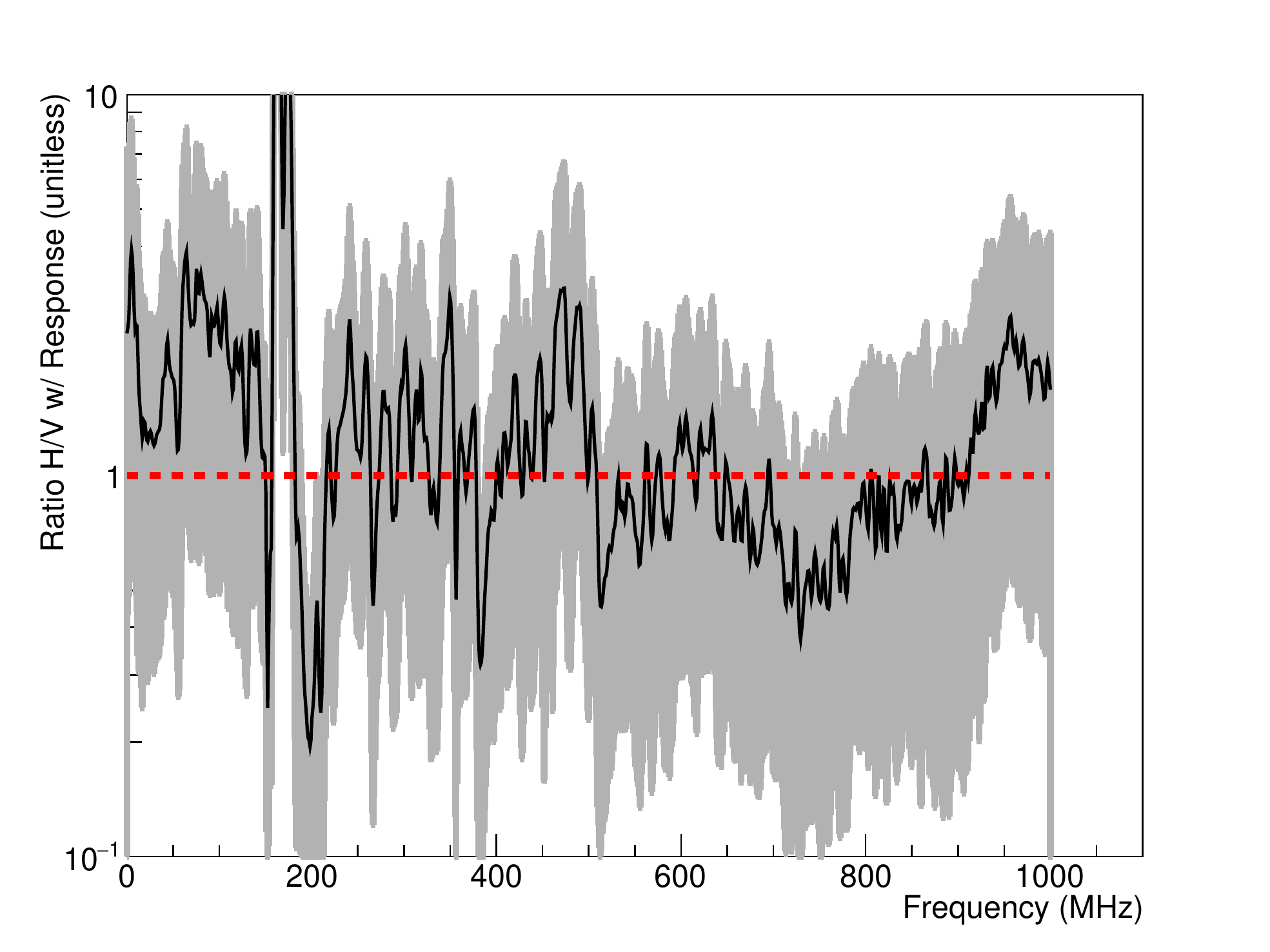}

\caption{H/V ratio of the solar flare spectra}
\end{subfigure}
\caption{\label{fig:weaker_hpol} (a) The antenna response of HPol antennas in ARA Testbed compared to that of VPol antennas, showing that the HPol antennas have lower gains by
about a factor of three.  (b) The observed emission from the solar flare in VPol and HPol after accounting for the different gains from the two types antennas, showing that the observed
emission during the flare was consistent with being observed equally in horizontal and vertical polarizations. Note the artifact around 150 MHz is related to the calibration and is visible in the left figure, and not due to the solar flare emission.}
\end{figure*}

For the solar flare sample--namely all 2323 events passing cuts--
we first build coherent sums of every event in both polarizations. 
The coherent sum waveforms are then Fourier transformed and averaged, 
accounting for Fresnel coefficients for the incidence angle of the sun (which is 13.3 $\degree$ above the surface). 
The ratio of the spectral averages is then corrected by the known relative 
antenna responses found through the ICL calibration procedure above. 
Fig.~\ref{fig:weaker_hpol} summarizes the results of this investigation of the 
polarization of the flare emission. 
The plots show (left) that for these shallow incidence angles, 
the HPol antenna response is intrinsically weaker than the VPol antenna 
response across the band from $\sim$220 MHz to 500 MHz by about a factor of three, 
and that (right) once this is accounted for, 
the emission from the flare is consistent with having equal contributions in HPol and VPol.
The anomalies at $\sim$150~MHz are likely due to a 
difference in the gains of the two types of antennas in this region, 
and can be seen as a calibration artifact in the left panel.

\section{Coordinate Systems}
\subsection{Solar Coordinates}
\label{SolarCoordinatesAppendix}
The location of the sun is computed with a publicly available C library \cite{PSAAlgorithm} 
developed by the Platforma Solar de Almer\'{i}a (PSA). 
For a given unixtime, latitude, and geographic longitude, 
the program returns the position of the sun (both azimuth and zenith angle) to within 0.5 arcminutes. 
The PSA algorithm is chosen for its simplicity of implementation while 
preserving sub-degree solar position accuracy for the years 1999-2015 \cite{BlancoMuriel2001431}.

The PSA method returns the sun location in a coordinate system where zero in 
azimuth is defined by the meridian of the observer, in this case, 
the longitude of the prototype station, which is $74.22 \degree$ W from 
Prime Meridian (rounded to the nearest hundredth). 
We first rotate this reported value to align with the geographic grid of the South Pole, 
where zero azimuth is defined by the Prime Meridian. 
This ``continent" coordinate system natively follows a \textit{clock-wise} 
convention where N(orth) = $0 \degree$ = Prime Meridian, E(ast) = $90 \degree$, S(outh) = $180 \degree$, and W(est)= $270 \degree$. We further transform coordinate systems to one which proceeds \textit{counter-clockwise} in the physics tradition, so E = $0 \degree$,  
N = $90 \degree$ = Prime Meridian, W = $180 \degree$, and S = $270 \degree$. 
We finally rescale the azimuth range from $[0,360]$ to $[-180,180]$ to 
match the output of the analysis interferometric map methods described in section \ref{MapAppendix}. 
The result is that the solar azimuth is reported as a counter-clockwise 
angle in the range of $[-180,180]$ in a coordinate system where E = $0 \degree$, N = $90 \degree$ = Prime Meridian, W = $180 \degree$, and S = $-90 \degree$.

\subsection{Interferometric Map Coordinates}
\label{MapAppendix}
The interferometric map measures azimuthal angles in a counter-clockwise 
coordinate system with zero azimuth corresponding to the direction of Antarctic continent ice-flow. 
At the location of the prototype station, this corresponds to $36.77 \degree$ W 
from Prime Meridian (rounded to the nearest hundredth). 
We rotate the map by this constant offset to align the  map with the 
same ``continent" coordinate system as used for solar positions in 
section \ref{SolarCoordinatesAppendix}, so that E = $0 \degree$, N = $90 \degree$ = Prime Meridian, 
W = $180 \degree$, and S = $-90 \degree$. 
The interferometry routines are already designed to report answers in the 
range of $[-180,180]$ as in \cite{Allison201562}, so unlike the PSA routine, a range adjustment is not required.

\subsection{A note On Convention Selection}
We note that the azimuthal ($\phi$) coordinate system utilized here 
(E = $0 \degree$, N = $90 \degree$ = Prime Meridian, W = $180 \degree$, and S = $-90 \degree$) 
is $90 \degree$ out of phase with the ISO 6709 definition of longitude. 
It does however agree with the methodology used by South Pole surveyors in determining coordinates for both ARA and IceCube.

\section{Other Hypotheses For Unique Reconstruction to the Sun}
\label{app:alt_hyp}
We concluded, as described in the main body of the paper, that the events
were consistent with bright thermal emission.
In this Appendix, we detail other studies we undertook to explicitly
rule out the hypotheses of CW emission, low signal-to-noise transients,
broad-band but non thermal emission produced by solar flares.

\subsection{Continuous Wave}

We looked for evidence of the emission being CW-like by 
looking for a strong peak or peaks at unique frequencies in the spectra.
Any impulse whose duration would be shorter than our waveforms of 
$\sim$250~ns in length when produced at the sun would be 
dispersed over several $\mu$s in the ionosphere, so we looked for a chirp-like behavior in the waveforms.

\begin{figure}[ht]
\centering
\includegraphics[width=\linewidth]{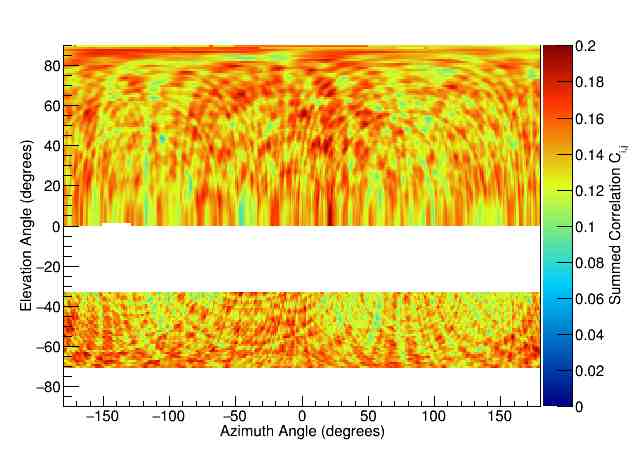}\hfill
\includegraphics[width=\linewidth]{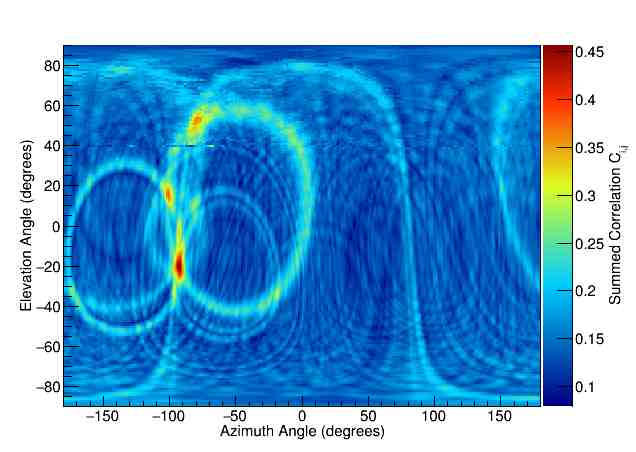}\hfill
\caption{\label{fig:reconstruction_cw_pulser} (Top) Reconstruction map from known continuous wave interference at $\sim$403~MHz (the radiosonde on South Pole weather balloons), showing peaks in the cross-correlation function  for time delays corresponding to many different directions. The behavior differs from the reconstruction maps seen for the flare as shown in Fig.~\ref{fig:reconstructions_flareevent_1}. The map was made assuming a source distance of 3000m, and the angles are all in the local Testbed coordinate system. (Bottom) Reconstruction map from an impulsive waveform from an in-ice calibration pulser. The map was made assuming a source distance of 30m, and the angles are all in the local Testbed coordinate system.}
\end{figure}

Continuous wave signal at frequencies higher than $\sim1/200 ns=5$~MHz will repeat over timescales less than $\sim200~ns$.
The ARA Testbed high-pass filters remove frequencies below 150~MHz, so the unique reconstruction to the sun that we are seeing cannot
be due to signals being at sufficiently low frequencies that their period is not contained in our waveforms.  Fig.~\ref{fig:reconstruction_cw_pulser}
shows a reconstruction map from continuous wave interference at $\sim$403~MHz (the radiosonde on South Pole weather balloons), where one can see peaks in the cross-correlation function corresponding
to directions all over the map.

\subsection{Transients}

Any signal that only begins, or ceases, to be observed at some time between 
the beginning and end of the waveforms would also give a unique reconstruction direction. 
For example, impulsive signals reconstruct cleanly, 
as demonstrated from a measured calibration pulser in Fig.~\ref{fig:reconstruction_cw_pulser}. 
However, any impulsive signal that might be produced by the flare 
would be dispersed as it exits the sun as well as in the earth's ionosphere.

\begin{figure}
\centering
\includegraphics[width=\linewidth]{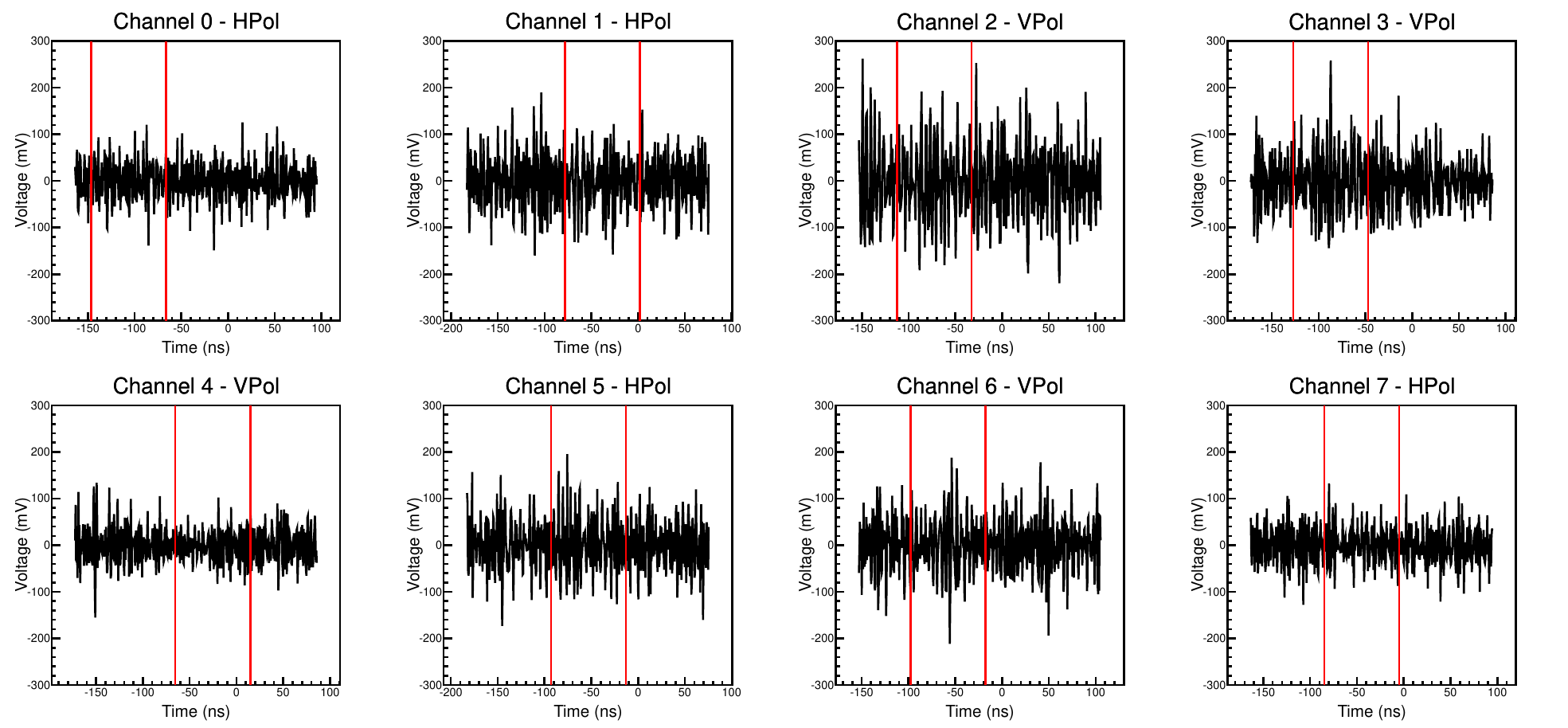}
\caption{The waveforms from an event that reconstructs to the Sun's direction. To test whether the reconstruction is caused by an impulsive signal, the impulsive portion is defined by identifying the largest $V_{peak}/$RMS values across all channels contributing to the map (i.e. VPol borehole antennas) and selecting an 80-ns window around that time. For other channels, the window is defined by shifting the window in the peak waveform by the delay consistent with the peak reconstruction direction. The ``impulsive" period is shown as the region between the red vertical lines.}
\label{fig:ImpulseRemovedWaveforms}
\end{figure}

\begin{figure}
\centering
\begin{subfigure}{0.4\textwidth}  \includegraphics[width=\linewidth]{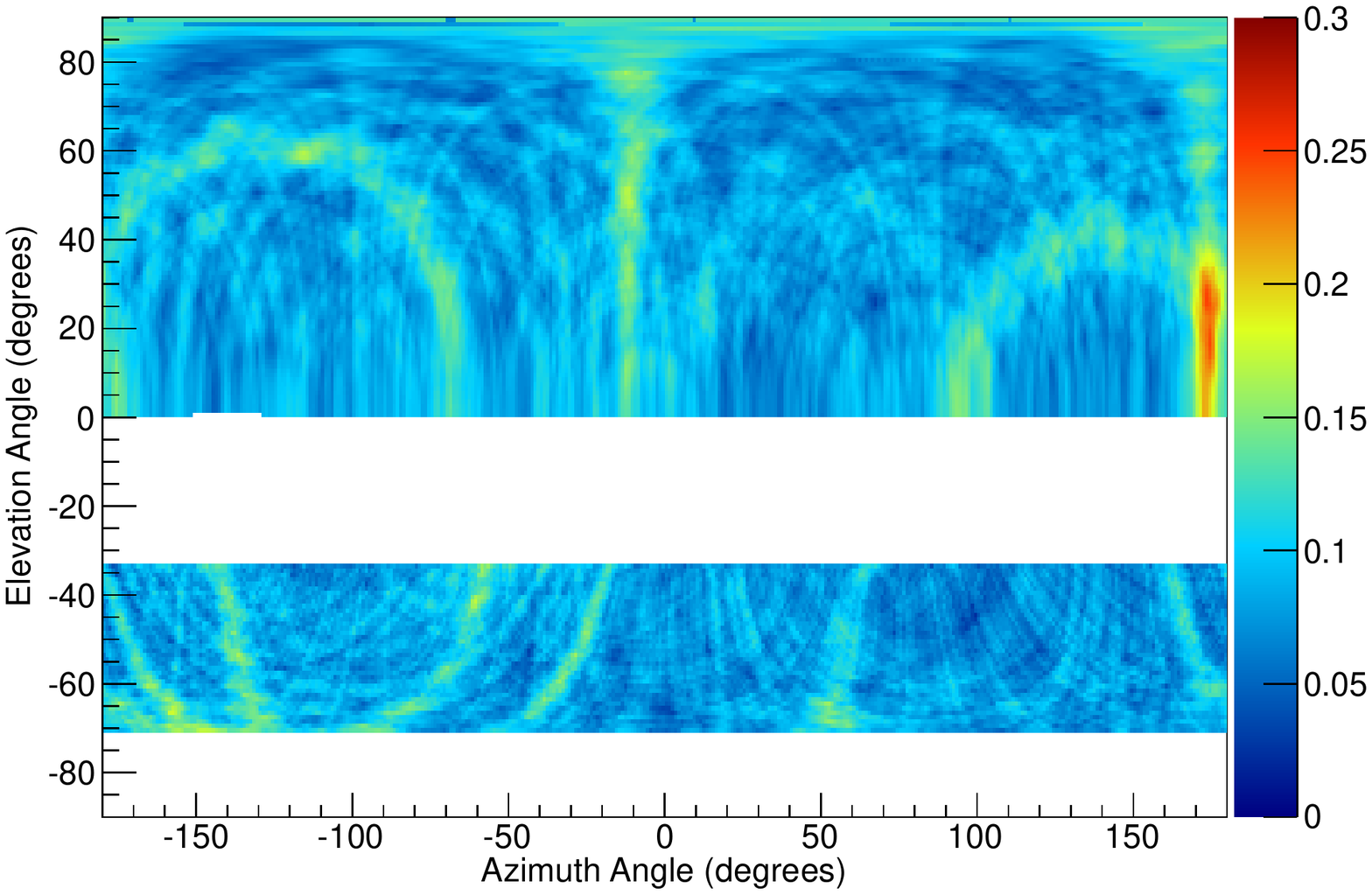}
\caption{Full waveform map}
\end{subfigure}
\hfill
\begin{subfigure}{0.4\textwidth}
\includegraphics[width=\linewidth]{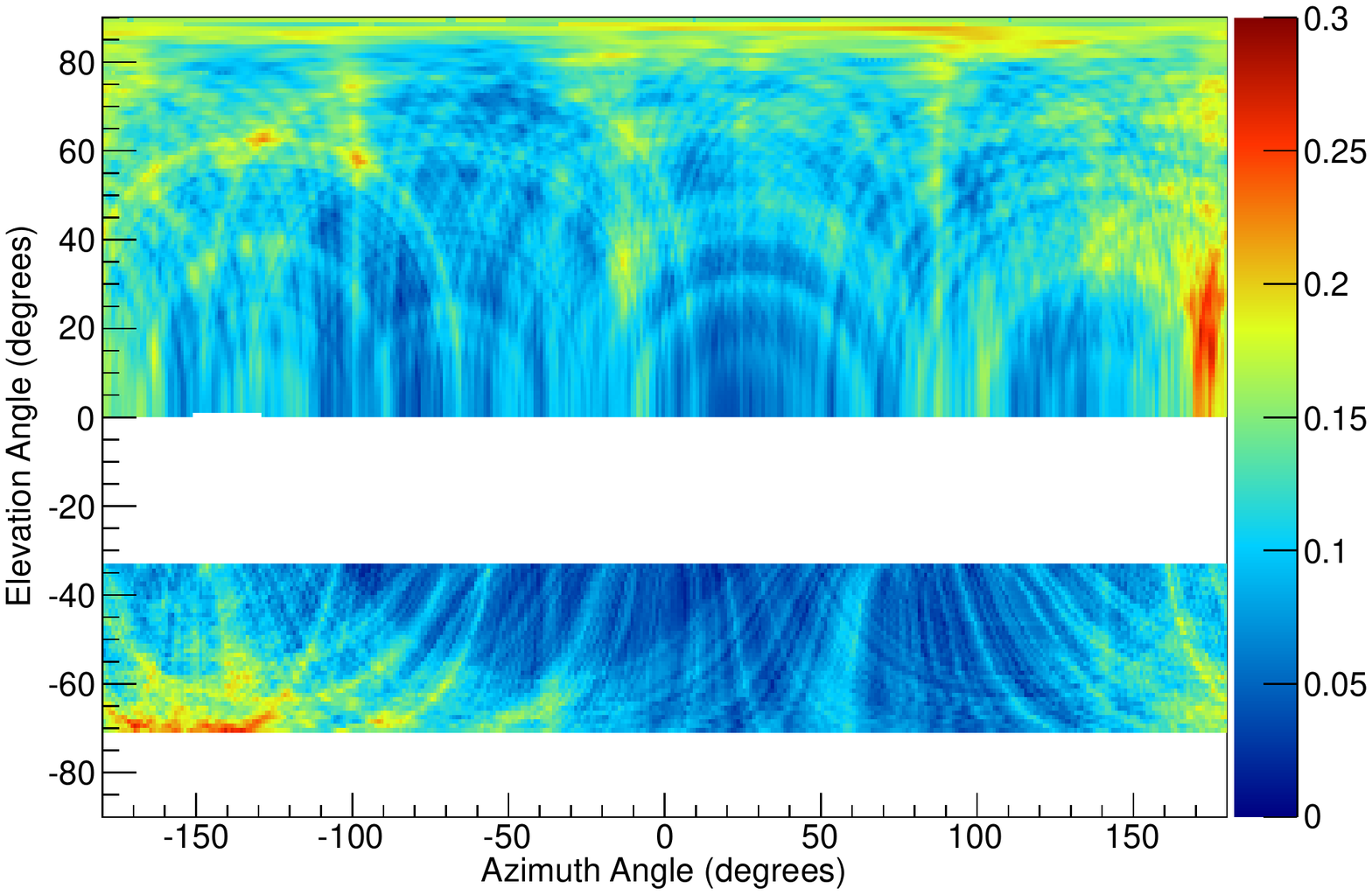}
\caption{``Before" map}
\end{subfigure}\hfill
\begin{subfigure}{0.4\textwidth}
\includegraphics[width=\linewidth]{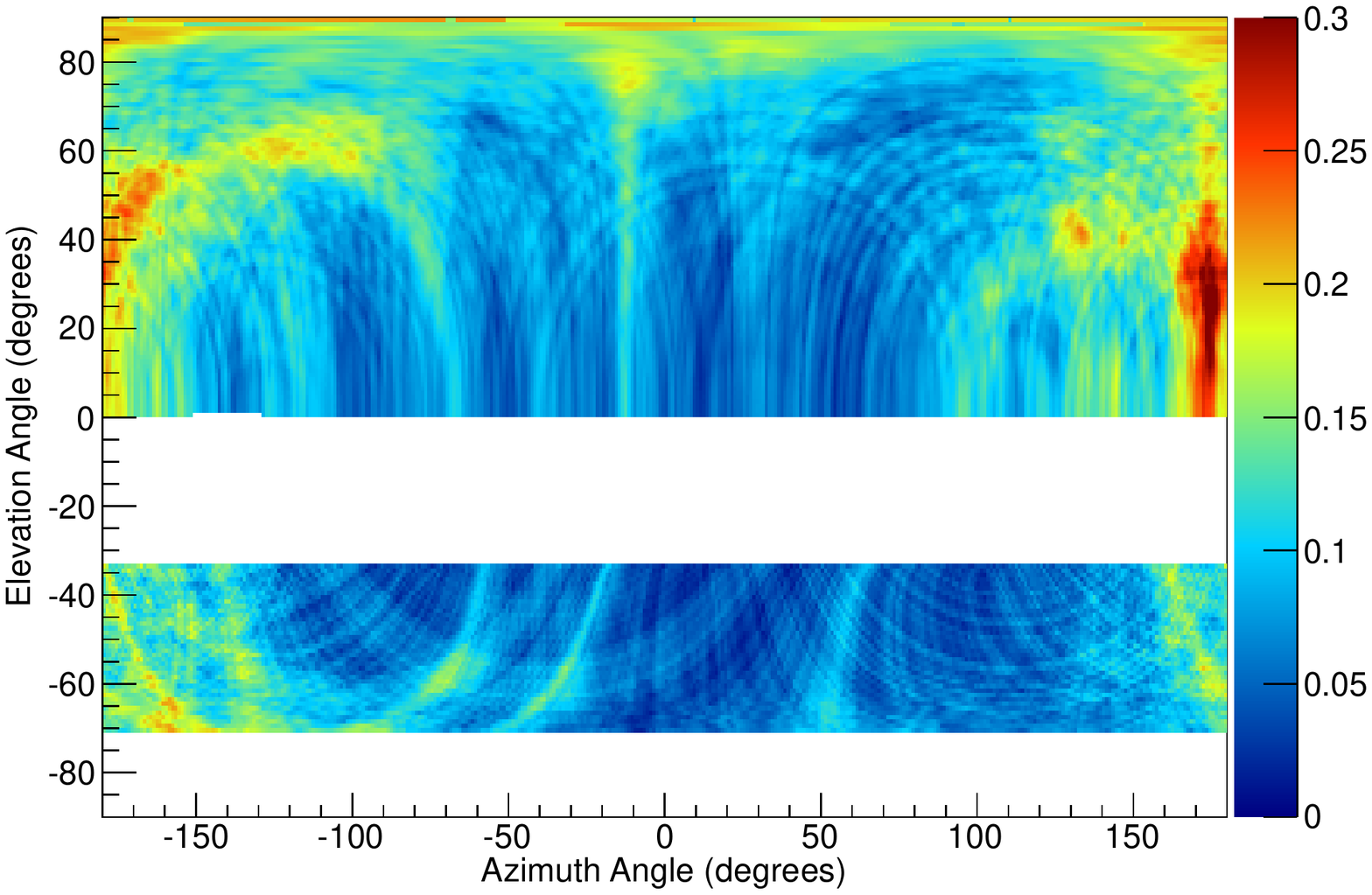}
\caption{``During" map}
\end{subfigure}\hfill
\begin{subfigure}{0.4\textwidth}
\includegraphics[width=\linewidth]{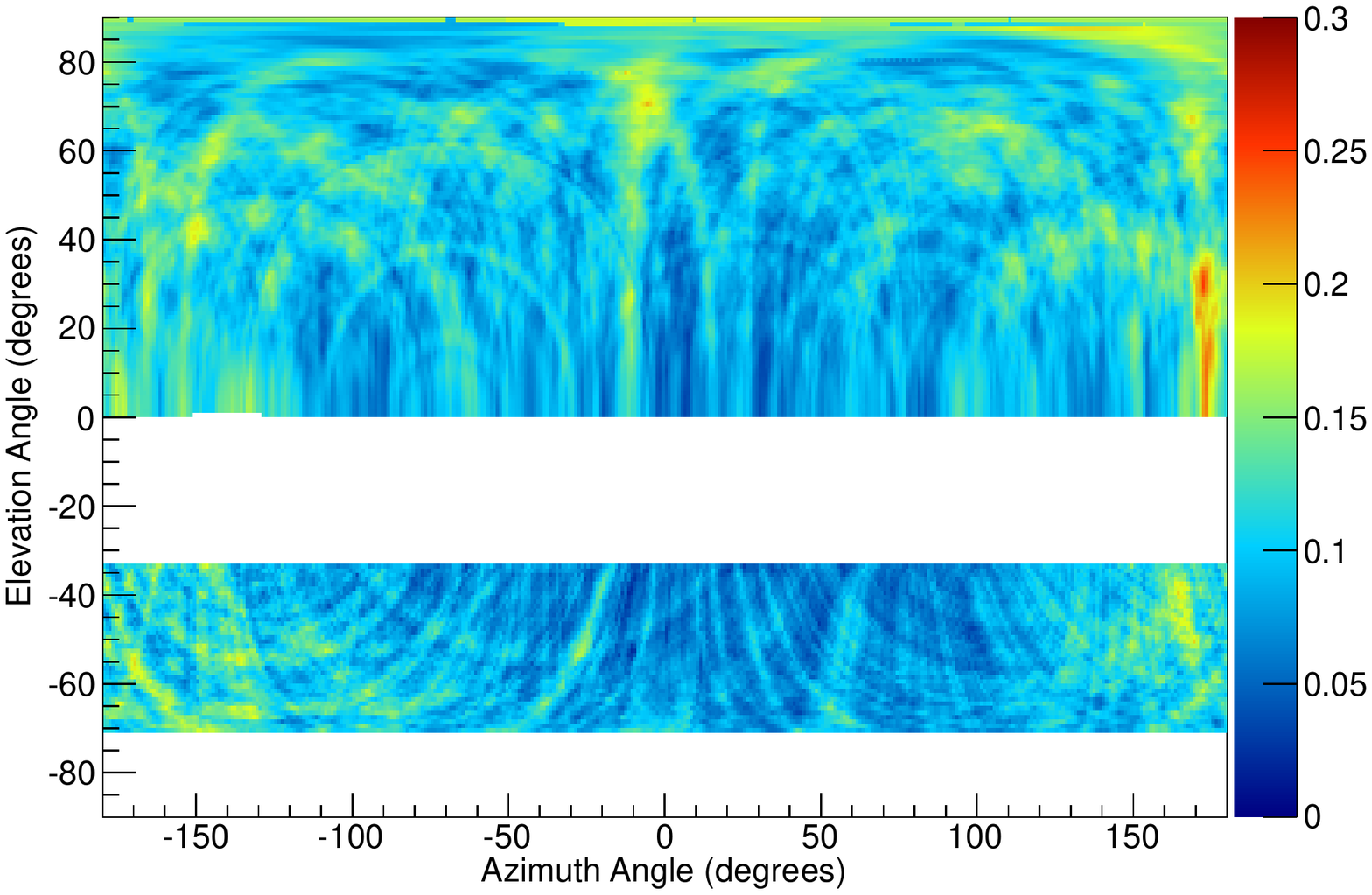}
\caption{``After" map}
\end{subfigure}
\caption{The reconstruction maps for the event shown in Fig. \ref{fig:ImpulseRemovedWaveforms} using (a) the full waveform, (b) only the ``before" section of the waveform, (c) only the ``during section, and (d) only the ``after" section. }
\label{fig:reconstruction_divided}
\end{figure}

To check whether the cause of the unique reconstruction of these events to the 
sun was some emission with start or end times observed in our waveforms, 
we considered the channel with the largest peak voltage compared to the RMS noise voltage, 
and divided up the waveforms in the event into three sections: an 80~ns window surrounding the time of the peak, and the periods before and after that window (See Fig.~\ref{fig:ImpulseRemovedWaveforms}).
The windows in the waveforms from each antenna are delayed to be 
consistent with the Sun's direction so that the windows in each waveform correspond with each other.
As seen in Fig.~\ref{fig:reconstruction_divided}, 
we find that the event reconstructs uniquely to the sun using any of the three time periods on their own. 
The same test applied to a calibration pulser event shows a strong 
reconstruction using only the ``impulsive" part of the waveform and no 
reconstructability to the calibration pulser's direction using the segments before and after the impulsive portion.
This suggests that the signal correlated with the solar flares is 
not impulsive in nature but rather that the signal correlating with the Sun's direction is spread across the waveform.

\subsection{Broadband, non thermal}

\subsubsection{Time-dependent spectral features expected from solar flares}

Some decimeter-wavelength (decimetric) radio emission associated with 
solar flares is due to excitations of the coronal plasma at the plasma frequency $f_p$ \cite{Kundu1965}. 
This plasma frequency depends on the square root of the electron density $n_e$, so that $f_p \propto \sqrt{n_e}$. 
The coronal electron density is in turn proportional to the inverse-square of the distance from the sun's center $R$ so that $n_e \propto 1/R^2$ \cite{1997ESASP.415..183R, WindWaves}. 
Therefore, as the electrons in the ejecta plasma are carried outward from the sun, 
the coronal density falls as $1/R^2$, leading to a chirp from higher to lower frequencies. 
This chirp is characterized by its so-called ``frequency drift rate'' $df/dt$, 
where $df/dt>0$ indicates the electrons are being ejected from the sun, and $df/dt<0$, or ``reverse drift'', indicates they are being dragged back in.

The drift rate is a measure of how quickly the density of the solar 
atmosphere changes from the perspective of an ejecting electron and divides 
flares with plasma emission into two categories \cite{Aschwanden}, 
type-II and type-III \footnote{The numbering on the flare types is arbitrary and historical. At the time of naming, type-II flares were ``slow-drift" bursts, and type-III were ``fast-drift bursts", referring to the time scale of the frequency drift. Type-I flares were ``noise-storms", type-IV were ``broadband-continuum emission," and type-V were ``continuum emission at meter wavelengths."}. 
The faster the ejecta, the more quickly the change in the plasma frequency, 
and the greater the magnitude of the frequency drift rate. Type-II 
flares are characterized by $df/dt$ on the order of $10^{-1}$ MHz/s, 
and are usually attributed to magneto-hydrodynamic shocks 
(shock waves in the solar plasma excited by the evolving magnetic fields) 
moving through the solar corona \cite{Wild1950}. 
Type-III flares drift significantly faster, upwards of $10^{3}$ MHz/s \cite{Melendez1999}, 
and are caused by electrons traveling outward along open field lines at speeds upward of $0.5c$.

\begin{figure}[ht]
\centering
\includegraphics[width=\linewidth]{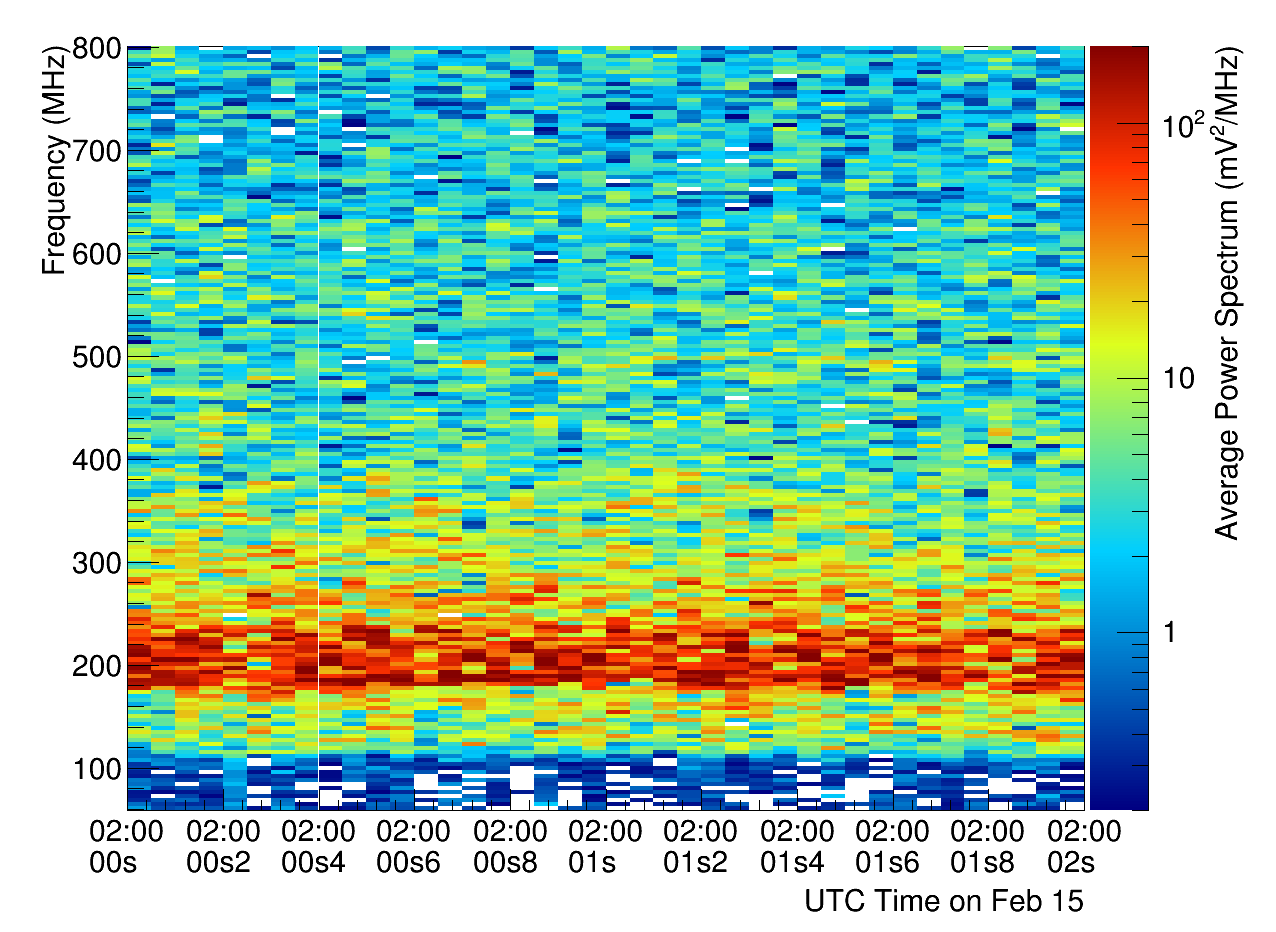}
\caption{A spectrogram produced over a two second-long scale using 48 sequential events during the flare period.  The events were recorded as RF triggers between 2:00:00 and 2:00:02 on Feb.~15$^{\rm{th}}$, 2011. }
\label{fig:second-scale_spectrogram}
\end{figure}

\subsubsection{Looking for features in spectrograms }
\label{sec:spectral_features_100ns}

The time scales of the chirps from both types of flares are orders of 
magnitude too long to be observable on our individual waveforms of length $\sim 250$ ns, 
and cannot explain the correlation that we are seeing between waveforms on an event by event basis. 
In our bandwidth (approximately 125-850 MHz), the fastest drift rates are those associated with type-III flares. 
Melendez \textit{et. al.} \cite{Melendez1999} suggest that for our center frequency of 350 MHz, $df/dt \sim 700$ MHz/s, 
which would generate roughly 170 Hz of drift over the waveform. 
Such a chirp should only be visible on a timescale of $\sim 1/ 170 \mathrm{Hz} ~\sim 6$ ms, which is over 10,000 times longer than our waveform. 

Impulsive events with time-scales of nanoseconds would be 
dispersed in the ionosphere due to the electron plasma frequency 
resonance in the 10-100 MHz range
--as has been observed by other impulsive radio experiments like FORTE \cite{Jacobson1999}.
The ionospheric plasma frequency is even greater during active solar periods 
thus producing an even greater dispersive effect with group delays potentially reaching 1$\mu$s at 300 MHz.
Therefore, if we were seeing the effect of impulsive signals that were of $\sim$ns or 10's of ns duration 
when produced at the sun, one may expect to find a chirp on the scale of 10's to 100's of nanoseconds.
Because of the 25-ms deadtime between triggered events, $\mu$s-scale and ms-scale spectrograms cannot be obtained.
However, we can still check for features in spectrograms from a single
event over a 250~ns time scale, and using $\sim$ 50~events we can produce spectrograms over timescales
of order a second.

Using $\sim$~50 events, we can produce second-scale spectrograms, an example of which can be seen in Fig. \ref{fig:second-scale_spectrogram}.
No clear pattern emerges from these spectrograms.

By examining subsections of each event, we produce event spectrograms,which demonstrate 
no clear chirp-like behavior on the time-scale of the event length ($\sim$250 ns).
We divide each waveform into many segments and calculate the spectrum for each segment.
Although the frequency resolution is made coarser by this 
process than when analyzing the full waveform, 
it allows the examination of the power spectrum as a function on time during the period covered by the waveform.

\begin{figure}
\centering
\begin{subfigure}{0.47\textwidth}
\includegraphics[width=\linewidth]{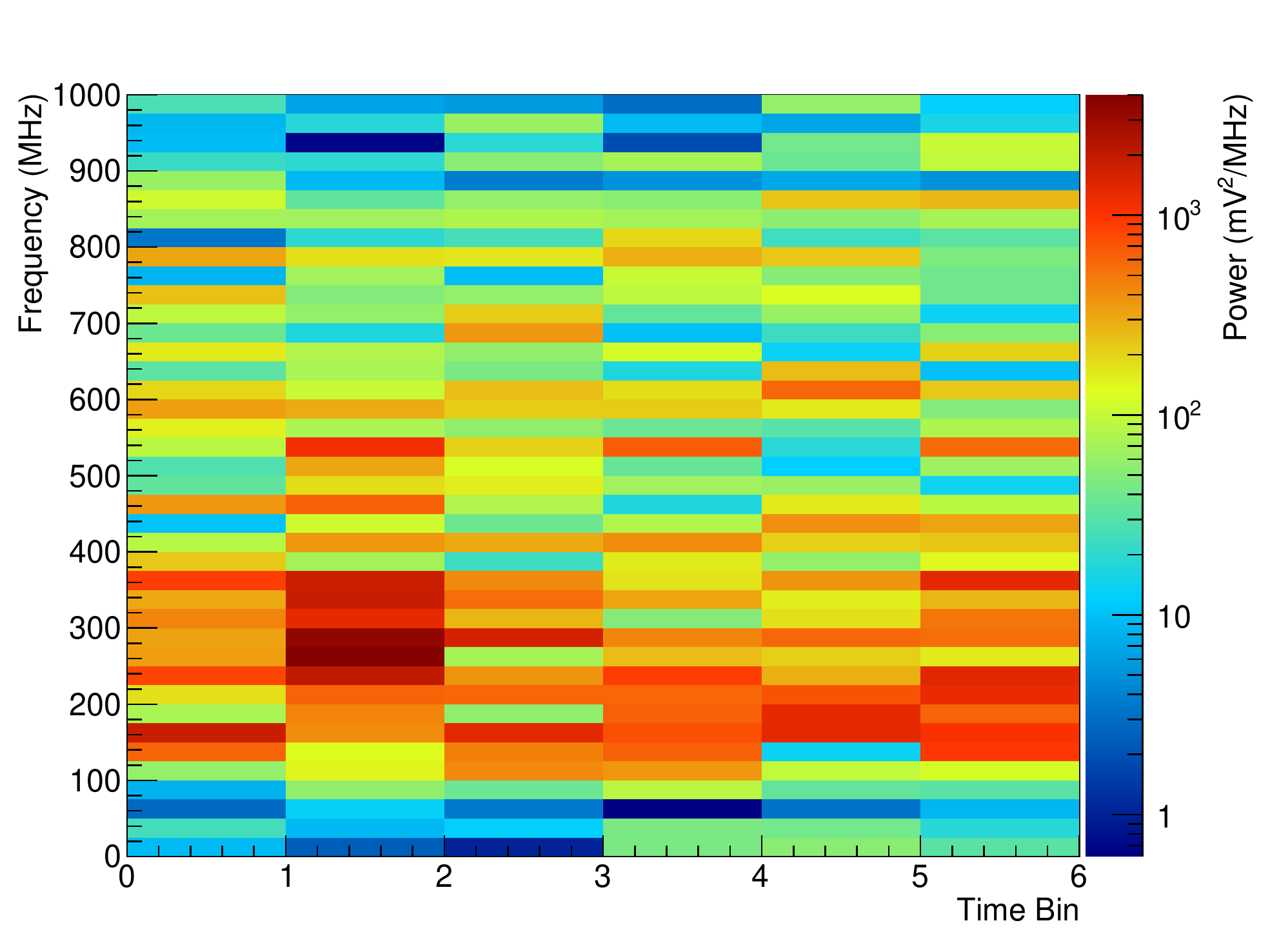}
\caption{Spectrogram with independent time bins}
\end{subfigure}
\hfill
\begin{subfigure}{0.47\textwidth}
\includegraphics[width=\linewidth]{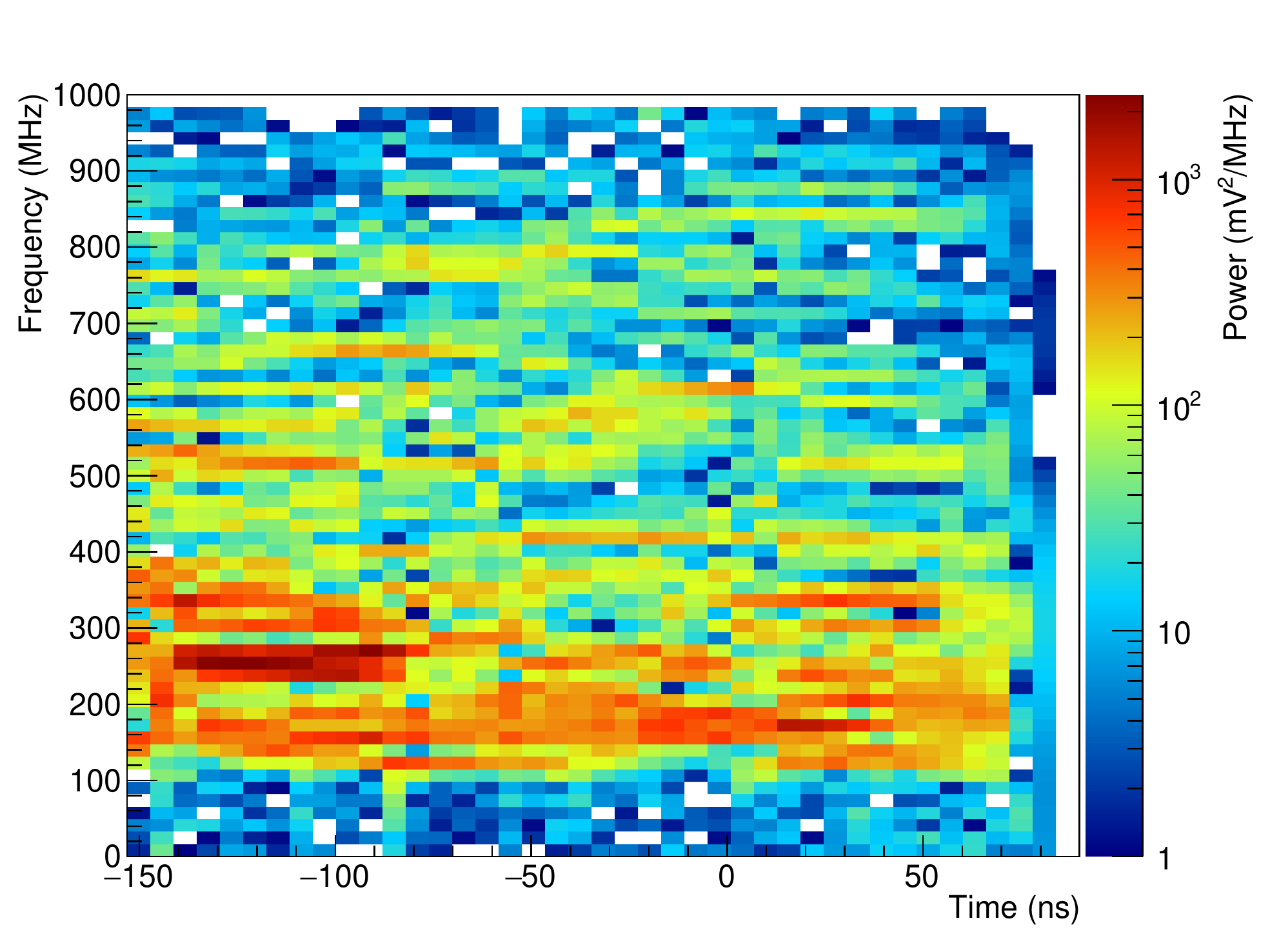}
\caption{Spectrogram from running sub-samples of the waveform}
\end{subfigure}
\caption{Spectrogram created from a single waveform, measured in Channel 2, from the same event as shown in Figs.~\ref{fig:wf_flareevent_1}, ~\ref{fig:sp_flareevent_1} and~\ref{fig:reconstructions_flareevent_1}Figure (a) shows the spectrogram from 6 independent sub-samples of the waveform. Figure (b) shows the spectrogram taken from a running sub-window of the waveform. Although each sample is correlated with its neighbors in time, this approach yields finer frequency resolution due to the greater number of samples allowed.}  \label{fig:singlewaveform_spectrograms_flareevent_1}
\end{figure}

We confirm the absence of single-waveform chirp behavior by plotting spectrograms for single waveforms as in Fig.~\ref{fig:singlewaveform_spectrograms_flareevent_1}.

\section{Other solar flares during the ARA livetime}
\label{app:otherflares}
In order for ARA to observe radio emission from a solar flare during the livetime of the instrument, 
the sun should be up in Antarctica (from mid-September to mid-March only, 
and 24 hours per day during that period), the flare must face the earth.  
Table~\ref{tab:flares_summary} summarizes other X-class flares that 
satisfied the above criteria for the ARA Testbed and three deep stations from 
2011 to 2013 using the RHESSI flare catalog.

\begin{table}
\centering
\begin{tabular}{c c c}
\hline \hline
Date & Time & Class\\
\hline
2011/02/15 & 01:43:44 & X2.2 \\
2011/03/09 & 23:10:28 & X1.5 \\
2011/09/22 & 10:53:16, 11:06:40, 11:36:04 & X1.4 \\
2011/09/24 & 09:19:28, 9:30:00 & X1.9 \\
2012/03/05 & 02:41:48, 03:12:20, 03:16:28, 03:56:04 & X1.1 \\
2012/10/23 & 03:14:36 & X1.8 \\
2013/10/25 & 07:52:44 & X1.7 \\
2013/10/25 & 14:56:16 & X2.1 \\
2013/10/28 & 01:45:24, 01:51:44 & X1.0 \\
\hline
\end{tabular}
\caption{Table summarizing X-class flares during the livetime of the ARA Testbed and three deep stations and the periods when the Sun is above the horizon at the South Pole. Flares with multiple times indicate an active period where X-ray emission rose and fell repeatedly within a short time frame but the X-ray emission is integrated over those sub-flares. This list of flares was extracted from the RHESSI catalog \cite{RHESSI}.}
\label{tab:flares_summary}
\end{table}

\subsection{The Feb.~13$^{\rm{th}}$ solar flare reported by ARA}
The ARA team observed emission from an M-class flare in the ARA Testbed that occurred on Feb.~13$^{\rm{th}}$, 2011,
just two days before the flare that was the subject of this paper, and  reported it in a paper about the
performance of the Testbed instrument in 2011~\cite{Allison2011wk}.
This Feb.~13$^{\rm{th}}$ flare  was sought by ARA analyzers after the Green Bank Solar Radio Burst
Spectrometer detected a series of strong type-II solar radio
bursts in the 10 MHz-1 GHz band~\cite{Cho}.
ARA reported a spectrogram from the Feb.~13$^{\rm{th}}$ flare with many features similar to ones in the spectrogram report by Green Bank, demonstrating that 
the ARA Testbed system was sensitive to low-level radio emission, and also that the noise
environment of the South Pole ARA site is sufficiently quiet to allow for such a detection.

\begin{figure}
\centering
\includegraphics[width=\linewidth]{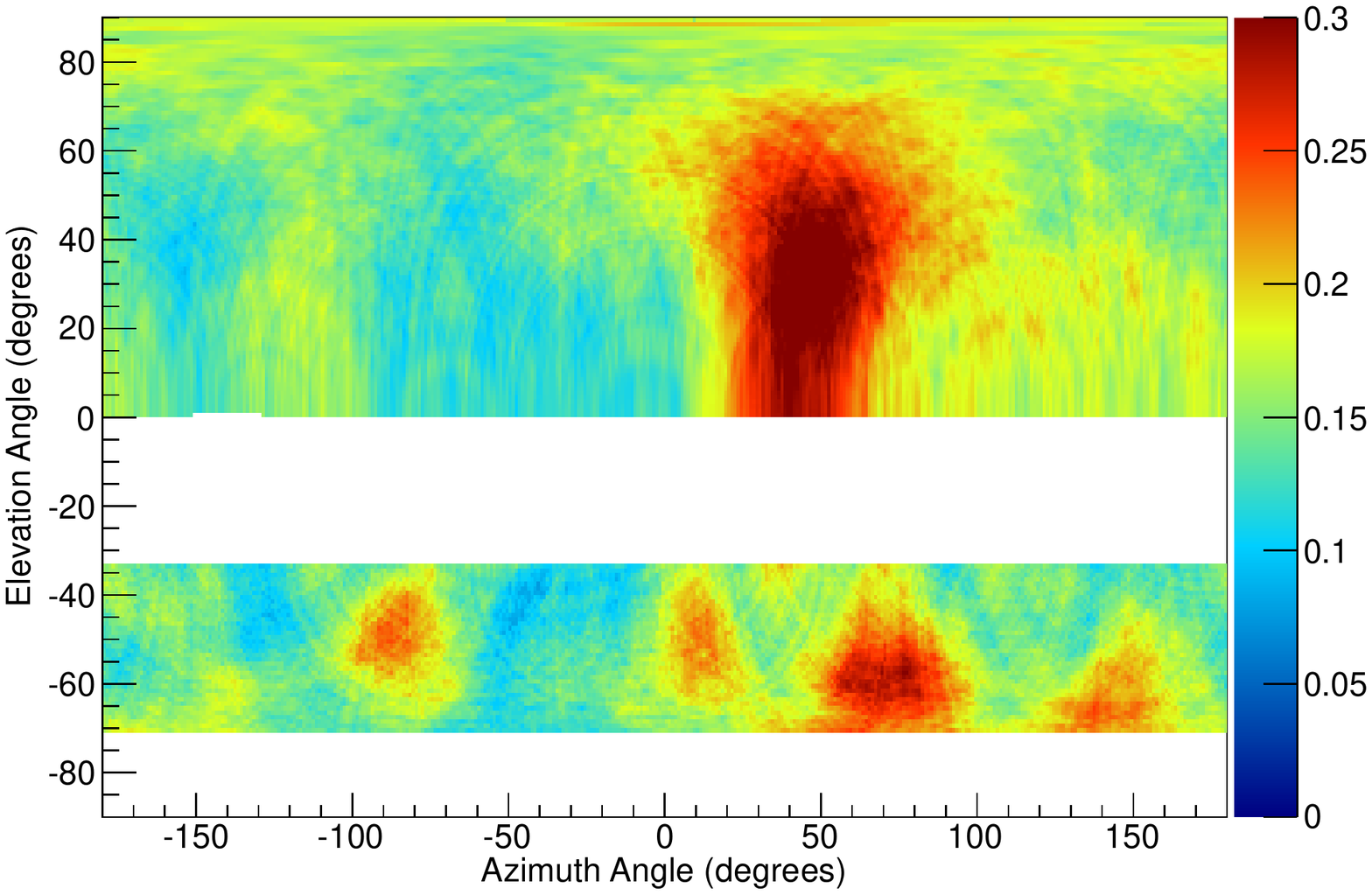}
\caption{Reconstruction map for a typical event during the time of the Feb.~13$^{\rm{th}}$ flare.
This map shows a broader range of directions that give cross-correlation values similar to the peak value, compared to the narrow peak typical of an event from the Feb.~15$^{\rm{th}}$ flare.}
\label{fig:reconstruction_feb13}
\end{figure}

The Feb.~13$^{\rm{th}}$ and Feb.~15$^{\rm{th}}$ flares had many differing characteristics 
that account for the different circumstances under which ARA analyzers encountered the two events.
The emission from the Feb.~13$^{\rm{th}}$ flare peaked at frequencies below about 150~MHz, 
and the ARA Testbed search for diffuse neutrinos rejected events with more than 10\% of the power 
accounted for below 150~MHz.  This cut was made to reduce anthropogenic background noise. 
In addition, the Feb.~13$^{\rm{th}}$ flare did not produce any events that survived the ``Reconstruction
Quality Cut'' imposed in the neutrino search, 
which required that the region in the sky that gives a high cross-correlation be narrow and unique. 
Fig.~\ref{fig:reconstruction_feb13} shows a typical event during the time of the Feb.~13$^{\rm{th}}$ flare. 
This map shows a broader range of directions that give cross-correlation values similar to the peak value, 
compared to the narrow peak typical of an event from the Feb.~15$^{\rm{th}}$ flare.
Additionally, the results shown in the previously published work on the 
Feb.~13$^{\rm{th}}$ flare~\cite{Allison2011wk} only used the surface antennas 
which operate in a lower frequency band (30-300 MHz) than the borehole antennas used in this work (150-850 MHz). 

The solar flare events from Feb. 13$^{\rm{th}}$ failed to pass the ``Reconstruction
Quality Cut," which requires that the reconstruction map of a ``well-reconstructed" event be both well-defined and unique.
The key parameter of the cut is the area of the 85\% contour around the absolute peak of the map.
The conditions of the cut are two-fold. First, this area must be less than 50 square degrees in magnitude. 
This indicates that the reconstruction is ``well-defined," and does not allow for a broad range of possible reconstruction locations within the peak contour.
Second, the area around this peak must be no less than two-thirds of the total area on the map with a value of 85\% of the peak or greater.
This condition forces the event to be ``unique," in that no other location on the sky reconstructs as well.
A fuller description of the requirement can be found in \cite{Allison201562}.

\end{appendices}
\end{document}